\pdfoutput=1

\documentclass[11pt,twoside,a4paper,cmspaper,final,collab]{cms-tdr}
\def\svnVersion{372314}\def\cmsCernNoTag{CERN-EP/2016-094}\def\cmsCernDate{\today}\def\cmsMessage{Published in the Journal of High Energy Physics as \href{http://dx.doi.org/10.1007/JHEP10(2016)129}{\doi{10.1007/JHEP10(2016)129.}}}

\begin{document}\cmsNoteHeader{SUS-15-010}

\hyphenation{had-ron-i-za-tion}
\hyphenation{cal-or-i-me-ter}
\hyphenation{de-vices}

\RCS$Revision: 372280 $
\RCS$HeadURL: svn+ssh://svn.cern.ch/reps/tdr2/papers/SUS-15-010/trunk/SUS-15-010.tex $
\RCS$Id: SUS-15-010.tex 372280 2016-10-29 20:28:17Z sbein $
\newlength\cmsFigWidth
\ifthenelse{\boolean{cms@external}}{\setlength\cmsFigWidth{0.98\columnwidth}}{\setlength\cmsFigWidth{0.75\textwidth}}
\ifthenelse{\boolean{cms@external}}{\providecommand{\cmsLeft}{top}}{\providecommand{\cmsLeft}{left}}
\ifthenelse{\boolean{cms@external}}{\providecommand{\cmsRight}{bottom}}{\providecommand{\cmsRight}{right}}
\newcommand{\MHT}{\ensuremath{H_\mathrm{T}^\text{miss}}\xspace}
\newcommand{\MTtwo}{\ensuremath{M_\mathrm{T2}}\xspace}
\newcommand{\stl}{\ensuremath{\PSQt_1}\xspace}
\newcommand{\scL}{\ensuremath{\PSQc_\mathrm{L}}\xspace}
\newcommand{\scR}{\ensuremath{\PSQc_\mathrm{R}}\xspace}
\newcommand{\stauOne}{\ensuremath{\PSgt_1}\xspace}
\newcommand{\smuL}{\ensuremath{\PSGm_\mathrm{L}}\xspace}
\newcommand{\preCMS}{non-DCS\xspace}
\newcommand{\mCMS}{\textrm{CMS}}
\newcommand{\mpreCMS}{{\textrm{non-DCS}}}
\newcommand{\T}{\rule{0pt}{2.5ex}}
\newcommand{\B}{\rule[-1.1ex]{0pt}{0pt}}

\cmsNoteHeader{SUS-13-020}
\title{Phenomenological MSSM interpretation of CMS searches in pp collisions at $\sqrt{s}=7$ and 8\TeV}

\date{\today}

\abstract{
Searches for new physics by the CMS collaboration are interpreted in the framework of the phenomenological minimal supersymmetric standard model (pMSSM). The data samples used in this study were collected at $\sqrt{s} = 7$ and 8\TeV and have integrated luminosities of 5.0\fbinv and 19.5\fbinv, respectively. A global Bayesian analysis is performed, incorporating results from a broad range of CMS supersymmetry searches, as well as constraints from other experiments. Because the pMSSM incorporates several well-motivated assumptions that reduce the 120 parameters of the MSSM to just 19 parameters defined at the electroweak scale, it is possible to assess the results of the study in a relatively straightforward way. Approximately half of the model points in a potentially accessible subspace of the pMSSM are excluded, including all pMSSM model points with a gluino mass below 500\GeV, as well as models with a squark mass less than 300\GeV. Models with chargino and neutralino masses below 200\GeV are disfavored, but no mass range of model points can be ruled out based on the analyses considered. The nonexcluded regions in the pMSSM parameter space are characterized in terms of physical processes and key observables, and implications for future searches are discussed.
}

\hypersetup{%
pdfauthor={CMS Collaboration},%
pdftitle={Phenomenological MSSM interpretation of CMS searches in pp collisions at sqrt(s) = 7 and 8 TeV},%
pdfsubject={CMS},%
pdfkeywords={CMS, physics, supersymmetry, pMSSM}}

\maketitle

\section{Introduction}
\label{sec:intro}

Supersymmetry (SUSY)~\cite{PhysRevD.3.2415,Golfand:1971iw,Volkov:1972jx,Wess:1974tw,Fayet:1974pd,Chung:2003fi} is a strongly
motivated candidate for physics beyond the standard model (SM).
Searches for the superpartner particles (sparticles) predicted by SUSY performed in a variety of channels at the CERN LHC
at $\sqrt{s} = 7$ and $8\TeV$ have been reported \cite{Aad:2015iea,SUS12011,SUS12003,SUS12006,SUS13012,Chatrchyan:2012jx,SUS12024,Khachatryan:2014rra,Khachatryan:2015wza,Khachatryan:2015lwa,SUS13013,SUS13006}.  The results,
found to be consistent with the SM, are interpreted as limits on SUSY
parameters, based mostly on models with restricted degrees of freedom,
such as the constrained minimal supersymmetric standard model
(cMSSM)~\cite{Chamseddine:1982jx,Barbieri:1982eh,Ibanez:1982ee,Hall:1983iz,Nath:2003zs,Kane:1993td,Baer:1994nc},
or, more recently, within the simplified model spectra (SMS)
approach~\cite{Alwall:2008ag,Alves:2011wf,Chatrchyan:2013sza}.
The cMSSM models feature specific relations among the soft-breaking terms at some mediation scale that translate into specific mass patterns typical for the model.
While this problem is avoided in the SMS approach, the signatures of realistic models cannot always be fully covered by SMS topologies. This holds true, for instance, in the case of long decay chains that do not correspond to any SMS, $t$-channel exchanges of virtual sparticles in production, or the presence of multiple production modes that overlap in kinematic distributions.

In the work reported here, data taken with the CMS experiment at the LHC are revisited with an alternative
approach that is designed to assess more generally the coverage of
SUSY parameter space provided by these searches.  The method is based on the minimal
supersymmetric standard model (MSSM) and combines several search
channels and external constraints.  Given the large diversity of decay modes leading to
multiple signatures, the potential benefit of such a combined limit is
to exclude parameter regions that would otherwise be allowed when
considering each analysis separately.

Specifically, we interpret the CMS results in terms of the
phenomenological MSSM (pMSSM)~\cite{Djouadi:1998di}, a 19-dimensional
parametrization of the $R$-parity conserving, weak-scale MSSM that
captures most of the latter's phenomenological features.
Here, $R$-parity is a $\mathbb{Z}_2$ symmetry ensuring the conservation
of lepton and baryon numbers \cite{Farrar:1982te}, which suppresses proton
decay and results in the lightest SUSY particle (LSP) being stable. In
the pMSSM, all MSSM parameters are specified at the electroweak (EW)
scale, and allowed to vary freely, subject to the requirement that the
model remain consistent with EW symmetry breaking (EWSB)
and other basic constraints.
Since the pMSSM incorporates neither relations among SUSY-breaking terms
at a high scale, nor large correlations among sparticle masses from
renormalization group evolution, it allows a much broader set of
scenarios than those in, for example,  the cMSSM and related grand
unified theories (GUTs). Many of these scenarios
are difficult to constrain using current LHC data, despite some having small sparticle masses.

To assess how the data obtained by CMS impact SUSY in the context of the pMSSM,
we use a representative subset of the results based on data corresponding to  integrated
luminosities of 5.0\fbinv at 7\TeV and  19.5\fbinv at 8\TeV.
We use results from hadronic searches, both general searches and those
targeting top squark production; also included are searches with
leptonic final states, both general and EW-targeted. For a selected set of pMSSM parameter points, event samples were simulated using the CMS fast detector simulation \cite{fastsim} and analyzed. Since the fast detector simulation does not accurately model the detector response to massive long-lived charged particles, and since it was not feasible to use the CMS full simulation \cite{Banerjee:2011zzc} given the large number of model points,
we work within a subspace of the pMSSM in which the chargino proper decay lifetime $c \tau(\PSGcpm)$ is less than 10\unit{mm}. This constraint restricts the class of final states considered to those with prompt decays.
The 7 and 8\TeV data are treated consistently; in particular, we use the same set of points in the pMSSM model phase space, chosen randomly from a larger set of points that are consistent with pre-LHC experimental
results and basic theoretical constraints.
This approach greatly facilitates the combination of the results from the 7 and 8\TeV (Run 1) data.

The statistical analysis follows closely the Bayesian approach of Refs. \cite{Bayes:1, Bayes:2}. The work is an extension of
Ref.~\cite{Sekmen:2011cz}, which interpreted three independent CMS
analyses based on an integrated luminosity of about 1\fbinv of
data~\cite{Chatrchyan:2011zy,Chatrchyan:2011gqa,Chatrchyan:2012te} in
terms of the pMSSM, confirming that the approach is both feasible and
more successful in yielding general
conclusions about SUSY than those based on constrained SUSY models.
Furthermore, the diversity of phenomena covered by the pMSSM is also helpful
in suggesting new approaches to searches for SUSY at the LHC. A similar study has been performed by the ATLAS experiment~\cite{Aad:2015baa}.

The paper is organized as follows.
The definition of the pMSSM is presented in Section~\ref{sec:pmssm}.
Section~\ref{sec:anl} describes the analysis,
which includes the construction of a statistical prior for the pMSSM model
and the calculation of likelihoods for the CMS searches.
The results of this study are presented in Section~\ref{sec:results},
including discussions of the impact of the Run 1 CMS searches and
their current sensitivity to the pMSSM. Section~\ref{sec:unexplored} discusses nonexcluded pMSSM phase space. A summary of the results is given in Section~\ref{sec:concl}.

\section{Definition of the phenomenological MSSM}
\label{sec:pmssm}

The weak-scale $R$-parity conserving MSSM~\cite{Djouadi:1998di} has 120 free parameters, assuming the gravitino is heavy.
This is clearly too large a parameter space for any phenomenological study.
However, most of these parameters are associated with CP-violating phases and/or flavor changing neutral currents (FCNC), which are severely constrained by experiment.  Therefore, a few reasonable assumptions about the flavor and CP structure allow a factor of six reduction in the number of free parameters, without imposing any specific SUSY
breaking mechanism. This has the virtue of avoiding relations, which need not hold in general, between the soft terms introduced
by models of SUSY breaking.

Strong constraints on CP violation are satisfied by taking all parameters to be real, and
FCNC constraints are satisfied by taking all sfermion mass matrices and trilinear couplings to be
diagonal in flavor. Moreover, the first two generations of sfermions are assumed to be degenerate.
The trilinear $A$-terms of the first two generations give rise to amplitudes that are proportional to very small Yukawa couplings and are thus not experimentally relevant. Only the third generation
parameters $A_{\text{t}}$, $A_{\text{b}}$, and $A_\tau$ have consequences that are potentially observable.

This leaves 19 real weak-scale SUSY Lagrangian parameters that define the pMSSM~\cite{Djouadi:1998di}. As noted above, the pMSSM captures most of the phenomenological
features of the $R$-parity conserving MSSM and, most importantly, encompasses and goes beyond a broad
range of more constrained SUSY models.
In addition to the SM parameters, the free parameters of the pMSSM are:
\begin{itemize}
   \item three independent gaugino mass parameters $M_1$, $M_2$, and $M_3$,
   \item the ratio of the Higgs vacuum expectation values $\tan\beta=v_2/v_1$,
   \item the higgsino mass parameter $\mu$ and the pseudoscalar Higgs boson mass $m_A$,
    \item 10 independent sfermion mass parameters $m_{\tilde{\text{F}}}$, where
         $\tilde{\text{F}} = \tilde{\text{Q}}_1$, $\tilde{\text{U}}_1$, $\tilde{\text{D}}_1$,
                      $\tilde{\text{L}}_1$, $\tilde{\text{E}}_1$,
                      $\tilde{\text{Q}}_3$,  $\tilde{\text{U}}_3$, $\tilde{\text{D}}_3$,
                      $\tilde{\text{L}}_3$,  $\tilde{\text{E}}_3$
(for the 2nd generation we take $m_{\tilde{\text{Q}}_2}\equiv m_{\tilde{\text{Q}}_1}$,  $m_{\tilde{\text{L}}_2}\equiv m_{\tilde{\text{L}}_1}$,
           $m_{\tilde{\text{U}}_2}\equiv m_{\tilde{\text{U}}_1}$, $m_{\tilde{\text{D}}_2}\equiv m_{\tilde{\text{D}}_1}$, and $m_{\tilde{\text{E}}_2}\equiv m_{\tilde{\text{E}}_1}$; left-handed up- and down-type squarks are by construction mass degenerate), and
   \item the trilinear couplings $A_{\text{t}}$, $A_{\text{b}}$ and $A_\tau$.
\end{itemize}

To minimize theoretical uncertainties in the Higgs sector, these parameters are conveniently defined
at a scale equal to the geometric mean of the top squark masses, $M_\mathrm{SUSY}\equiv \sqrt{m_{\tilde{\text{t}}_1}m_{\tilde{\text{t}}_2}}$, often also referred to as the EWSB scale.

The pMSSM parameter space is constrained by a number of theoretical requirements.
First, the sparticle spectrum must be free of tachyons (particles with negative physical mass) and cannot lead to color or charge breaking minima in the scalar potential.
We also require that
EWSB be consistent and that the Higgs potential be bounded from below.
Finally, in this study, we also require that the lightest SUSY particle (LSP) be the lightest neutralino, $\PSGczDo$.
These requirements yield a model that is an excellent proxy for the full MSSM
with few enough parameters that an extensive exploration is possible.

It is of interest to note the generic properties of sparticle mass spectra of the pMSSM.
By definition, each first generation sfermion is exactly degenerate in mass with the corresponding second generation sfermion.
Other generic properties of pMSSM mass spectra are actually MSSM properties;
in the first and second generations, spartners of left-handed down-type quarks
are nearly mass-degenerate with the corresponding up-type squarks.
Likewise, first and second generation spartners of left-handed charged leptons
are nearly degenerate with the corresponding sneutrinos.
The nature of the spectrum of neutralinos and charginos depends on the relative magnitudes and separation of the pMSSM parameters $M_1$, $M_2$ and $\mu$.  If these scales are well separated, then the approximate eigenstates will divide into a single bino-like state with mass of order $M_1$, a wino-like triplet consisting of two charginos and one neutralino with masses of order $M_2$, and a higgsino-like quartet of two charginos and two neutralinos with masses of order $\mu$.  The LSP will then be primarily composed of the neutral member(s) of the lightest of these three.  If the parameters above are not well separated, then the LSP will be a mixture of the neutral states.

\section{Analysis}
\label{sec:anl}
The purpose of this work is to assess how the current data constrain the MSSM using the more tractable pMSSM as a proxy.  We use the results from several CMS analyses, which cover a
variety of final states, to construct posterior densities of model
parameters, masses, and observables. The posterior density of the
model parameters, which are denoted by $\theta$, is given by
\begin{equation}
	p(\theta|D^\textrm{CMS}) \propto L(D^\textrm{CMS} | \theta ) \, p^\textrm{non-DCS}(\theta),
	\label{eq:posterior}
\end{equation}
where $D^\textrm{CMS}$ denotes the data analyzed by the direct CMS
SUSY searches, $L(D^\textrm{CMS} | \theta )$ is the associated CMS
likelihood that
incorporates the impact of these direct CMS searches, and
$p^\textrm{non-DCS}(\theta)$ is the prior density constructed from
results not based on direct CMS SUSY searches (non-DCS results).
The posterior density for an observable $\lambda$ is obtained as follows,
\begin{equation}
	p(\lambda|D^\textrm{CMS}) = \int \delta[\lambda- \lambda^\prime(\theta)] \, p(\theta| D^\textrm{CMS}) \, \rd\theta,
	\label{eq:lambda}
\end{equation}
where $\lambda^\prime(\theta)$ is the value of the observable as
predicted by model point $\theta$ ($\theta$ identifies the model point). Equation~\ref{eq:lambda} is approximated using Monte Carlo (MC) integration.  In the following, we describe the construction of the prior density and CMS likelihoods.

\subsection{Construction of the prior}
\label{sec:prior}
If the posterior density for a given parameter differs significantly from its prior density (or prior, for short), then we may conclude that the data have provided useful information about the parameter; otherwise, the converse is true. However, for such conclusions to be meaningful, it is necessary to start with a prior that encodes as much relevant information as
possible. In this study, the prior $p^\textrm{non-DCS}(\theta)$
encodes several constraints: the parameter space boundary, several
theoretical conditions, the chargino lifetimes, and most importantly
the constraints from non-DCS data, such as precision measurements and pre-LHC new physics searches. We choose not to include data from dark matter (DM) experiments in the prior, which avoids any bias from cosmological assumptions (e.g., DM density and distribution, assumption of one thermal relic, no late entropy production, etc.).

The prior $p^\textrm{non-DCS}(\theta)$ is formulated as a product of four factors,
\begin{align}
	p^\textrm{non-DCS}(\theta) &\propto \left [ \prod_j L(D_j^\textrm{non-DCS} | \lambda_j(\theta)) \right] \
		 p\bigl(c\tau(\PSGcpm) {<} \textrm{10\mm}\, | \theta \bigr) \ \
		 p(\textrm{theory}|\theta) \ \
		 p_0(\theta).
		\label{eq:prior}
\end{align}
The initial prior $p_0(\theta)$ is taken to be uniform in the pMSSM subspace,
\begin{align}
	-3 \le& \ \, M_1, M_2 \, \le 3\TeV, \nonumber\\
	0 \le&  \ \ \ \ \, M_3 \ \ \ \ \, \le 3\TeV,			 \nonumber\\
	-3 \le& \ \ \ \ \  \ \mu \ \ \ \ \ \ \, \le 3\TeV,		\nonumber\\
	0 \le& \ \ \ \  \ m_{\rm A} \ \ \ \ \, \le 3\TeV,			 \nonumber\\
	2 \le& \ \ \ \tan\beta \ \ \ \le 60,		 \nonumber\\
	0 \le m_{{\tilde{\text{Q}}_{1,2}}}, m_{{\tilde{\text{U}}_{1,2}}}, m_{{\tilde{\text{D}}_{1,2}}}, & m_{{\tilde{\text{L}}_{1,2}}}, m_{{\tilde{\text{E}}_{1,2}}}, m_{{\tilde{\text{Q}}_3}}, m_{{\tilde{\text{U}}_3}}, m_{{\tilde{\text{D}}_3}},m_{{\tilde{\text{L}}_3}}, m_{{\tilde{\text{E}}_3}} \le 3\TeV,	 \nonumber \\
	-7 \le&  \, A_{\text{t}}, A_{\text{b}}, A_\tau \le 7\TeV,
\label{eq:subspace}
\end{align}
and the formally unbounded
SM subspace defined by $m_{\PQt}$, $m_{\PQb}(m_{\PQb})$, and $\alpha_\mathrm{s}(m_{\PZ})$; the non-DCS measurements, which are listed in Table~\ref{tab:preCMS}, constrain these parameters within narrow ranges.  A point in this subspace is denoted by $\theta$.
The subspace defined in Eqs. (\ref{eq:subspace})
covers the
phenomenologically viable parameter space for the LHC and
is large enough to cover sparticle masses to which the LHC might
conceivably be
ultimately sensitive.  The lower bound of 2 for $\tanb$ evades non-perturbative effects in the top-quark Yukawa coupling after evolution up to the GUT scale. These effects typically become a very serious issue for $\tanb\lesssim 1.7$~\cite{Das:2000uk}.
The term $p(\text{theory} | \theta)$  imposes the theoretical constraints listed at the end of Section~\ref{sec:pmssm}, while $p(c\tau(\PSGcpm)<10\mm | \theta)$ imposes the prompt chargino constraint. Both  $p(\textrm{theory} | \theta)$
and $p(c\tau(\PSGcpm)<10\mm | \theta)$ are unity if the
inequalities are satisfied and zero otherwise.

The product of likelihoods $L(D^\mpreCMS|\lambda(\theta))$ in
Eq. (\ref{eq:prior}) over measurements $j$ is associated with \preCMS data $D^\mpreCMS$,
which imposes constraints from precision measurements and a selection
of pre-LHC searches for new physics. The measurements used and their associated likelihoods are listed in Table~\ref{tab:preCMS}.

Since the explicit functional dependence of the prior $p^\textrm{\preCMS}(\theta)$ on
$\theta$ is not available \emph{a priori}, but the predictions $\lambda(\theta)$ are available point by point, it is natural to represent the prior as a set of points sampled from it.  Owing to the complexity of the parameter space, the sampling is performed
using a Markov chain Monte Carlo (MCMC)
method~\cite{MCMC1,MCMC2,MCMC3,MCMC4,Bayes:2}.

All data in Table~\ref{tab:preCMS} except the Higgs boson signal strengths $\mu_{\Ph}$ were used in the
original MCMC scan. The $\mu_{\Ph}$ measurements were incorporated into the prior post-MCMC.
A number of measurements, marked ``reweight" in the last
column, were updated during the course of this study as new results
became available.  The weights, applied to the subset of scan points which were selected for simulation, were computed as the likelihood ratio of the new measurements shown in
Table~\ref{tab:preCMS} to the previously available measurements.

\begin{table}[htb]
\topcaption{The measurements that form the basis of the \preCMS prior
  $p^\mpreCMS(\theta)$ for the pMSSM parameters, their observed values and likelihoods. The observables are the decay branching fractions $\mathcal{B}(\PQb \to\PQs\gamma)$ and $\mathcal{B}(\PBs \to \mu \mu)$,
 the ratio of the measured branching fraction of the decay $\PB\to \tau \nu$ to that predicted by the standard model, $R(\PB\to \tau \nu)$,
  the difference in the muon anomolous magnetic moment from its SM
  prediction $\Delta a_\mu$, the strong coupling constant
  at the Z boson mass $\alpha_\mathrm{s}(m_{\PZ})$,  the top and bottom quark masses
  $m_{\PQt}$ and $m_{\PQb}(m_{\PQb})$, the Higgs boson mass $m_{\Ph}$ and signal strength $\mu_{\Ph}$, and sparticle mass limits from LEP.  All data except $\mu_{\Ph}$ were used in the initial MCMC scan. Details are given in the text.}
\label{tab:preCMS}
\renewcommand{\arraystretch}{1.2}
\resizebox{\textwidth}{!}{
\begin{tabular}{c|r|c|c|c}
\hline
\multirow{2}{*}{$i$}     & \multicolumn{1}{c|}{Observable}    & Constraint   & Likelihood function & \multirow{2}{*}{Comment} \\
        & \multicolumn{1}{c|}{$\mu_i(\theta)$}  & $D^\mpreCMS_i$  &  $L[D^\mpreCMS_i|\mu_i(\theta)]$ & \\
\hline
1 & $\mathcal{B}(\PQb\to \PQs\gamma)$~\cite{Amhis:2014hma} & $(3.43 \pm 0.21^\text{stat} \pm 0.24^\text{th} \pm 0.07^\text{sys})\times 10^{-4}$ & Gaussian & reweight \\

2 & $\mathcal{B}(\PBs \to \mu \mu)$~\cite{CMS:2014xfa} & $(2.9 \pm
0.7 \pm 0.29^\text{th})\times 10^{-9}$ & Gaussian & reweight \\

3 & $R(\PB\to \tau \nu)$\cite{Amhis:2014hma, Altmannshofer:2010zt} & $1.04\pm 0.34$ & Gaussian & reweight \\

4 & $\Delta a_\mu$~\cite{Hagiwara:2011af} & $(26.1 \pm 6.3^\text{exp}\pm 4.9^\mathrm{SM} \pm 10.0^\mathrm{SUSY})\times 10^{-10}$ & Gaussian & \\

5 & $\alpha_\mathrm{s}(m_{\PZ})$~\cite{Agashe:2014kda} & $0.1184 \pm 0.0007$ & Gaussian & \\

6 & $m_{\PQt}$~\cite{CDF:2013jga} & $173.20\pm0.87^\text{stat}\pm1.3^\text{sys}$\GeV & Gaussian & reweight \\

7 & $m_{\PQb}(m_{\PQb})$~\cite{Agashe:2014kda} & $4.19^{+0.18}_{-0.06}$\GeV & Two-sided Gaussian & \\
\hline
\multirow{2}{*}{8} & \multicolumn{1}{c|}{ \multirow{2}{*}{ $m_{\Ph}$ } } & \multirow{2}{*}{LHC: $m_{\Ph}^\text{low} = 120$\GeV, $m_{\Ph}^\text{high} = 130$\GeV }& 1 if
$m_{\Ph}^\text{low} \le m_{\Ph} \le m_{\Ph}^\text{high}$ & \multirow{2}{*}{reweight} \T\B\\
   &             &                               & 0 if $m_{\Ph} <
   m_{\Ph}^\text{low}$ or $m_{\Ph} > m_{\Ph}^\text{high}$ & \B\\
\hline
9 & \multicolumn{1}{c|}{$\mu_{\Ph}$} & CMS and ATLAS in LHC Run 1, Tevatron & \textsc{Lilith} 1.01~\cite{Bernon:2014vta,Bernon:2015hsa} & post-MCMC \\
\hline
\multirow{2}{*}{10} & \multicolumn{1}{c|}{ \multirow{2}{*}{sparticle masses } } & LEP \cite{lepsusy} & 1 if allowed & \\
   &  & (via {\sc micrOMEGAs}~\cite{Belanger:2001fz,Belanger:2004yn,Belanger:2008sj}) & 0 if excluded &  \\
\hline
\end{tabular}}
\end{table}

For a given point $\theta$, the predictions $\lambda(\theta)$ --- including those needed
to calculate the likelihoods $L(D^\mpreCMS|\lambda(\theta))$ --- are obtained as follows.
The physical masses and interactions are calculated
using the SUSY spectrum generator \textsc{SoftSUSY} 3.3.1~\cite{Allanach:2001kg},
with the input parameters $\theta$ defined at $M_\mathrm{SUSY}$.
This calculation includes 1-loop corrections for sparticle masses and mixings,
as well as 2-loop corrections for the small Higgs boson mass.
Low-energy constraints are calculated with \textsc{SuperIso}
v3.3~\cite{Mahmoudi:2008tp}. {\sc micrOMEGAs} 2.4.5~\cite{Belanger:2001fz,Belanger:2004yn,Belanger:2008sj} is used
to check the compatibility of pMSSM points with sparticle mass limits from LEP and other pre-LHC experiments. {\sc micrOMEGAs} is also used to compute the DM relic density, and the spin-dependent and spin-independent DM-nucleon scattering cross sections; these observables are not used in the construction of the prior, but we study how they are impacted by the CMS searches. The program  \textsc{sdecay} 1.3~\cite{Muhlleitner:2003vg} is used to generate sparticle decay tables and
\textsc{hdecay} 5.11~\cite{Djouadi:1997yw} to generate Higgs boson decay tables.
For evaluating the Higgs boson signal likelihood based on the latest ATLAS~\cite{ATLAS-CONF-2014-009}
and CMS~\cite{Khachatryan:2014jba} measurements, we use \textsc{Lilith} 1.01~\cite{Bernon:2014vta,Bernon:2015hsa}, following the approach
explained in Section~2.3 of Ref.~\cite{Dumont:2013npa}. The experimental results used in \textsc{Lilith} are the signal strengths
of the Higgs boson decay modes $Y=(\gamma\gamma,\,{\PW\PW}^*,\,{\PZ\PZ}^*,\,\bbbar,\tau\tau)$ in terms of the primary Higgs boson production modes
gluon-gluon fusion (ggF), vector boson fusion (VBF), associated
production with a W or Z boson (Wh and Zh, commonly denoted as
Vh), and associated production with a top-quark pair ($\ttbar\Ph$) as
published by ATLAS, CMS,
and Tevatron experiments.
When these signal strengths are given as 2-dimensional (2D) confidence level (CL) contours in, e.g., the
$\mu_{\Pg\Pg\text{F}+\PQt\PQt\Ph}(Y)$ versus $\mu_\mathrm{VBF+Vh}(Y)$ plane, the
likelihood is reconstructed by fitting a 2D Gaussian function to the 68\%~CL
contour provided by the experiments.
For each experiment, the likelihood is then given by $- 2 \log L_Y
= \chi_Y^2$ for each decay mode $Y$, and the combined likelihood is
then obtained by summing over all the individual $\chi_Y^2$ values.
Additional information on signal strengths (and invisible decays) in
one dimension is included analogously, using the published likelihood function
when available or else the Gaussian approximation.

The uncertainty in the anomalous magnetic moment of the muon includes a component that accounts for theoretical uncertainties in the SUSY calculations.

The large window on the Higgs boson mass of 120--130\GeV covers the theoretical uncertainty in the Higgs boson
mass calculation in the MSSM. All tools use the SUSY Les Houches accord~\cite{Skands:2003cj} for
data entry and output.  Approximately 20 million points are sampled from $p^\textrm{non-DCS}(\theta)$ using multiple MCMC chains, but omitting the prompt chargino requirement. When that requirement is imposed, the number of sampled points is reduced by 30\%, and the fraction of bino-like LSPs is enhanced from about 40 to 50\%. A random subsample of 7200 points is selected for simulation studies. Given the large dimensionality of the model, this is a rather sparse scan. However, the scan density is sufficient to learn much about the viability of the pMSSM model space. Distributions of model parameters in this subsample were compared with distributions from independent subsamples of similar size, as well as distributions from the original large sample, and consistency was observed within statistical uncertainties.

\subsection{Incorporation of the CMS data}
\label{sec:cmslhd}
We consider the analyses given in Table~\ref{tab:CMS}, which explore final-state topologies
characterized by a variety of event-level observables:
the scalar sum of the transverse momenta of jets (\HT{}); the magnitude of the vector sum of
the transverse momenta of final-state particles (\MET{} or \MHT{}); a measure of the transverse
mass in events with two semi-invisibly decaying particles (\MTtwo{}); the multiplicity of \cPqb-tagged jets (\cPqb-jets); and a
range of lepton multiplicities, including opposite-sign (OS) and
like-sign (LS) lepton pairs.  Other analyses that were not included in this study but which may impose additional constraints on the model space include  searches for SUSY in the single lepton channel with one or multiple \cPqb-jets~\cite{Chatrchyan:2013iqa} and searches for top squark production~\cite{Chatrchyan:2013xna} in the single lepton channel. The searches considered together comprise hundreds of signal regions and address a large diversity of possible signal topologies.

\begin{table}[htb]
\topcaption{The CMS analyses considered in this study.  Each row gives the analysis description, the center-of-mass energy at which data were collected,
  the associated integrated luminosity,
  the likelihood used,
  and the reference to the analysis documentation.
  }
\begin{center}
\begin{tabular}{l|c|c|c}
\hline
\multicolumn{1}{c|}{Analysis} & \multicolumn{1}{|c|}{$\sqrt{s}$ [TeV]} & \multicolumn{1}{|c|}{$\mathcal{L}$ [fb$^{-1}$]} & \multicolumn{1}{|c}{Likelihood} \\
\hline
Hadronic \HT{} + \MHT{} search \cite{SUS12011} & 7  & 4.98 & counts \\
Hadronic \HT{} + \MET{} + \cPqb-jets search  \cite{SUS12003} & 7  & 4.98 & counts\\
Leptonic search for EW prod. of $\widetilde{\chi}^0$,
$\widetilde{\chi}^{\pm}$, $\tilde{\rm l}$ \cite{SUS12006} & 7  & 4.98 & counts \\
\hline
Hadronic \HT{} + \MHT{} search \cite{SUS13012} & 8  & 19.5 & counts\\
Hadronic \MTtwo{} search \cite{Chatrchyan:2012jx} & 8 & 19.5 & counts\\
Hadronic \HT{} + \MET{} + \cPqb-jets search \cite{SUS12024} & 8 & 19.4 & $\chi^2$\\
Monojet searches \cite{Khachatryan:2014rra} & 8  & 19.7 & binary\\
Hadronic third generation squark search \cite{Khachatryan:2015wza}  & 8 & 19.4 & counts\\
OS dilepton (OS ll) search \cite{Khachatryan:2015lwa} & \multirow{2}{*}{8} & \multirow{2}{*}{19.4} & \multirow{2}{*}{counts} \\
\phantom{xx}(counting experiment only) & & & \\
LS dilepton (LS ll) search \cite{SUS13013} & \multirow{2}{*}{8} & \multirow{2}{*}{19.5} & \multirow{2}{*}{counts} \\
\phantom{xx}(only channels w/o third lepton veto)& & & \\
Leptonic search for EW prod. of $\widetilde\chi^0$,
$\widetilde{\chi}^{\pm}$, $\tilde{\rm l}$ \cite{SUS13006} & \multirow{2}{*}{8} & \multirow{2}{*}{19.5} & \multirow{2}{*}{counts} \\
\phantom{xx}(only LS, 3 lepton, and 4 lepton channels)& & &\\
\hline
Combination of 7\TeV searches & 7 & --- & binary  \\
Combination of 7 and 8\TeV searches & 7, 8 & --- & binary  \\
\hline
\end{tabular}
\\
\label{tab:CMS}
\end{center}
\end{table}

The CMS likelihoods $L(D^\mCMS|\theta)$ are calculated for each of these analyses (or combinations of analyses), using different forms of likelihood depending on the nature of the results that are available.
The first form of likelihood (\emph{counts}) uses observed counts,
$N$, and associated background estimates, $B \pm \delta B$; the second
($\chi^2$) uses profile likelihoods, $T(\mu, \theta)$, where $\mu =
\sigma / \sigma^\textrm{SUSY}(\theta)$ is the signal strength modifier
and $\sigma$ and $\sigma^\textrm{SUSY}(\theta)$ are the observed and
predicted SUSY cross sections, respectively; while the third
(\emph{binary}) joins either of the first two kinds of result together
with a signal significance measure $Z$, and is used for combining
results from overlapping search regions. In the following, we describe the three forms of the
likelihood used and the signal significance measure $Z$.

\paragraph*{Counts likelihood}
For a single-count analysis, the likelihood is given by
\begin{equation}
L(D^\mCMS|\theta) = \int \textrm{Poisson}(N | s(\theta) + b) \, p(b|B, \delta B) \rd b,
\label{eq:lhd_counts}
\end{equation}
where $N$ is the observed count,
$s(\theta)$ and $b$ are the expected number of signal and background counts, respectively,
and $B \pm\delta B$ is the estimated number of background event counts and its uncertainty.
The prior density for $b$, $p(b|B, \delta B)$, is modeled as a gamma density,
$\textrm{gamma}(x;\alpha,\beta) = \beta \exp(-\beta x)  (\beta x)^{\alpha-1}/\Gamma(\alpha)$,
with $\alpha$ and $\beta$ defined such that the mode and variance of the gamma density are  $B$ and $(\delta B)^2$, respectively.
For analyses that yield multiple independent counts, the likelihood is
the product of the likelihoods of the individual counts. For analyses
with multiple counts, we treat
the background predictions for the different search regions as uncorrelated.  Systematic effects on the signal counts are taken into account by varying the signal yield by multiplying it with a signal strength modifier $\mu$ with values $1-\delta\mu, 1, 1+\delta\mu$, where $\delta\mu$ is the fractional value of the systematic uncertainty.

\paragraph*{$\chi^2$ likelihood}
This likelihood is used for CMS searches that provide profile
likelihoods, $T(\mu,\theta) \equiv
L(D^\mCMS|\mu,\theta,\hat\nu(\mu,\theta))$, for the signal strength
modifier $\mu$, where $\nu$ represents the nuisance parameters and
$\hat\nu(\mu,\theta)$ their conditional maximum likelihood
estimates. Taking $\hat\mu$ to be the signal strength modifier that
maximizes $T(\mu,\theta)$, it can be shown that the quantity $t =
-2\ln\left[T(1,\theta)/T(\hat\mu,\theta)\right]$ follows a $\chi^2$
density with one degree of freedom in the
asymptotic limit~\cite{Wilks:1938dza},
\begin{eqnarray}
L(D^\mCMS|\theta) & = &\exp(-t/2) / \sqrt{2\pi t},
  ~\label{eq:lhd_chi2}
\end{eqnarray}
which we adopt as the CMS likelihood in this case. The systematic uncertainties in the signal yield can again be incorporated by varying the value of $\mu$.

\paragraph*{$Z$-significance}
This study uses a signal significance measure defined by
\begin{align}
  Z(\theta) = \textrm{sign} [\ln B_{10}(D, \theta )] \sqrt{2 | \ln B_{10}(D, \theta )|} ,
  \label{eq:Zsingle}
\end{align}
where
\begin{equation}
  B_{10}(D, \theta) = \frac{L(D | \theta, H_1)}{L(D | H_0)}
  \label{eq:B10}
\end{equation}
is the local Bayes factor for data $D$, at point $\theta$, and $L(D | \theta, H_1)$ and $L(D  | H_0)$
are the likelihoods for the signal plus background ($H_1$) and background only ($H_0$) hypotheses, respectively.  The function $Z(\theta)$ is a signed Bayesian analog of the frequentist ``$n$-sigma".
The case $Z \gg 0$ would indicate the presence of a signal
at a significance of $Z$ standard deviations, while the case $Z \ll 0$ would indicate the absence of signal, i.e., an \emph{exclusion} at a significance of $Z$ standard deviations.  The $Z$-significance is the basis of the binary likelihood.

\paragraph*{Binary likelihood}
This likelihood is used for combining results from search regions in
which data may not be independent, for example, multiple counts from
overlapping search regions.  We first divide the data into subsets for
which either a count or $\chi^2$ likelihood can be calculated.  For
each subset $j$, with data $D_j$, we compute $Z_j(\theta)$ using Eq.
(\ref{eq:Zsingle}).  An overall significance measure that includes all subsets under consideration is defined by
\begin{equation}
  Z(\theta) \equiv  Z_{j_{\rm max}}(\theta),
  \label{eq:Zmulti}
\end{equation}
where $j_{\rm max}$ is the index of the maximum element in the set \{$|Z_j(\theta)|$\}.
This quantity is used to define the binary likelihood as follows,
\begin{equation}
  L(D^\mCMS|\theta) =
  \begin{cases}
    1 & \text{if } Z(\theta) > -1.64, \\
    0 & \text{if } Z(\theta) \leq -1.64,
  \end{cases}
  \label{eq:lhd_binary}
\end{equation}
where $Z(\theta)=-1.64$ corresponds to the frequentist threshold for
exclusion at the 95\% CL. Systematic uncertainties are incorporated by computing each $Z_j(\theta)$ by varying the value of $\mu$, and using these recalculated $Z_j(\theta)$ to compute the binary likelihood.
Although use of the binary likelihood entails a loss of information, it is a convenient approach in cases of non-disjoint data, where a proper likelihood calculation is not feasible without more information.  In this study, we use binary likelihoods for monojet searches, which have overlapping search regions, and for combining the 7\TeV, and $7{+}8\TeV$ results, where the considered analyses use nondisjoint data.

To compute likelihoods and $Z$-significances,  expected signal counts for
the search regions of each analysis are computed
for the 7200 pMSSM points. The simulated events for each model point, which were generated using {\sc pythia} 6.4~\cite{Sjostrand:2006za} and processed with the CMS fast detector simulation program~\cite{fastsim}, are passed through the analysis procedures in order to determine the counts.  For each pMSSM point, 10,000 events have been simulated.

\section{Results}
\label{sec:results}

We present the results of our study using three
different approaches to assess the implications of the analyses for the pMSSM parameter space.  In the first approach, we compare the distributions of the $Z$-significances.
In the second approach, we compare the prior and posterior densities of the pMSSM parameters.
In the third approach, we use a measure of the parameter space that remains after inclusion of the CMS search results.  This measure, the survival probability in a region $\Theta$ of the pMSSM parameter space, is defined by
\begin{equation}
  \frac{\int_\Theta p^\textrm{non-DCS}(\theta)H(Z(\theta) + 1.64)\rd\theta}{\int_\Theta p^\textrm{non-DCS}(\theta) \, \rd\theta},
  \label{eq:Survive}
\end{equation}
where $H$ is the Heaviside step function with a threshold value $Z = -1.64$, which
again is the threshold for exclusion at the 95\% CL.

\subsection{Global significance}
\begin{figure}
\centering
  \includegraphics[width=0.38\textwidth]{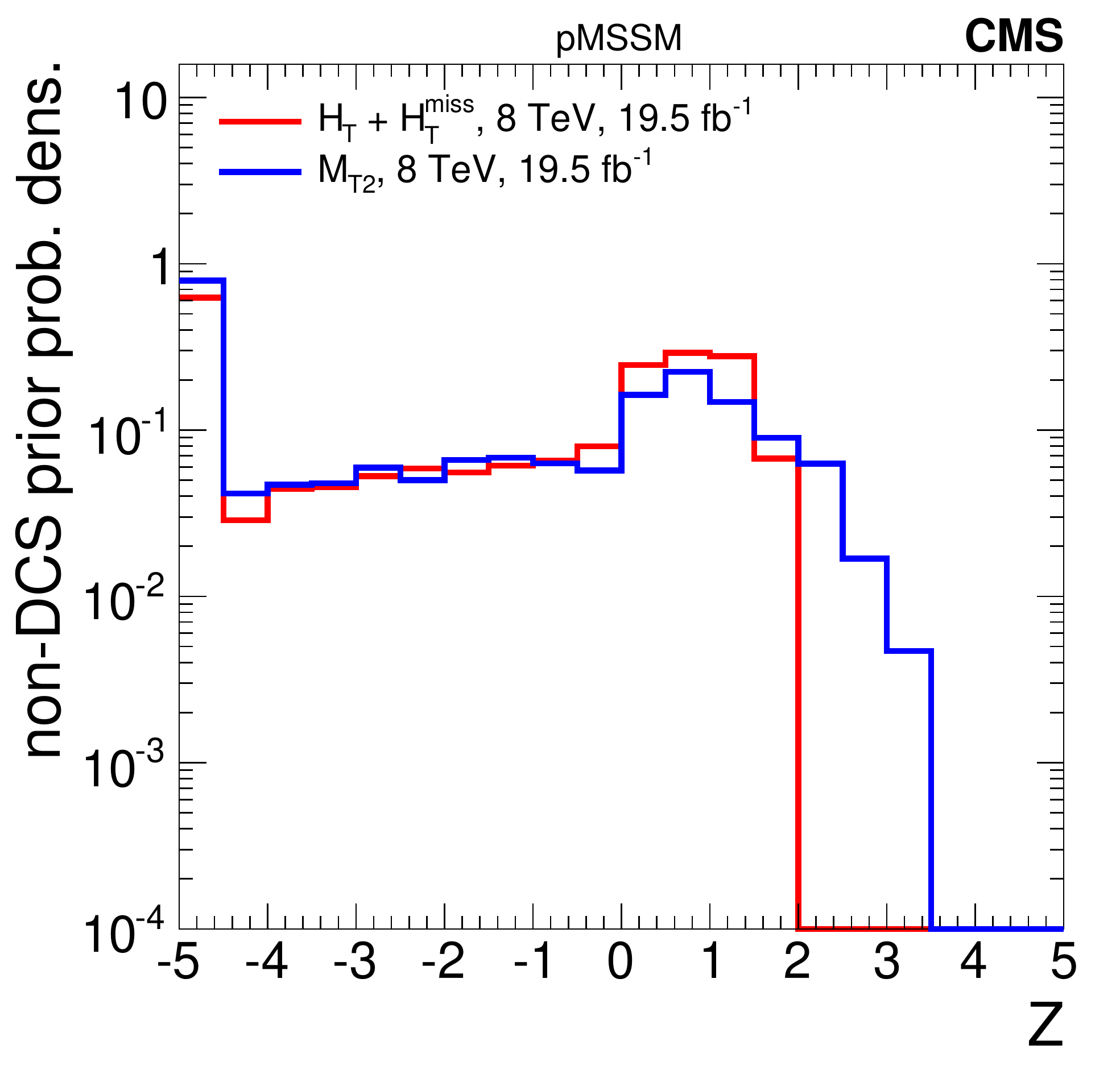}
  \includegraphics[width=0.38\textwidth]{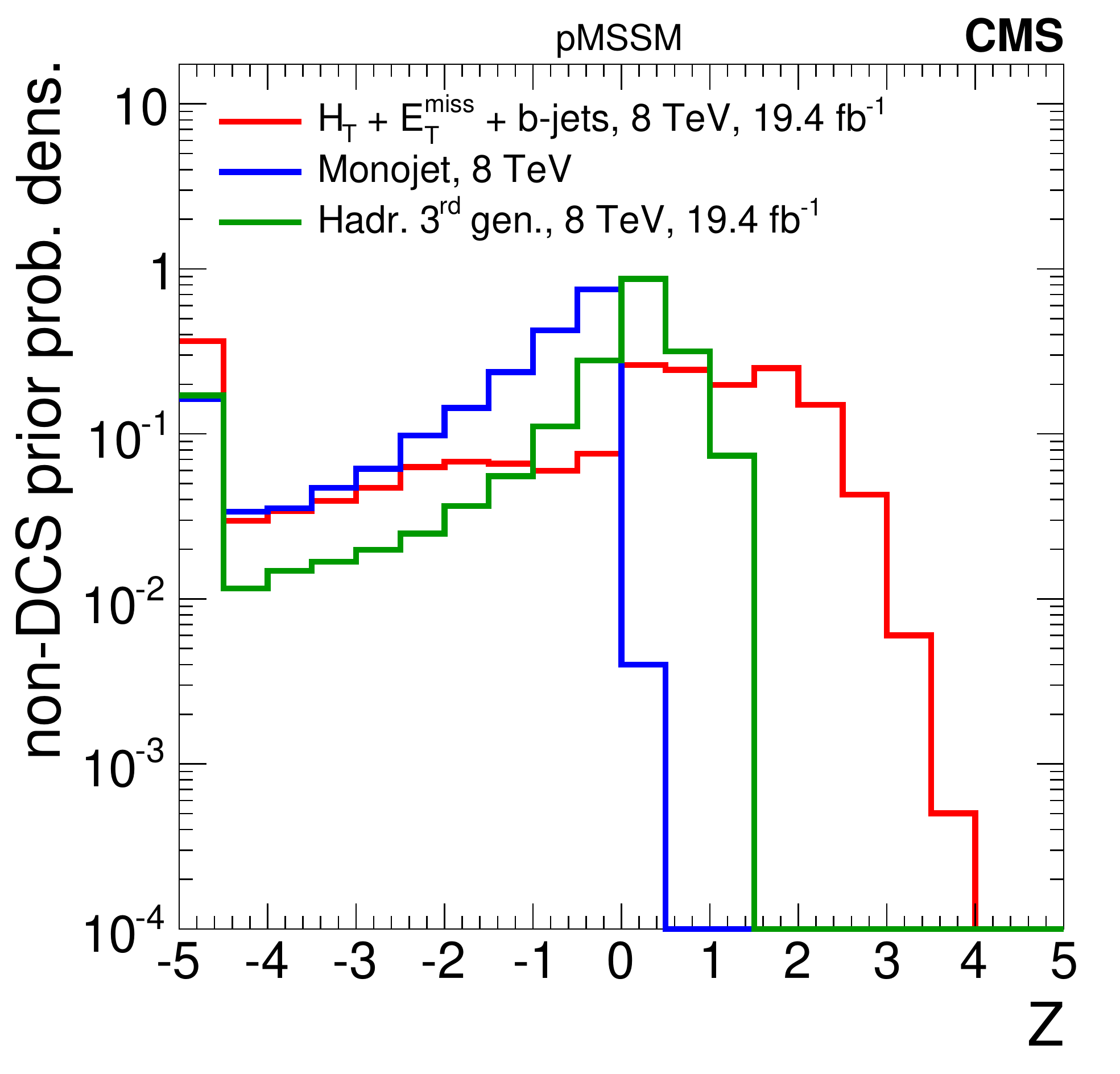}
  \includegraphics[width=0.38\textwidth]{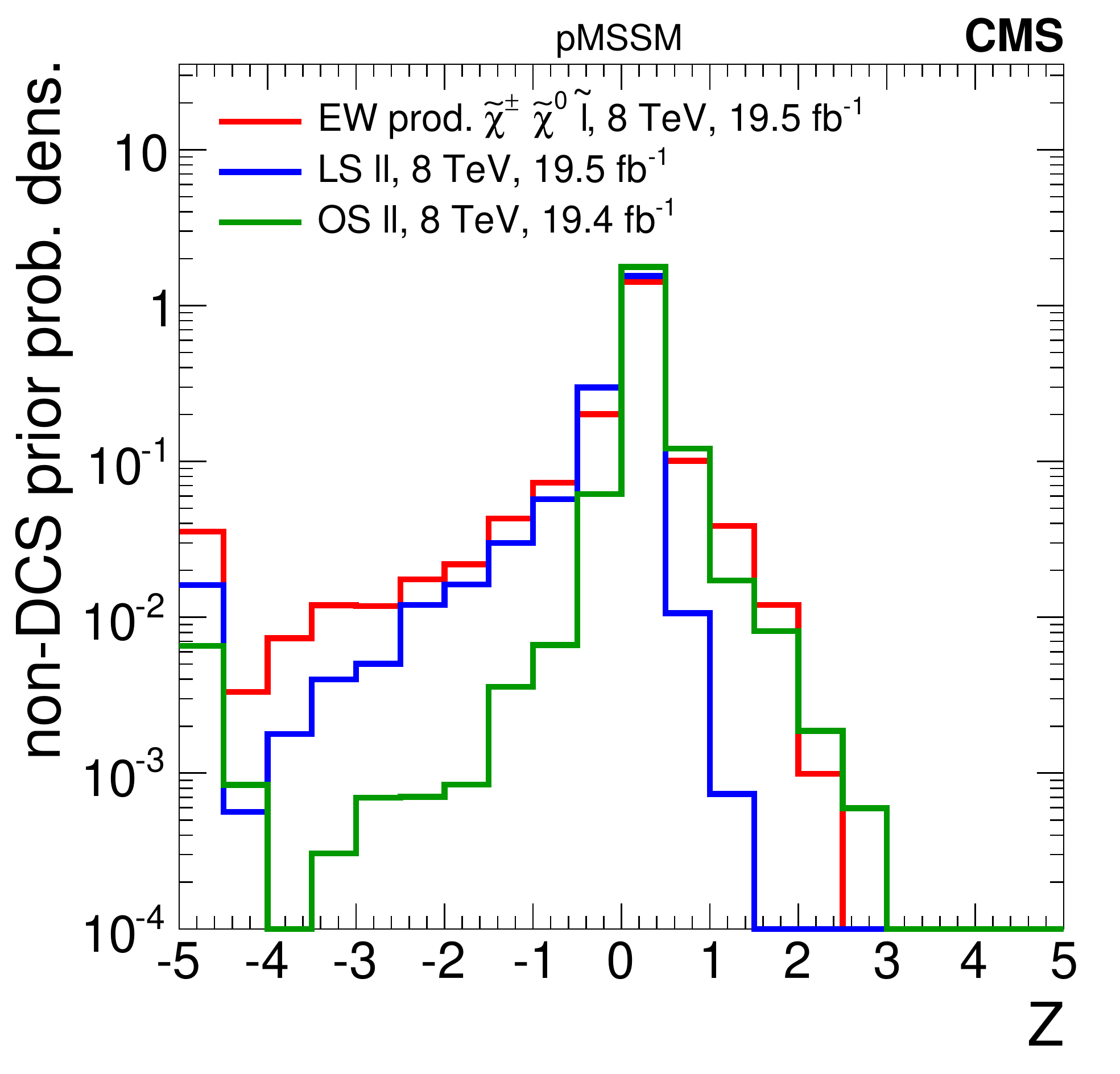}
  \includegraphics[width=0.38\textwidth]{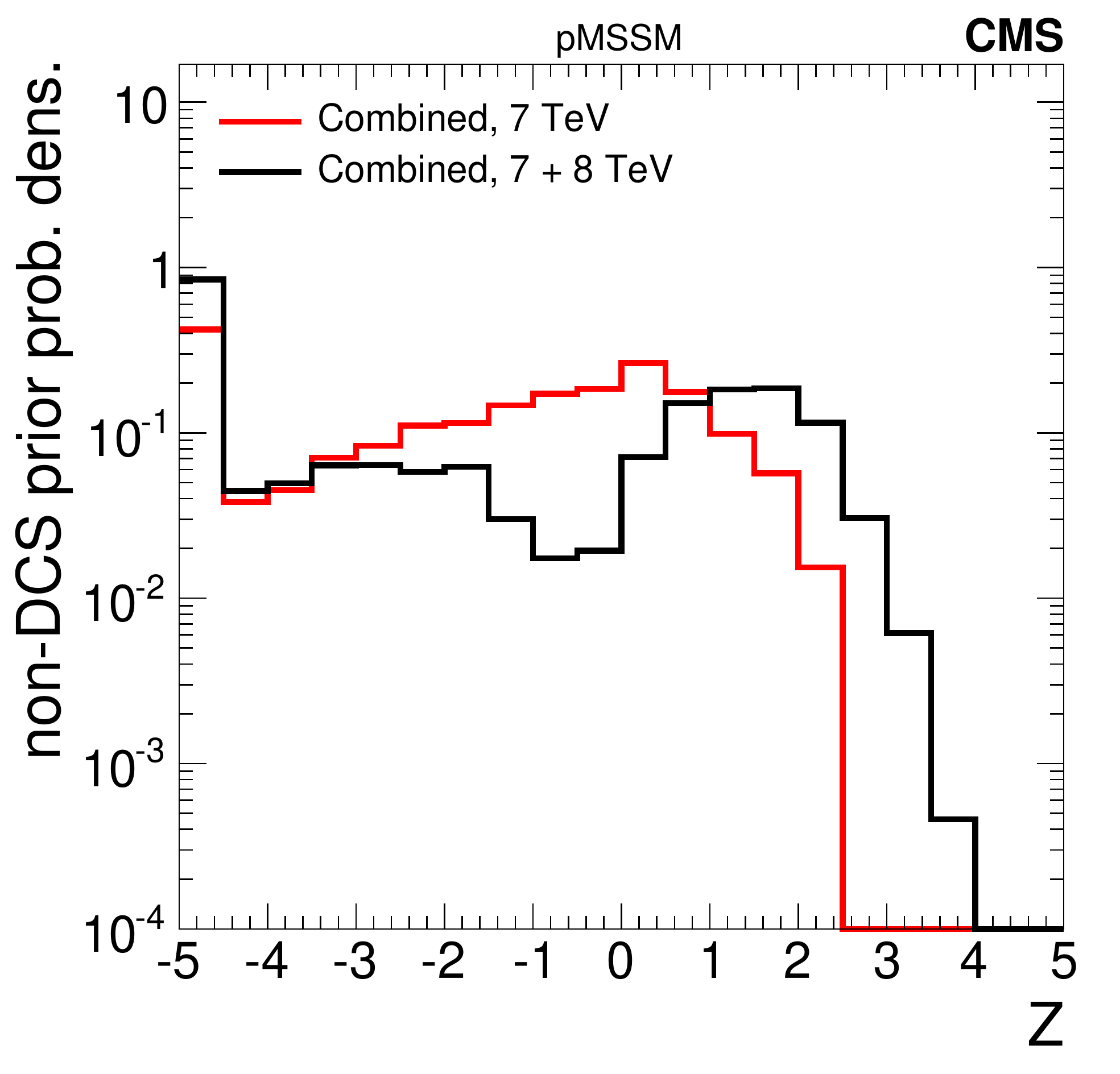}
\caption{The distribution of the $Z$-significance of model points, weighted by the \preCMS prior density of each model point, for the individual 8\TeV searches (top left, top right and bottom left), and for 7\TeV combined and $7{+}8\TeV$ combined searches (bottom right). The leftmost bins contain the underflow entries.}
\label{fig:Z}
\end{figure}

Distributions of $Z$-significance are shown in Fig. \ref{fig:Z} for all the CMS
searches included in this study: 8\TeV searches, combinations of 7\TeV searches, and combinations of
$7{+}8\TeV$ searches. The farther a $Z$ distribution is from zero, the
greater the impact of the analysis on the pMSSM parameter space.   As
noted in Section \ref{sec:anl}, negative and positive values indicate a preference for the background only ($H_0$) and the signal plus background ($H_1$) hypotheses, respectively.

All 8\TeV searches lead to distributions with negative tails,
indicating that each disfavors some region of the parameter space.
The searches making the greatest impact are the \HT{}$+$\MHT{} and \MTtwo{}
searches, which disfavor a significant portion of the parameter space.
The \MTtwo{}, \HT{}$+$\MET{}$+$\cPqb-jets, EW, and OS dilepton searches,
which yield modest excesses over the SM predictions, have
$Z$-significances up to 4.

As expected, the combined $7{+}8\TeV$ result has a greater impact than any individual analysis.
Overall, the impact of the 7\TeV combined result is relatively small as indicated by the high peak around zero.  The dip around zero in the combined $7{+}8\TeV$ distribution arises from the way we combine $Z$-significances. As expressed in Eq.~(\ref{eq:Zmulti}), the maximum $Z$-significance values are used in the combination.

\subsection{Impact on parameters}
\begin{figure*}[tb]
  \includegraphics[width=0.33\textwidth]{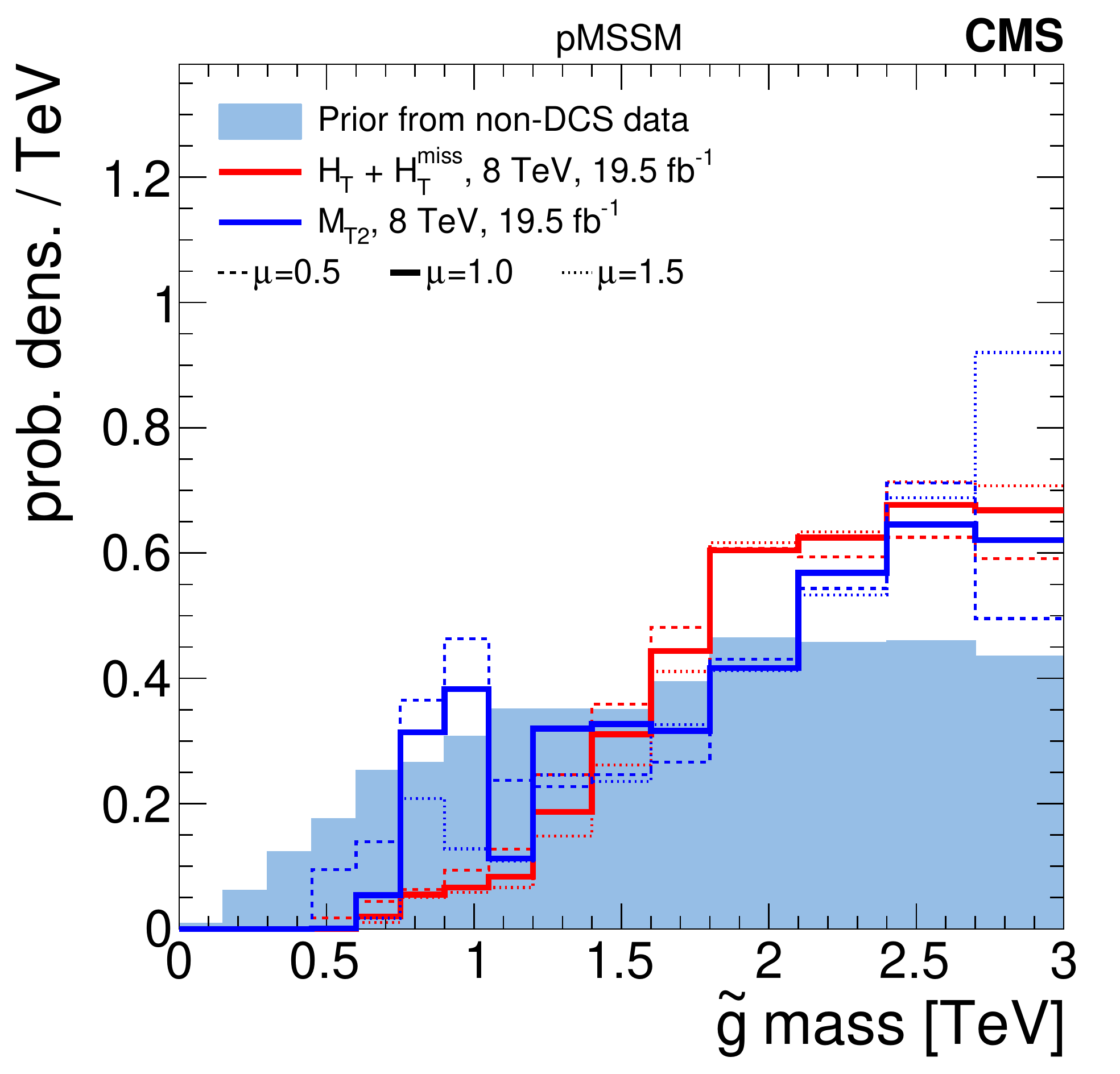}
  \includegraphics[width=0.33\textwidth]{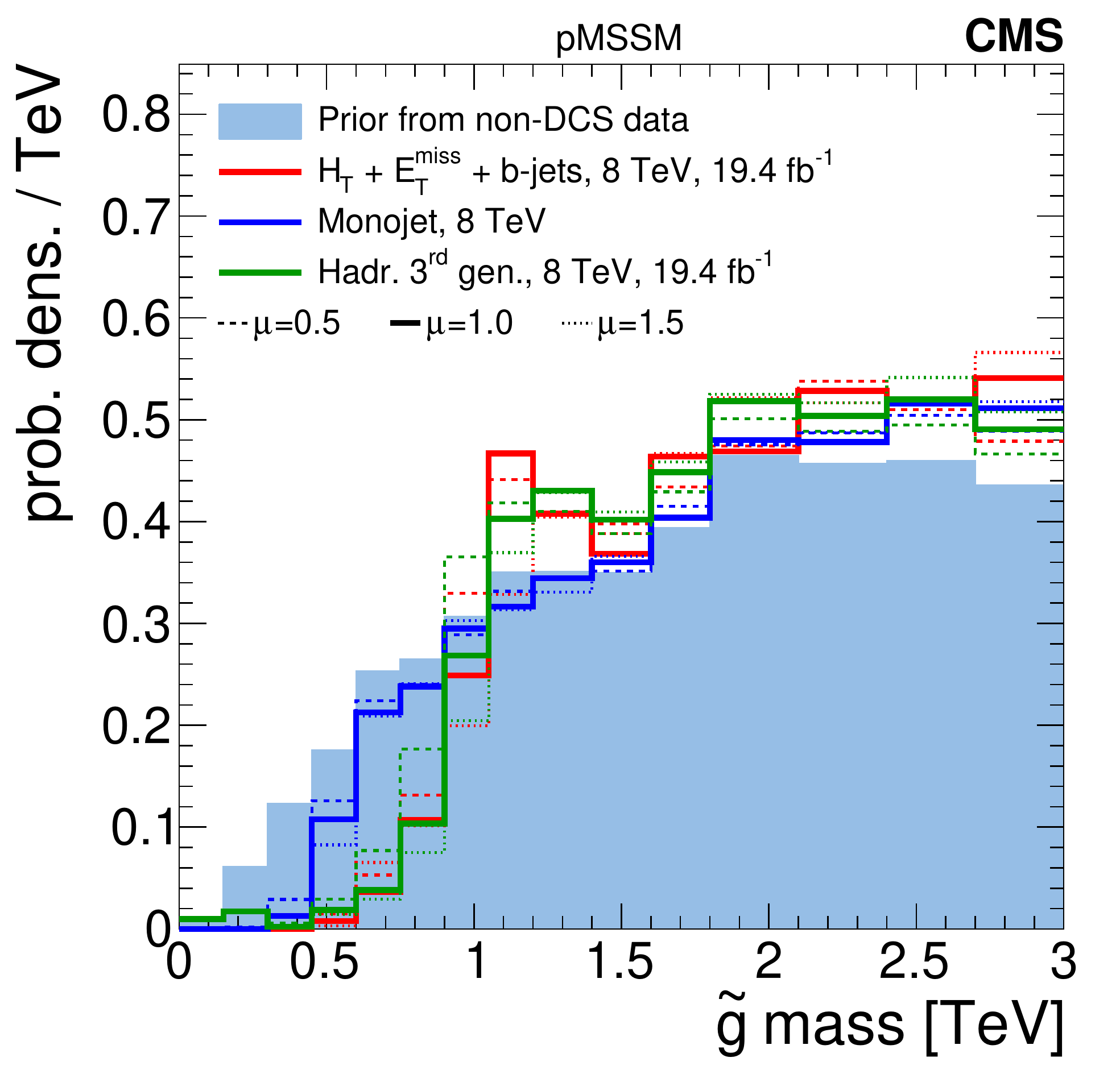}
  \includegraphics[width=0.33\textwidth]{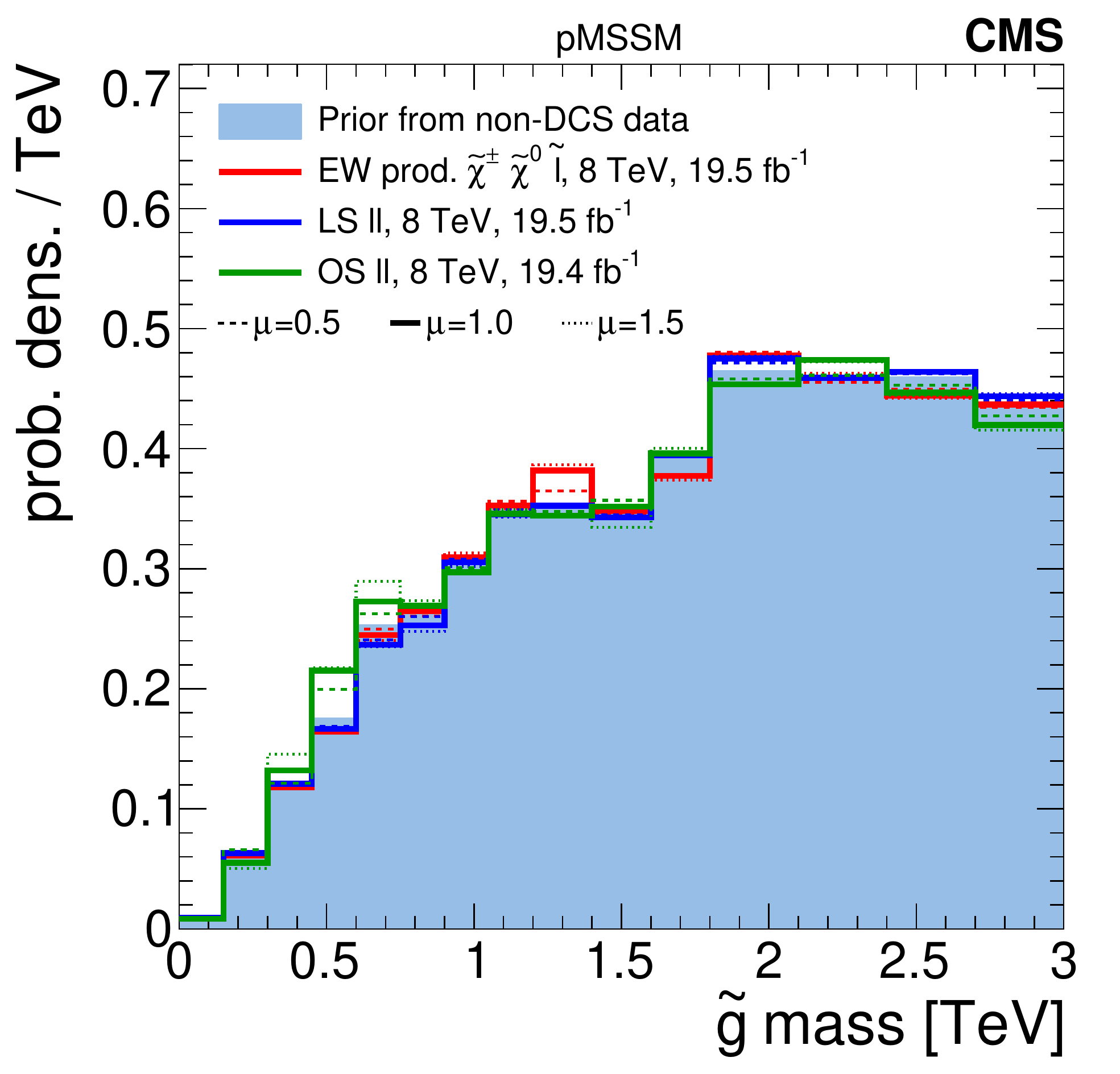}
  \includegraphics[width=0.33\textwidth]{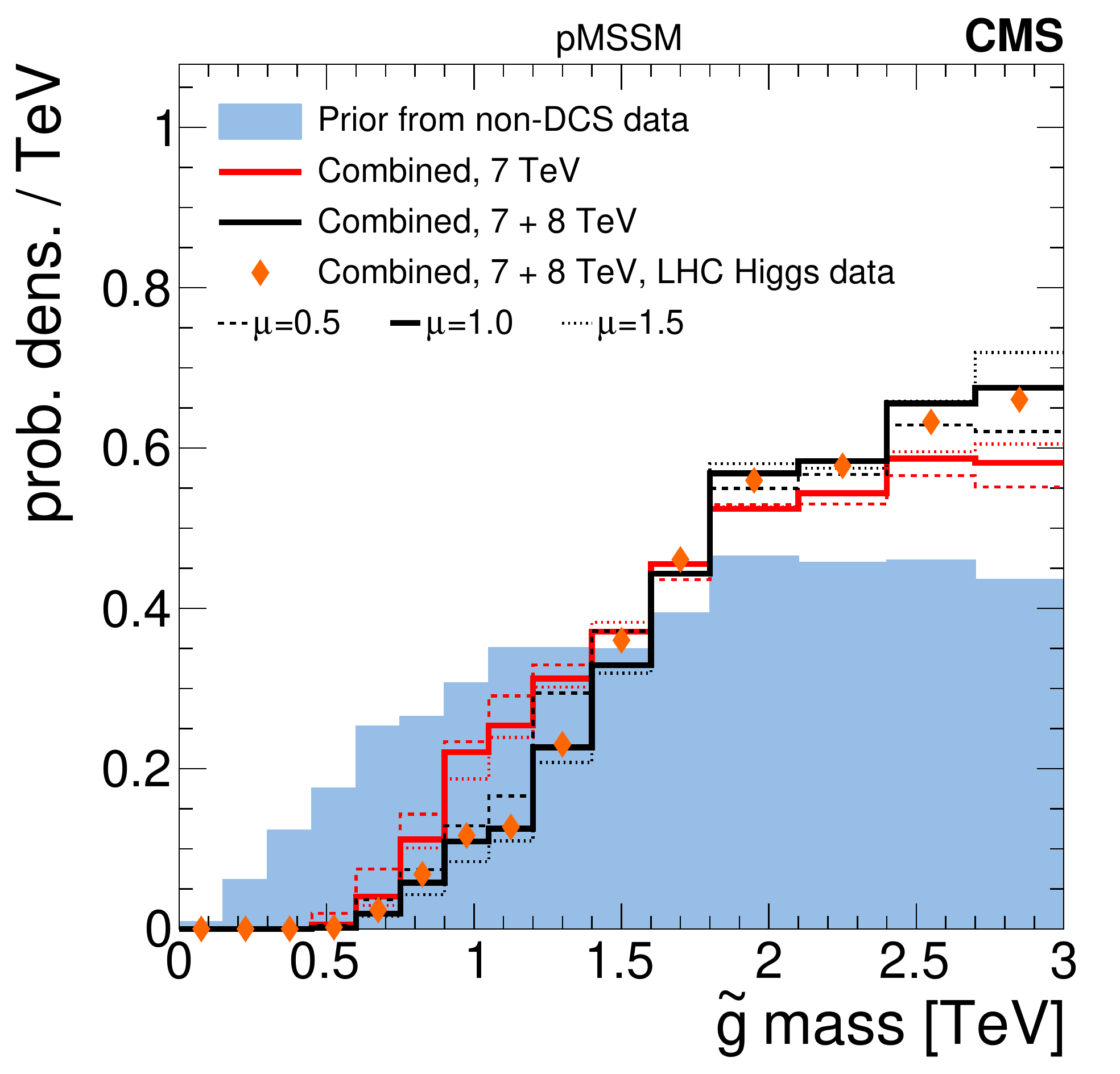}
  \includegraphics[width=0.33\textwidth]{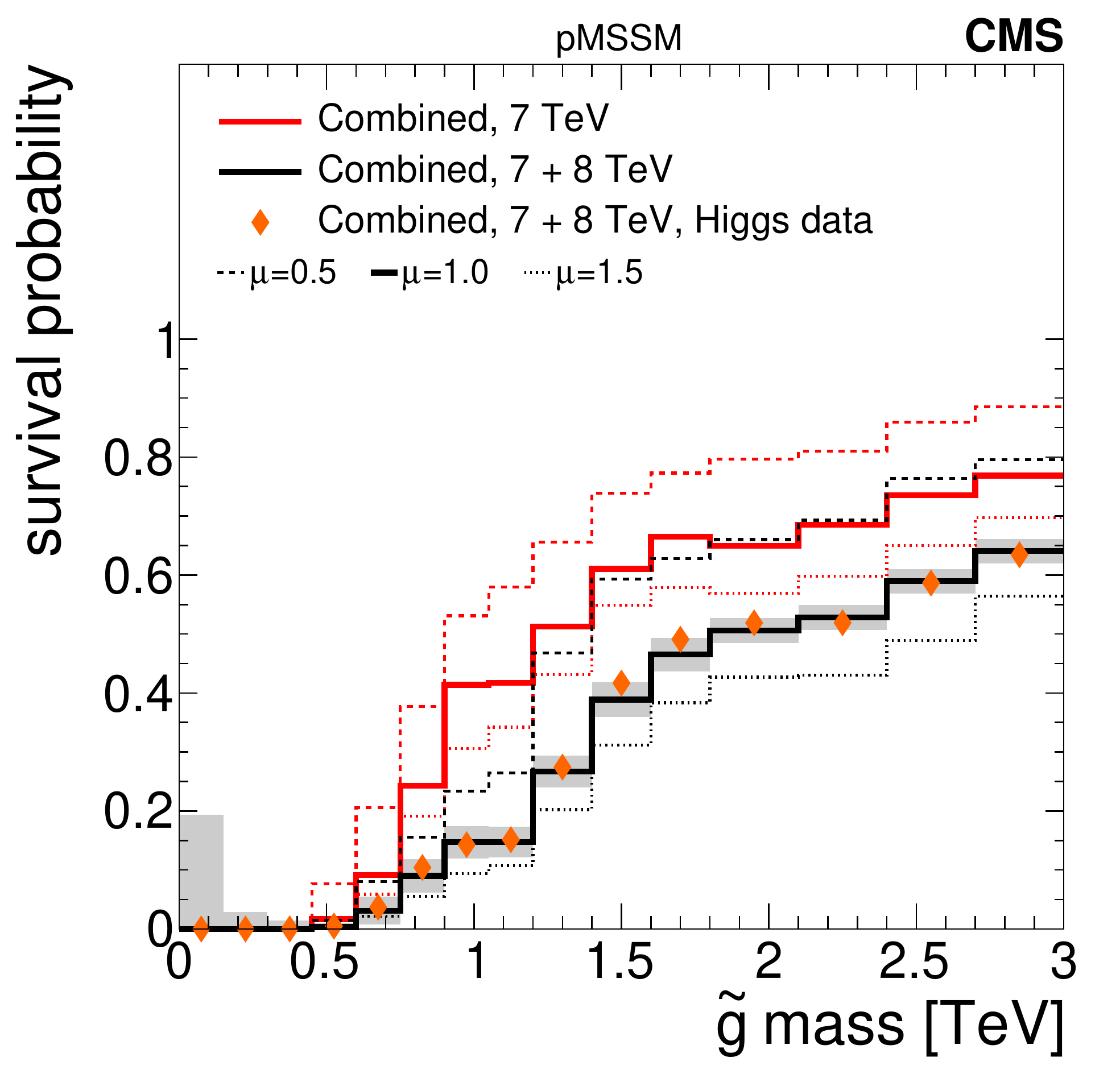}
  \includegraphics[width=0.33\textwidth]{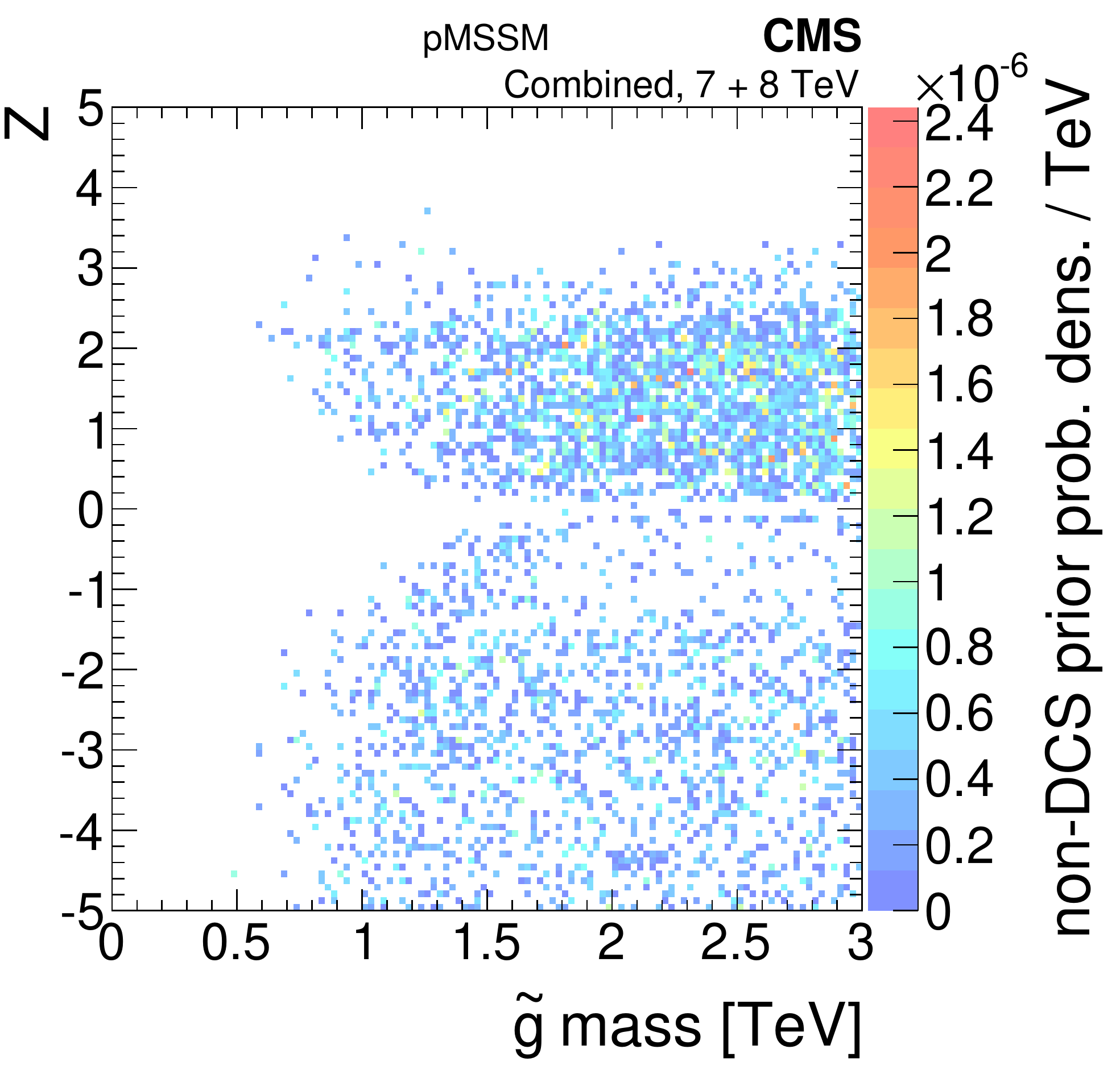}
    \caption{
    A summary of the impact of CMS searches on the probability density of the gluino mass in the pMSSM parameter space.
    The first-row and bottom-left plots compare the \preCMS prior distribution of the gluino mass (blue filled histograms) to posterior distributions after data from various CMS searches (line histograms), where the bottom-left plot shows the combined effect of CMS searches and the Higgs boson results.
    The bottom-center plot shows survival probabilities as a function of the gluino mass for
    various combinations of CMS data and data from Higgs boson
    measurements, where the shaded grey band gives the statistical uncertainty on the
    black histogram.
    The bottom-right plot shows the distribution of the gluino mass versus the $Z$-significance calculated from the combination of all searches.}
    \label{fig:mg}
\end{figure*}

Figure \ref{fig:mg} shows the impact of the CMS searches on our
knowledge of the gluino mass. Figures \ref{fig:mg} (top left, top right and bottom left) show marginalized
distributions of the gluino mass.  Posterior distributions obtained
using three  signal strength modifier values $\mu = 0.5, 1.0, 1.5$
illustrate the effect of a $\pm50$\% systematic uncertainty in the
predicted SUSY signal yields.  Since the uncertainty in the signal efficiency typically varies between 10 and 25\%, and the uncertainty in the signal cross section ranges between 30 and 50\%, this prescription is considered to be conservative. Figure \ref{fig:mg} (top-left) shows the strong impact of
the inclusive analyses on the gluino mass distribution.  The
\HT{}$+$\MHT{} search strongly disfavors the region below 1200\GeV,
while the \MTtwo{} search leads to a distribution with two regions of peaking probability, one at relatively low mass, around 600 to 1000\GeV, and one
above 1200\GeV.  In Fig. \ref{fig:mg} (top-center) we observe that the other hadronic
analyses also disfavor the low-mass region, though to a lesser degree,
 and two of these analyses (the \HT{}$+$\MET{}$+\cPqb$-jets and the
 hadronic third generation) also exhibit secondary
 preferred regions around 1100\GeV, while Fig. \ref{fig:mg} (top-right) shows that the EW, OS dilepton, and LS dilepton searches have little impact on the gluino mass distribution.
Figure \ref{fig:mg} (bottom-left) compares the  prior distribution to posterior distributions after inclusion of the combined 7\TeV and combined $7{+}8\TeV$ data.  The 7\TeV data already have sufficient sensitivity to exclude much of the low-mass gluino model space, and the 8\TeV data further strengthen this result.

 The enhancements induced by the hadronic
searches in the 800\textendash1300\GeV range disappear in the combination
since the observed excesses driving the enhancements are not
consistent with a single model point or group of model points.

Figure \ref{fig:mg} (bottom-center) shows the survival probability (Eq. \ref{eq:Survive}) as a function of gluino mass
for the combined 7\TeV, and $7{+}8\TeV$ results.
The CMS searches exclude all the pMSSM points with a gluino mass below 500\GeV, and can probe scenarios up to the highest masses covered in the scan.
While the direct production of gluinos with masses of order 3\TeV is beyond the reach of these searches, such gluinos are probed indirectly
due the production of other lighter sparticles.
In some cases, the production of lighter sparticles is enhanced by the
presence of heavy gluinos, such as in the case of $t$-channel squark pair production.

Finally, Fig. \ref{fig:mg} (bottom-left) shows the $Z$-significance versus gluino mass.  A
slight negative correlation for positive $Z$ values and gluino masses
is observed below 1200\GeV; $Z$ declines slightly as mass
increases, which indicates that some small excesses of events observed by the various searches are consistent with models with light gluinos.

\begin{figure*}[t]
  \includegraphics[width=0.33\textwidth]{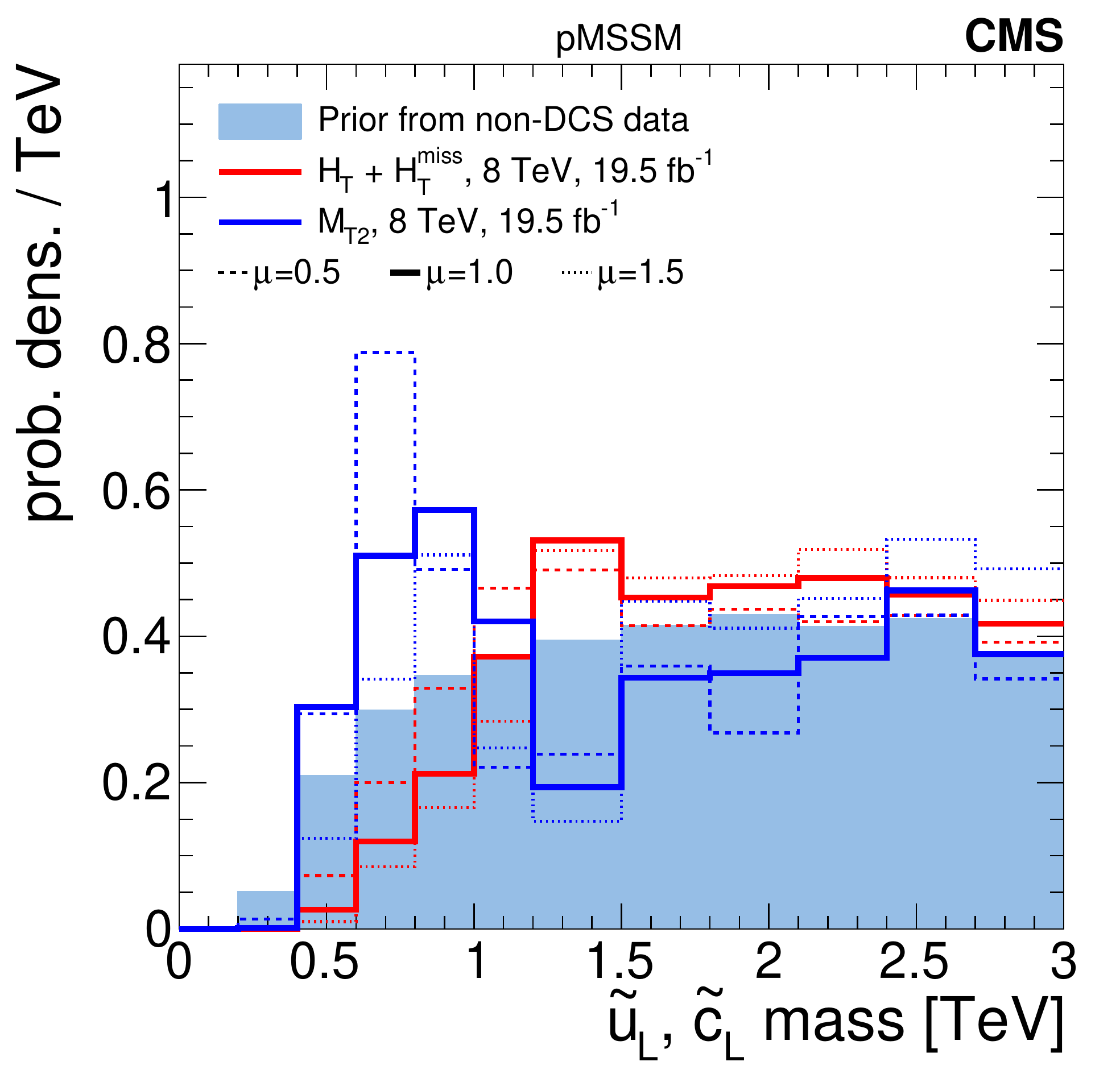}
  \includegraphics[width=0.33\textwidth]{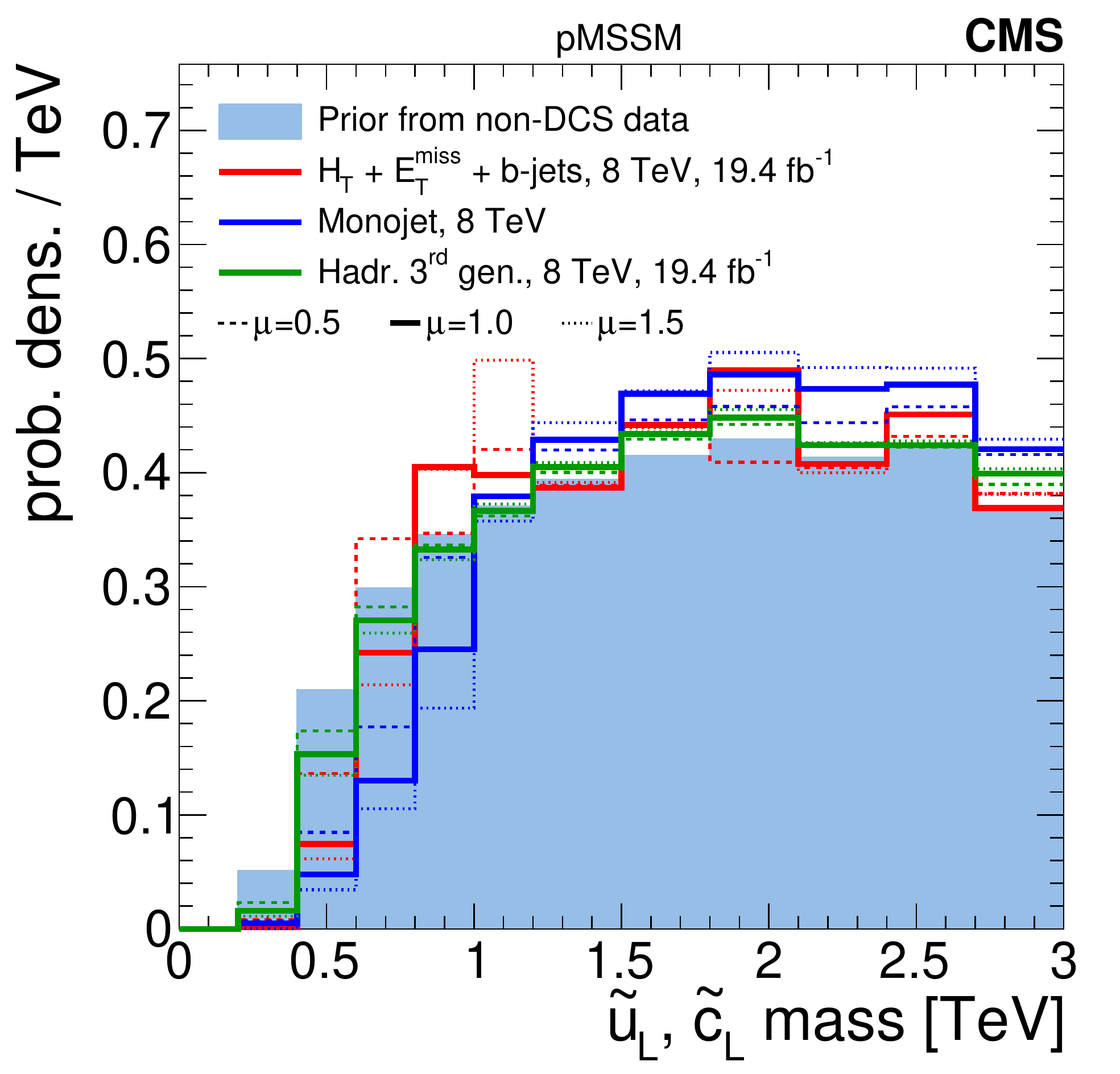}
  \includegraphics[width=0.33\textwidth]{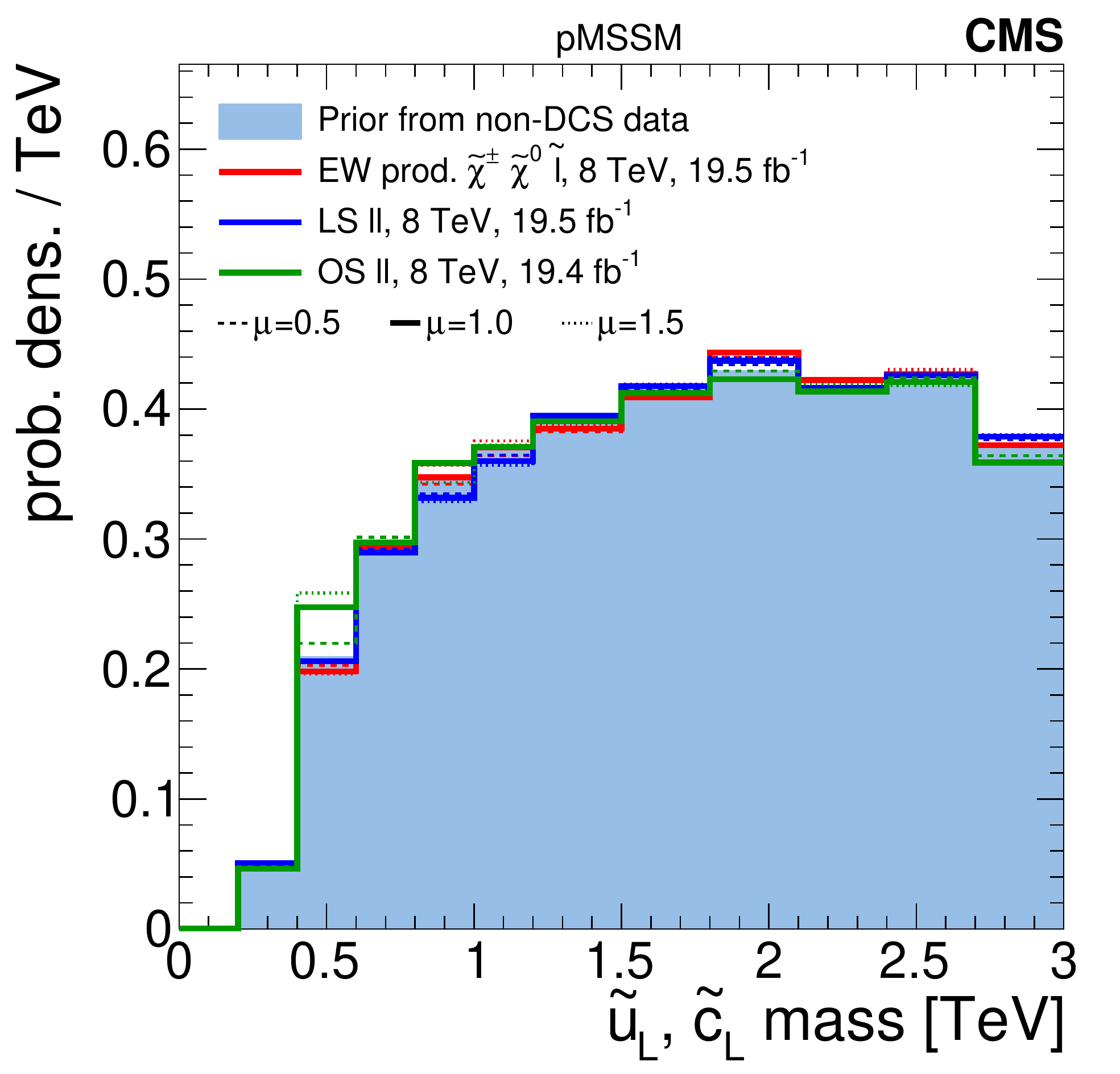}
  \includegraphics[width=0.33\textwidth]{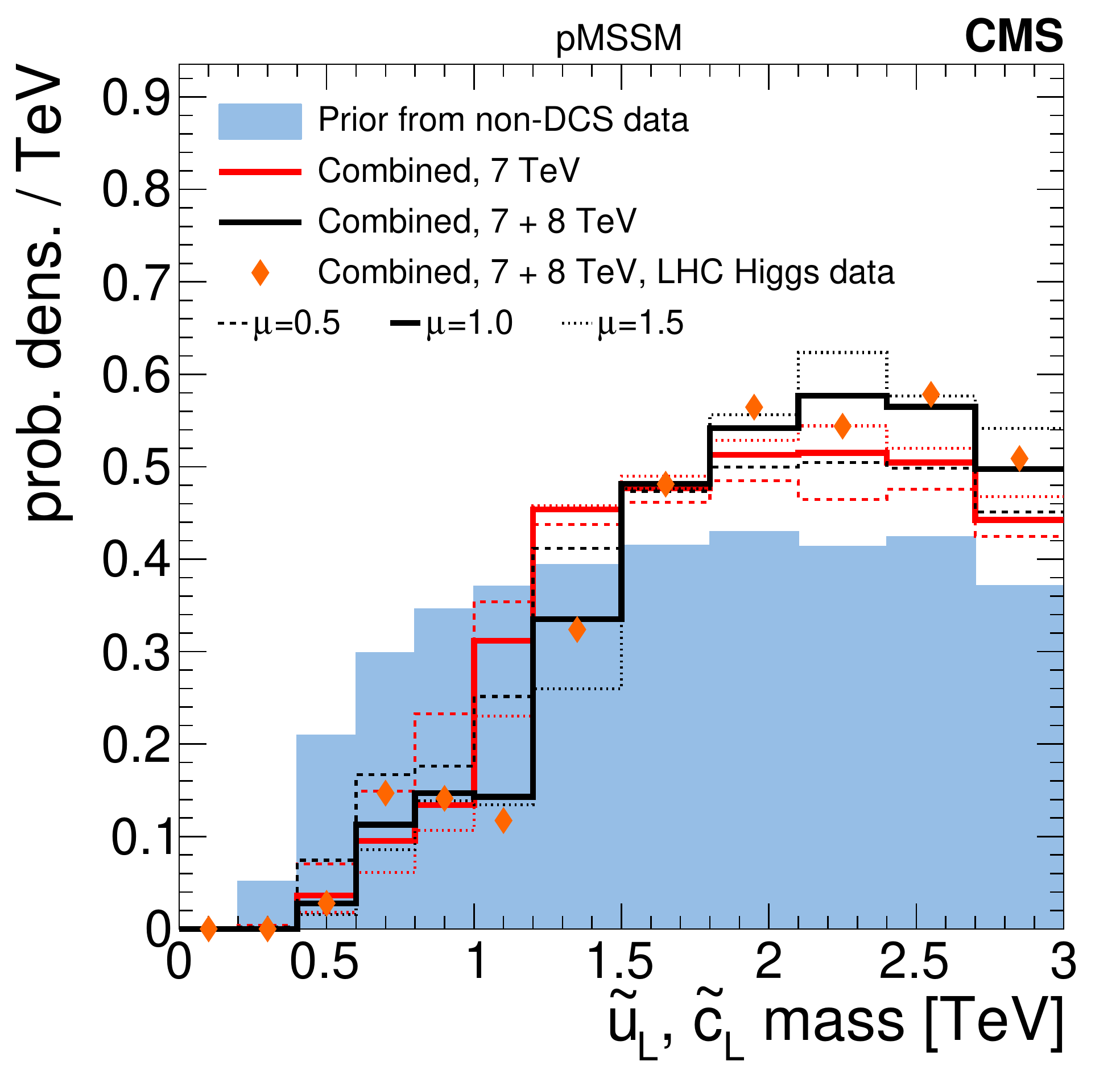}
  \includegraphics[width=0.33\textwidth]{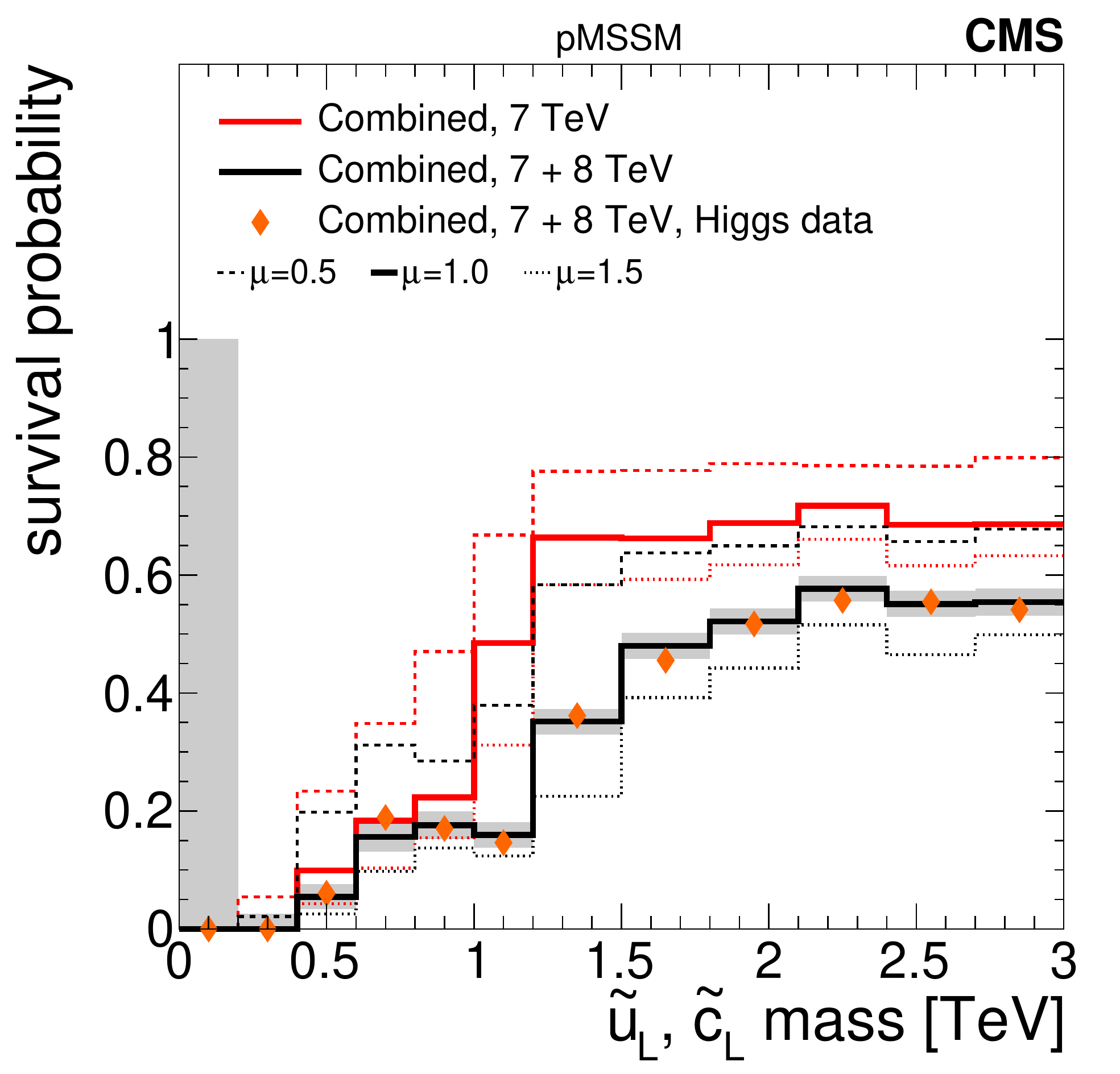}
  \includegraphics[width=0.33\textwidth]{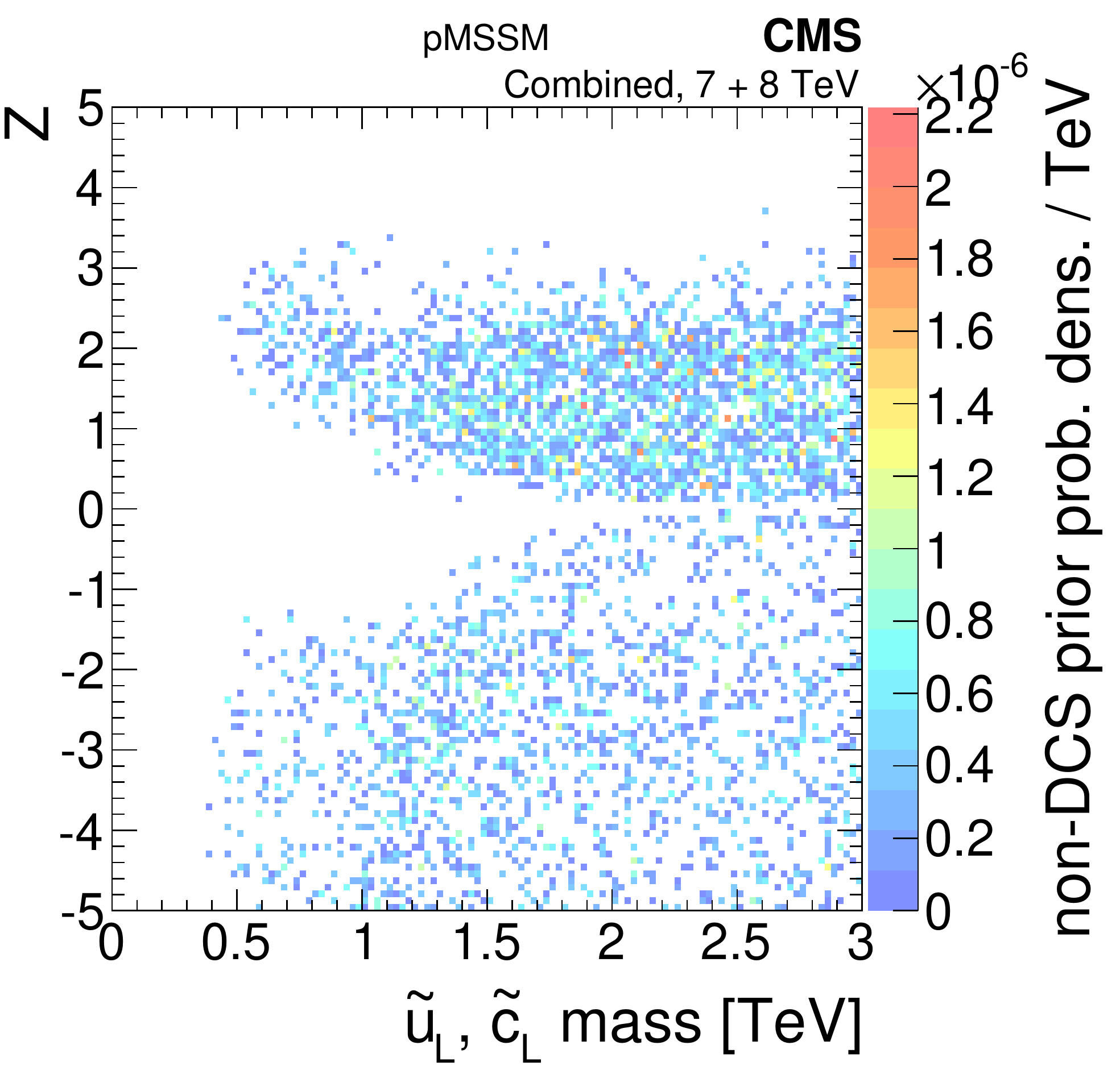}
    \caption{A summary of the impact of CMS searches on the probability density of the $\suL$ mass (equivalently, the $\scL$ mass) in the pMSSM parameter space.
    The first-row and bottom-left plots compare the \preCMS prior distribution of the $\suL$ mass to posterior distributions after data from various CMS searches, where the bottom-left plot shows the combined effect of CMS searches and the Higgs boson results.
    The bottom-center plot shows survival probabilities as a function of the $\suL$ mass for various combinations of CMS data and data from Higgs boson measurements.
    The bottom-right plot shows the distribution of the $\suL$ mass versus the $Z$-significance calculated from the combination of all searches. See Fig. \ref{fig:mg} for a description of the shading.}
    \label{fig:mq}
\end{figure*}

\begin{figure*}[t]
  \includegraphics[width=0.33\textwidth]{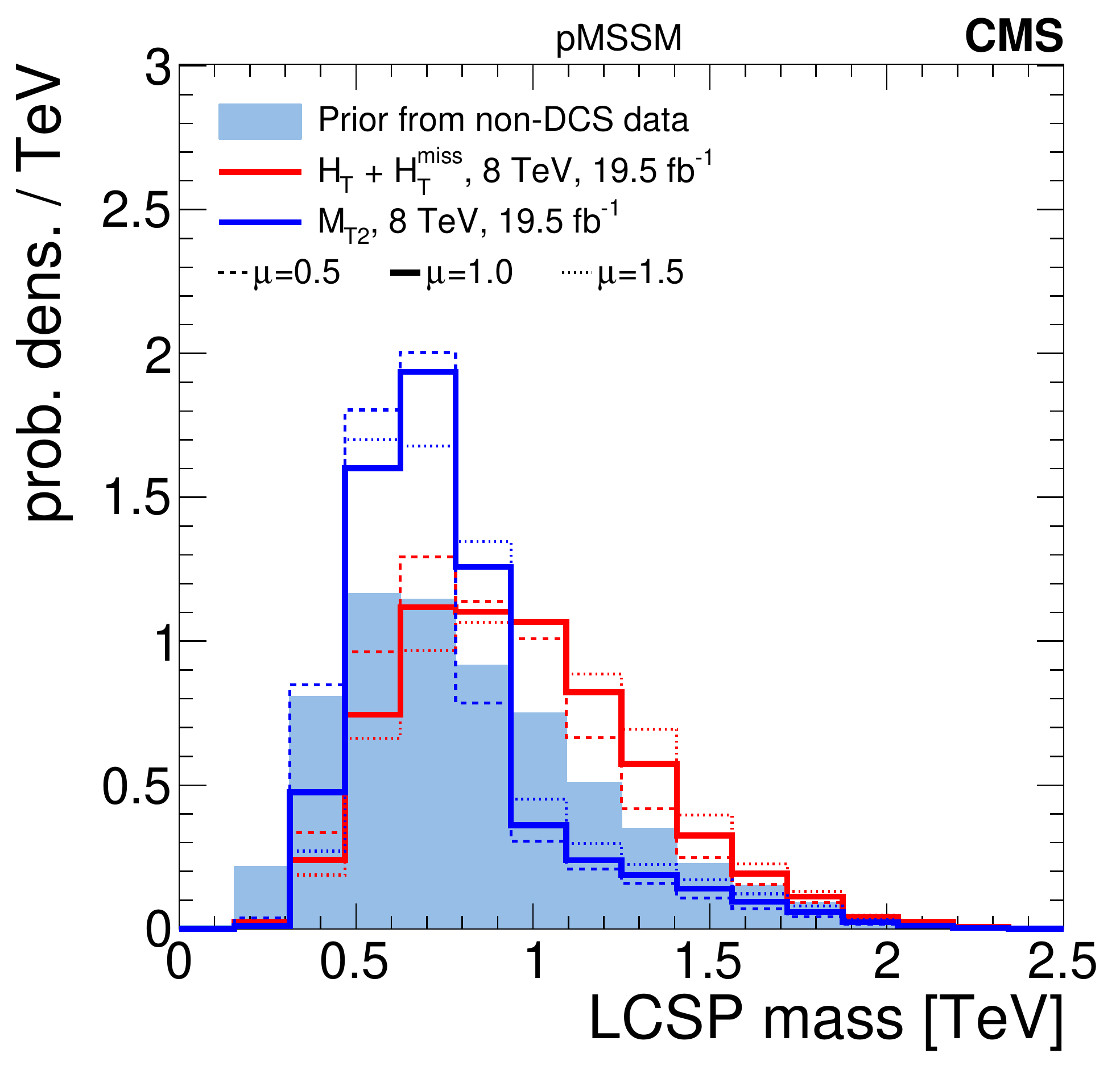}
  \includegraphics[width=0.33\textwidth]{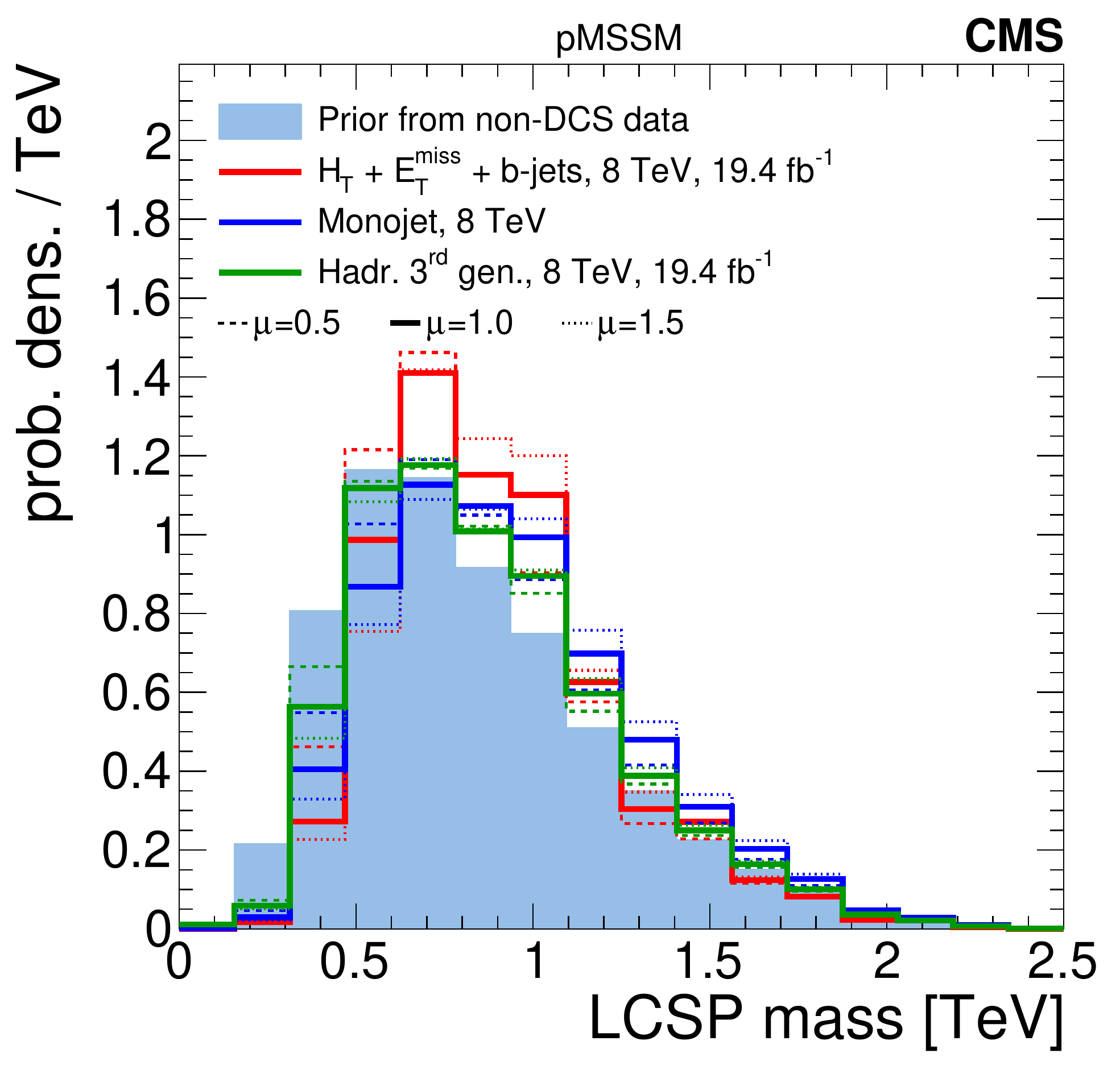}
  \includegraphics[width=0.33\textwidth]{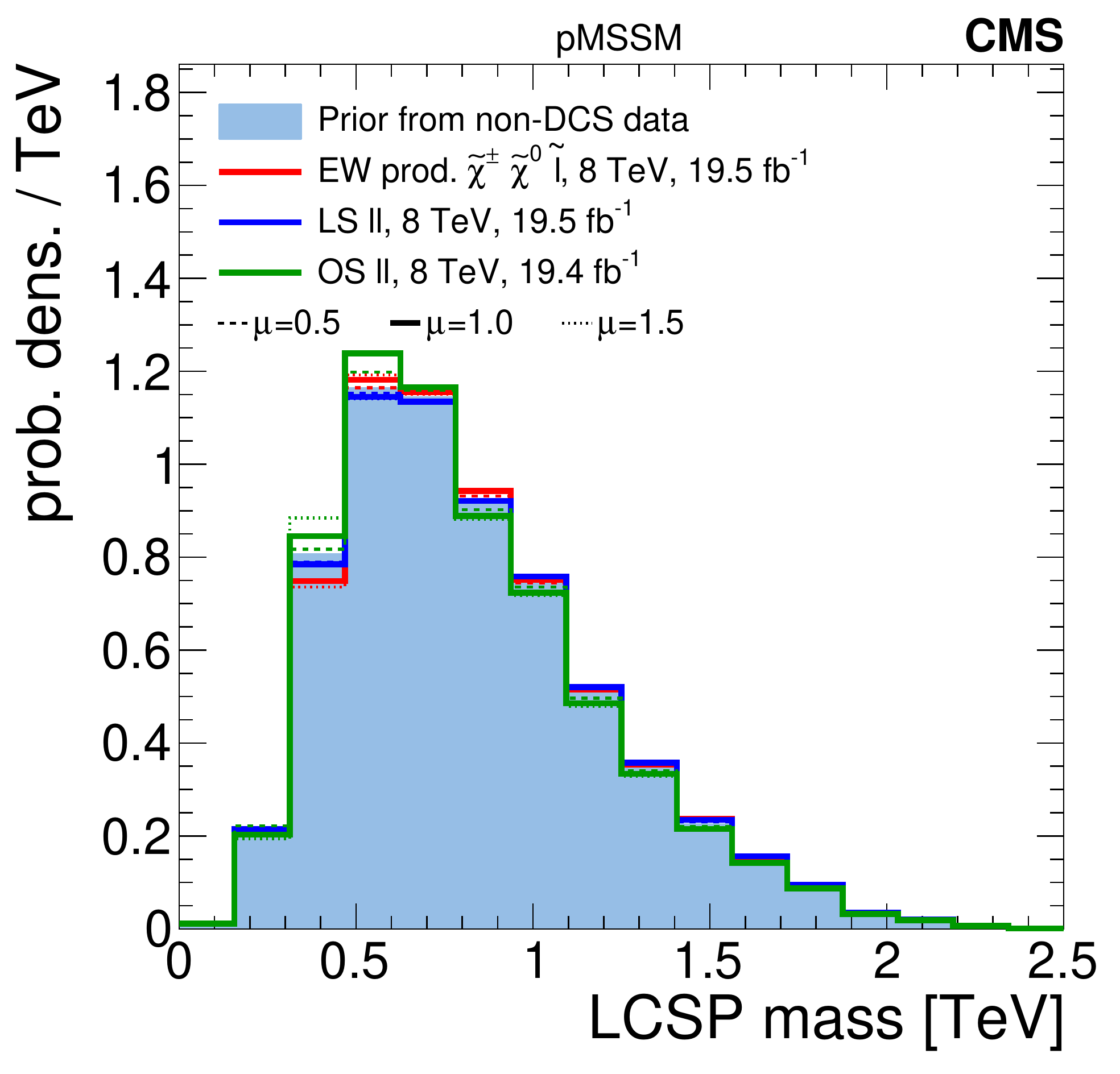}
  \includegraphics[width=0.33\textwidth]{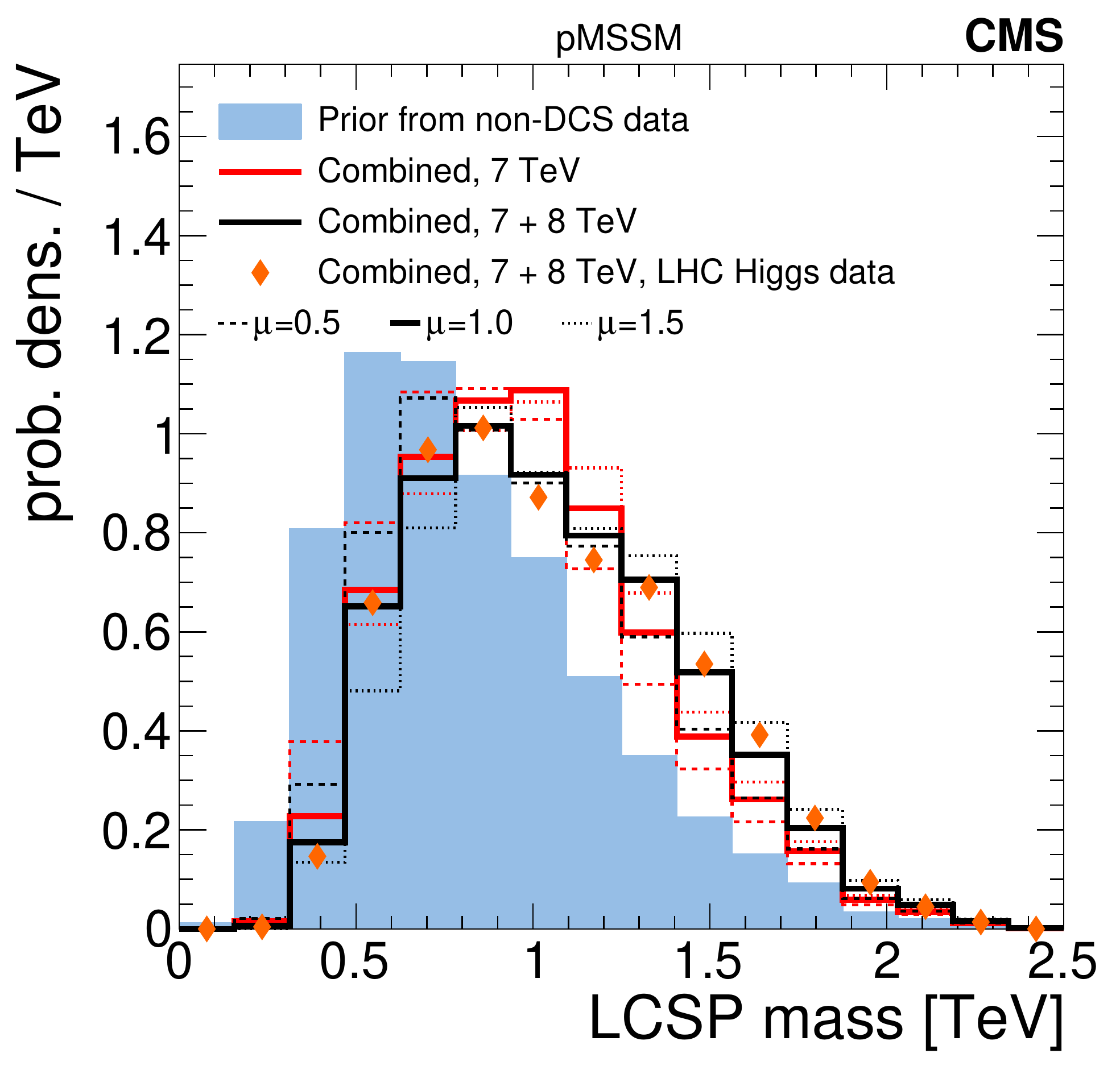}
  \includegraphics[width=0.33\textwidth]{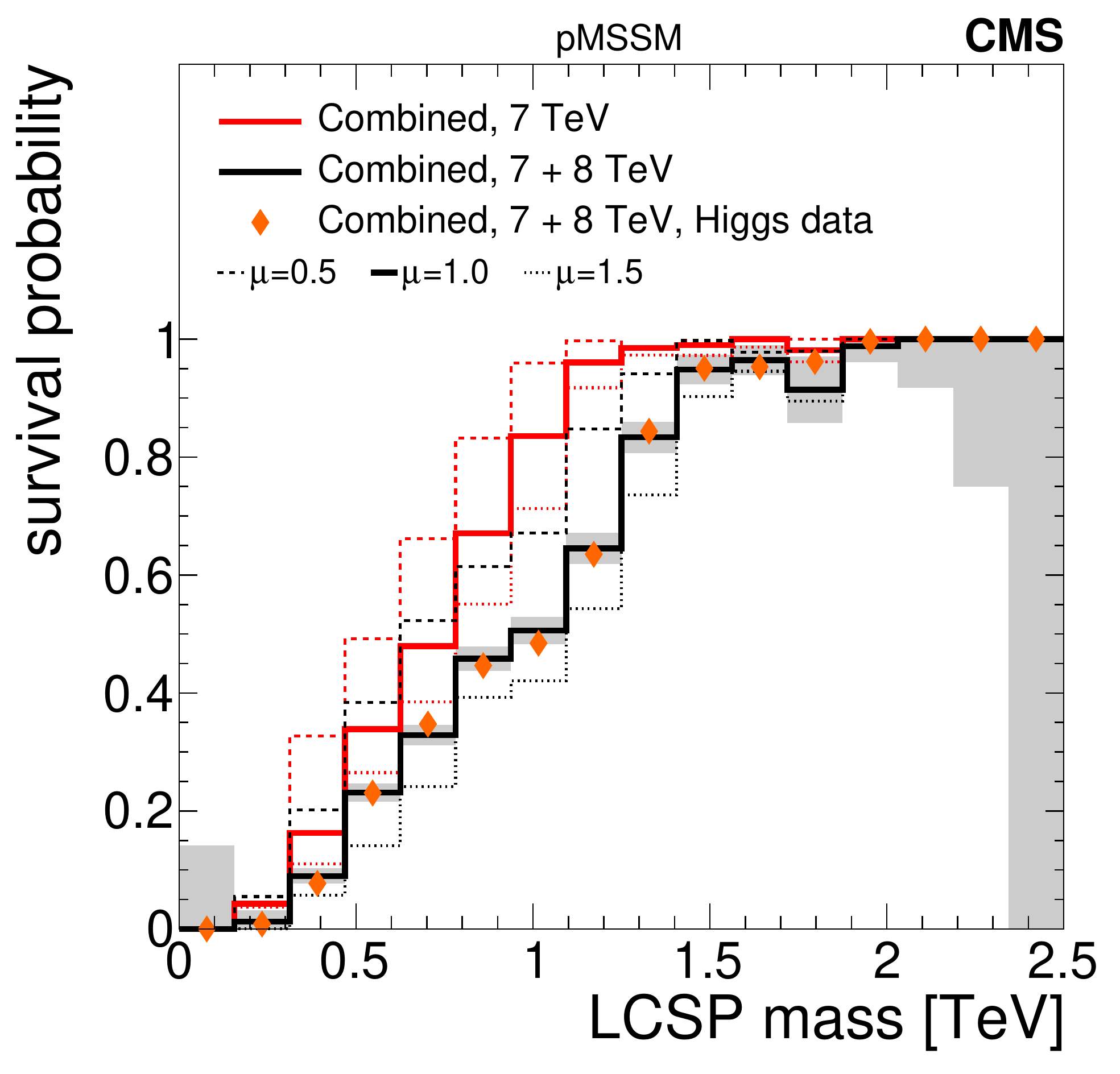}
  \includegraphics[width=0.33\textwidth]{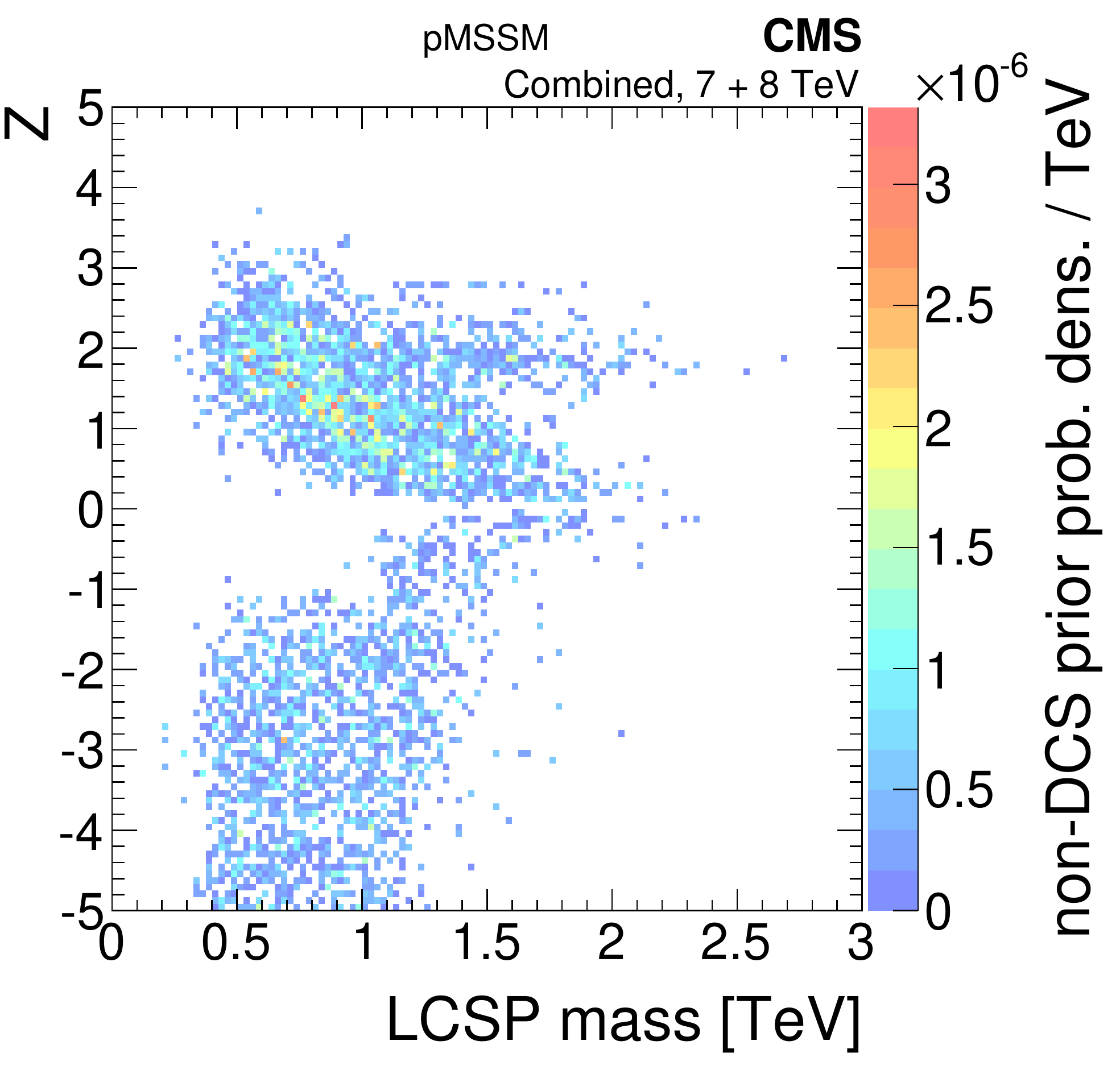}
    \caption{    A summary of the impact of CMS searches on the probability density of the  mass of the lightest colored SUSY particle (LCSP) in the pMSSM parameter space.
    The first-row and bottom-left plots compare the \preCMS prior distribution of the LCSP mass to posterior distributions after data from various CMS searches, where the bottom-left plot shows the combined effect of CMS searches and the Higgs boson results.
    The bottom-center plot shows survival probabilities as a function of the LCSP mass for various combinations of CMS data and data from Higgs boson measurements.
    The bottom-right plot shows the distribution of the LCSP mass versus the $Z$-significance calculated from the combination of all searches. See Fig. \ref{fig:mg} for a description of the shading.}
    \label{fig:mLCSP}
\end{figure*}

Figures \ref{fig:mq} and~\ref{fig:mLCSP} similarly summarize the impact
of searches on the first- and second-generation left-handed up squark mass and the mass of the lightest colored SUSY particle (LCSP), respectively.  The picture is similar to that for the gluino mass.  For both $\suL$ and the LCSP, the \MTtwo{} search shows a preference for masses from 500 to 1100\GeV.  The overall impact of the searches on $\suL$ is less than the impact on the gluino mass owing to the more diverse gluino decay structure that can be accessed by a greater number of searches.  For the LCSP, the overall impact is the least  because the LCSP has the fewest decay channels; nevertheless, CMS searches exclude about 98\% of the approximately 3000 model points with an LCSP mass below 300\GeV; in the surviving 2\% of these model points, the LCSP is the $\tilde{\text{D}}_{\rm R}$.  We also see that the searches can be sensitive to scenarios with LCSP masses up to $\sim$1500\GeV.  Again we find that the
Higgs boson results  make a negligible contribution. In each case we find a negative correlation between the $Z$-significance and the sparticle mass for positive $Z$ values and masses below 1200\GeV; this is most pronounced for the LCSP.

\begin{figure*}[t]
  \includegraphics[width=0.33\textwidth]{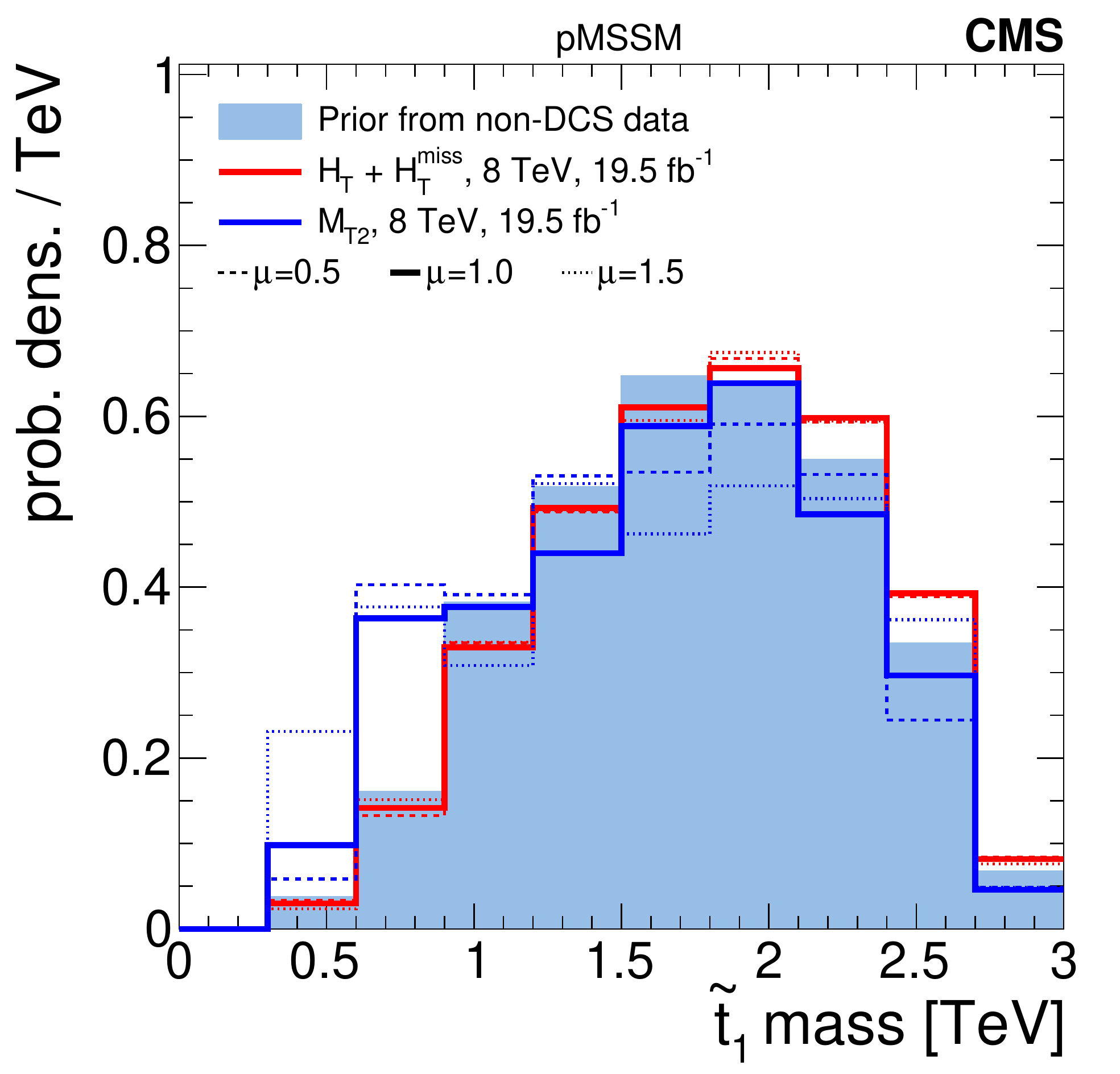}
  \includegraphics[width=0.33\textwidth]{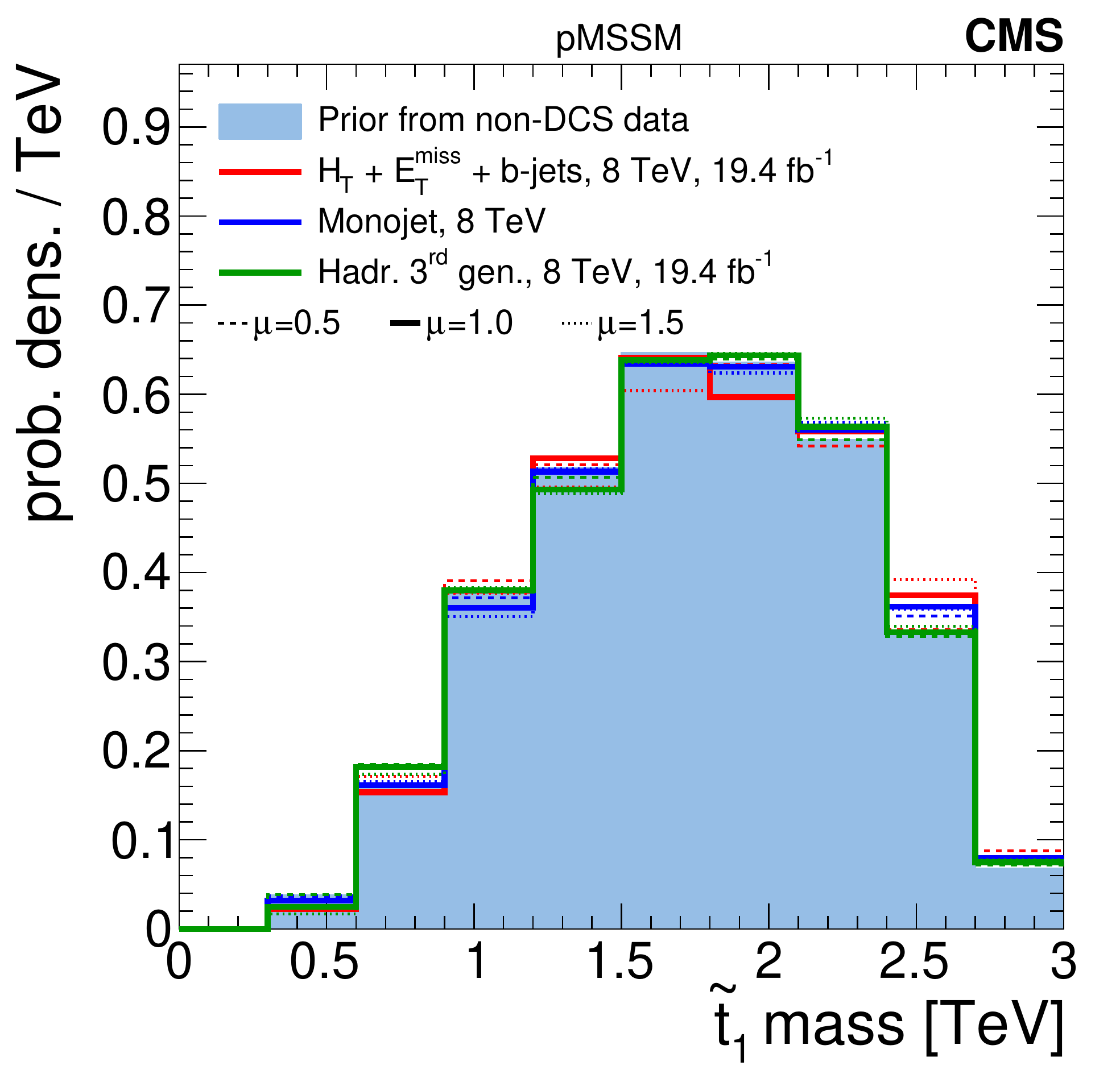}
  \includegraphics[width=0.33\textwidth]{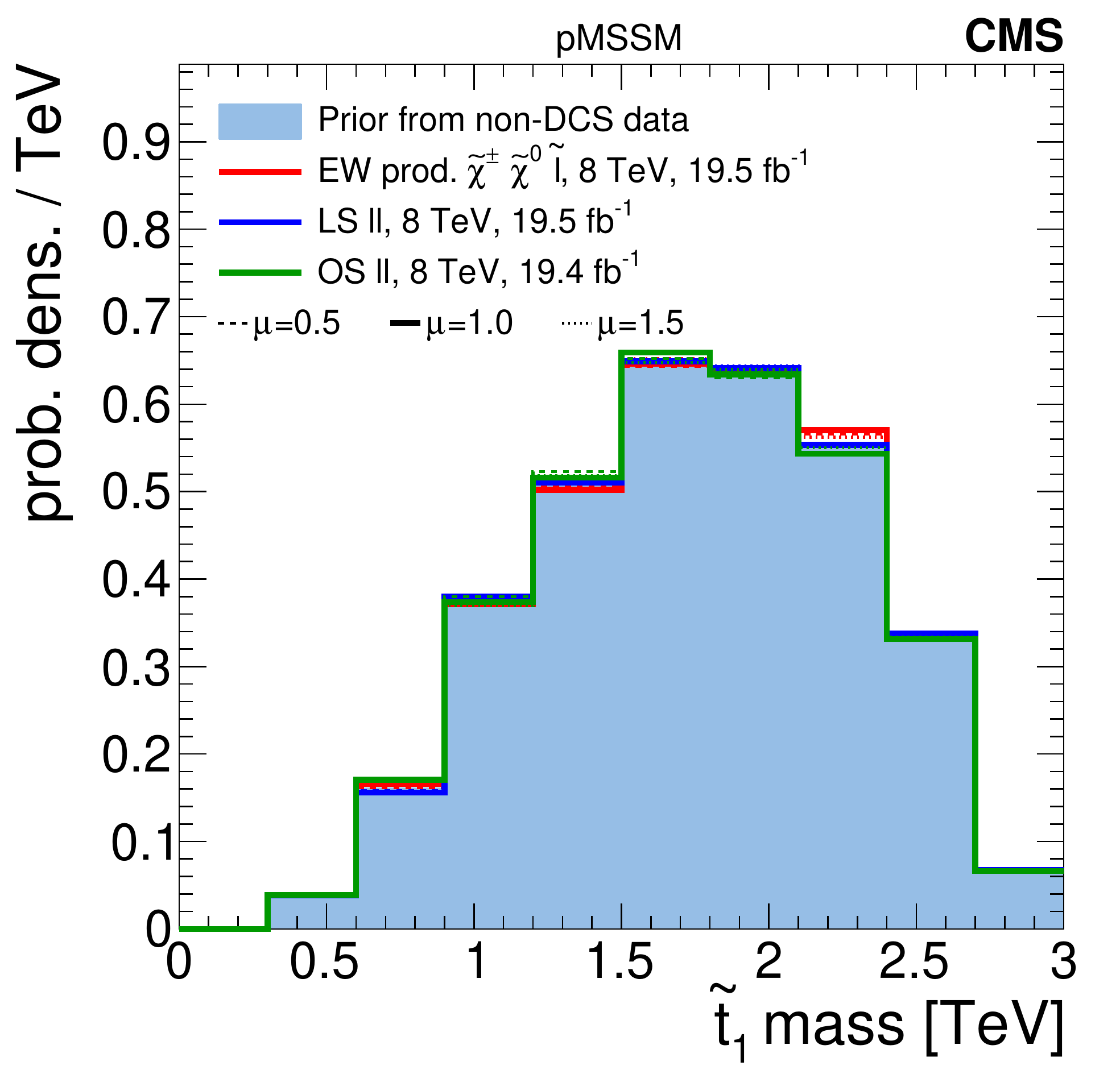}
  \includegraphics[width=0.33\textwidth]{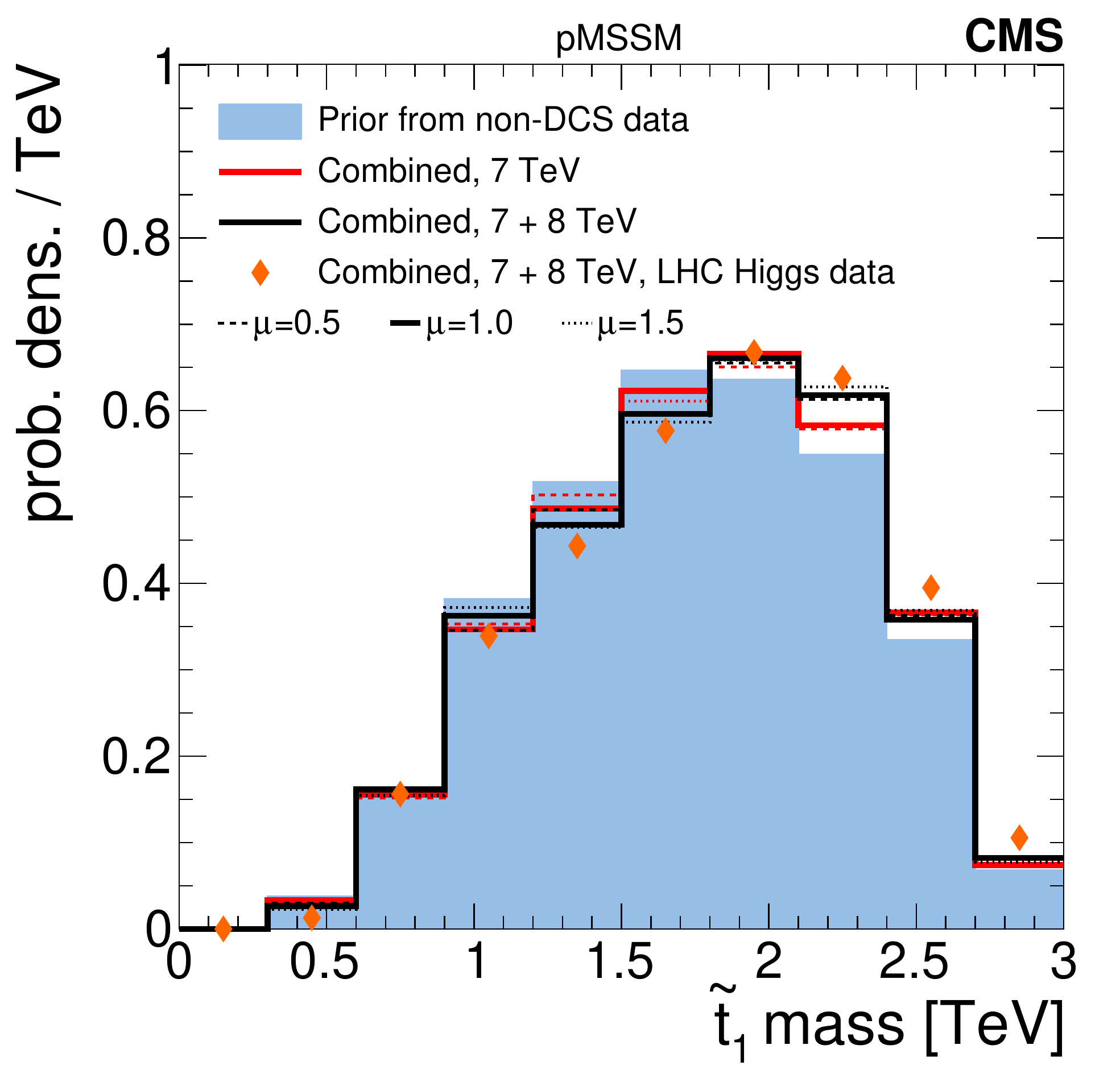}
  \includegraphics[width=0.33\textwidth]{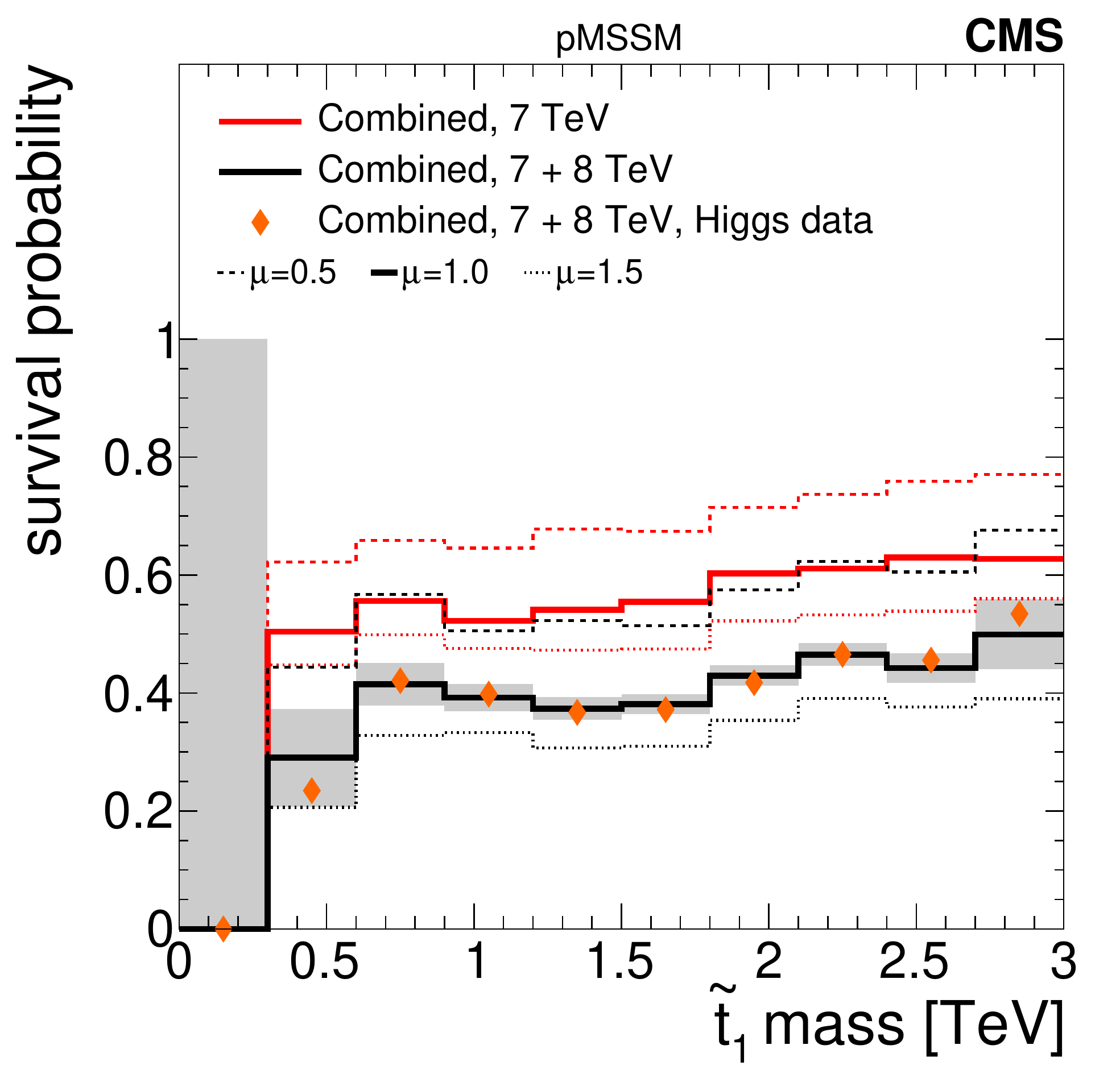}
  \includegraphics[width=0.33\textwidth]{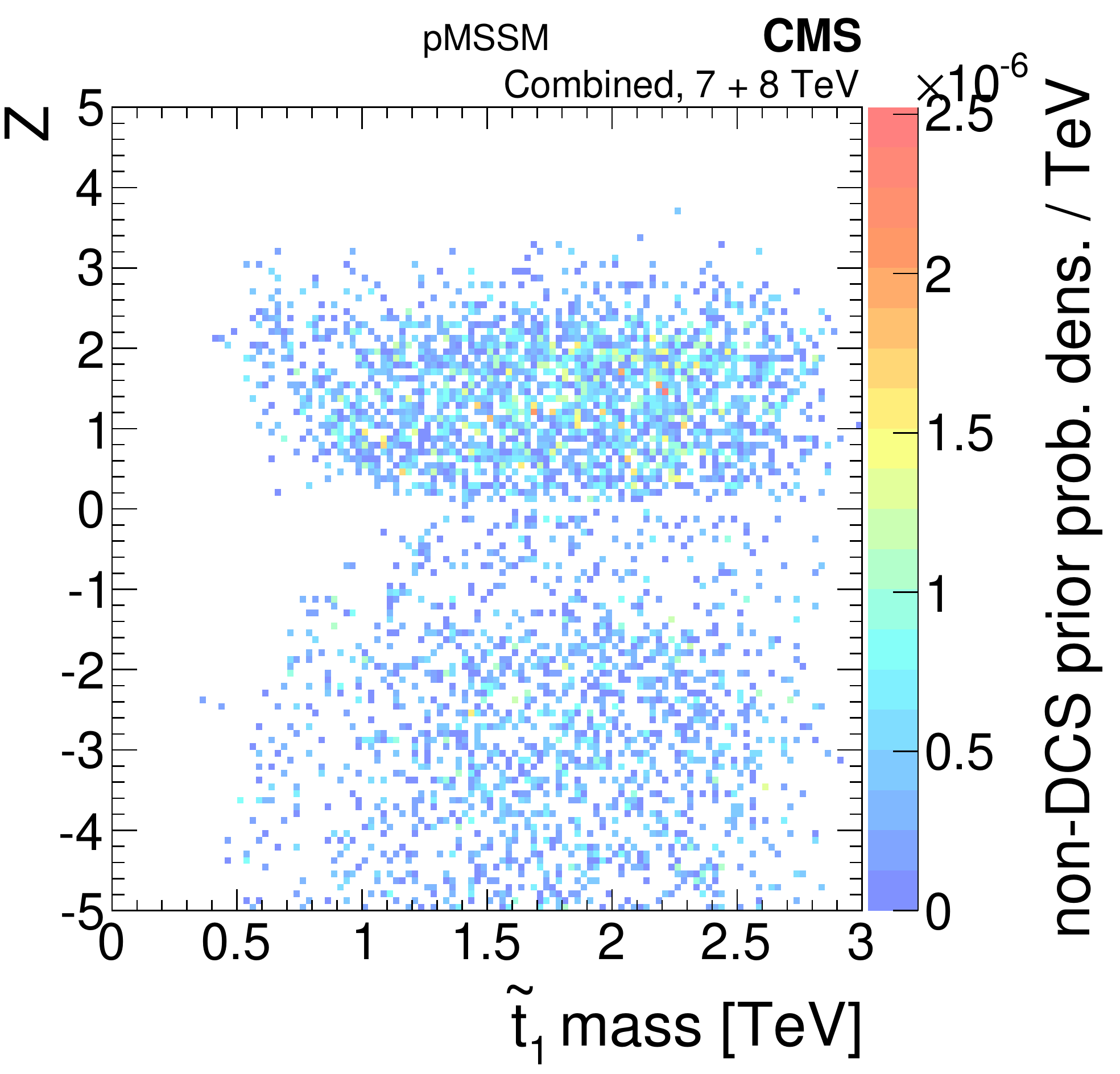}
    \caption{
        A summary of the impact of CMS searches on the probability density of the $\PSQt_1$ mass in the pMSSM parameter space.
    The first-row and bottom-left plots compare the \preCMS prior distribution of the $\PSQt_1$ mass to posterior distributions after data from various CMS searches, where the bottom-left plot  shows the combined effect of CMS searches and the Higgs boson results.
    The bottom-center plot shows survival probabilities as a function of the $\PSQt_1$ mass for various combinations of CMS data and data from Higgs boson measurements.
    The bottom-right plot shows the distribution of the $\PSQt_1$ mass versus the $Z$-significance calculated from the combination of all searches. See Fig. \ref{fig:mg} for a description of the shading.}
    \label{fig:mt1}
\end{figure*}

Figure \ref{fig:mt1} illustrates what information this set of searches provides about the mass of the lightest top squark $\stl$.
The difference between the prior and posterior distributions is minor. The reason is that the low-energy measurements like
 the b $\to$ s $\gamma$ branching fraction (see
Table \ref{tab:preCMS}) impose much stronger constraints on the mass of the $\stl$ than do the considered analyses.
This is not to say the CMS analyses are insensitive to top squark masses. The posterior distribution for the \MTtwo{}
search exhibits an enhancement at $m_{\stl}<1\TeV$ relative to the \preCMS distribution. This enhancement does not
appear in the combined posterior density because is suppressed by observations of other more sensitive searches.
 In the distribution of $m_{\stl}$ versus $Z$, the positive (negative) $Z$ values have a slight negative (positive)
correlation with the $\stl$ mass below 1\TeV, indicating that the CMS analyses considered have some
direct sensitivity to top squarks with masses up to 1\TeV. The overall conclusion is that light top squarks
with masses of the order of 500\GeV cannot be excluded.

\begin{figure*}[t]
  \includegraphics[width=0.33\textwidth]{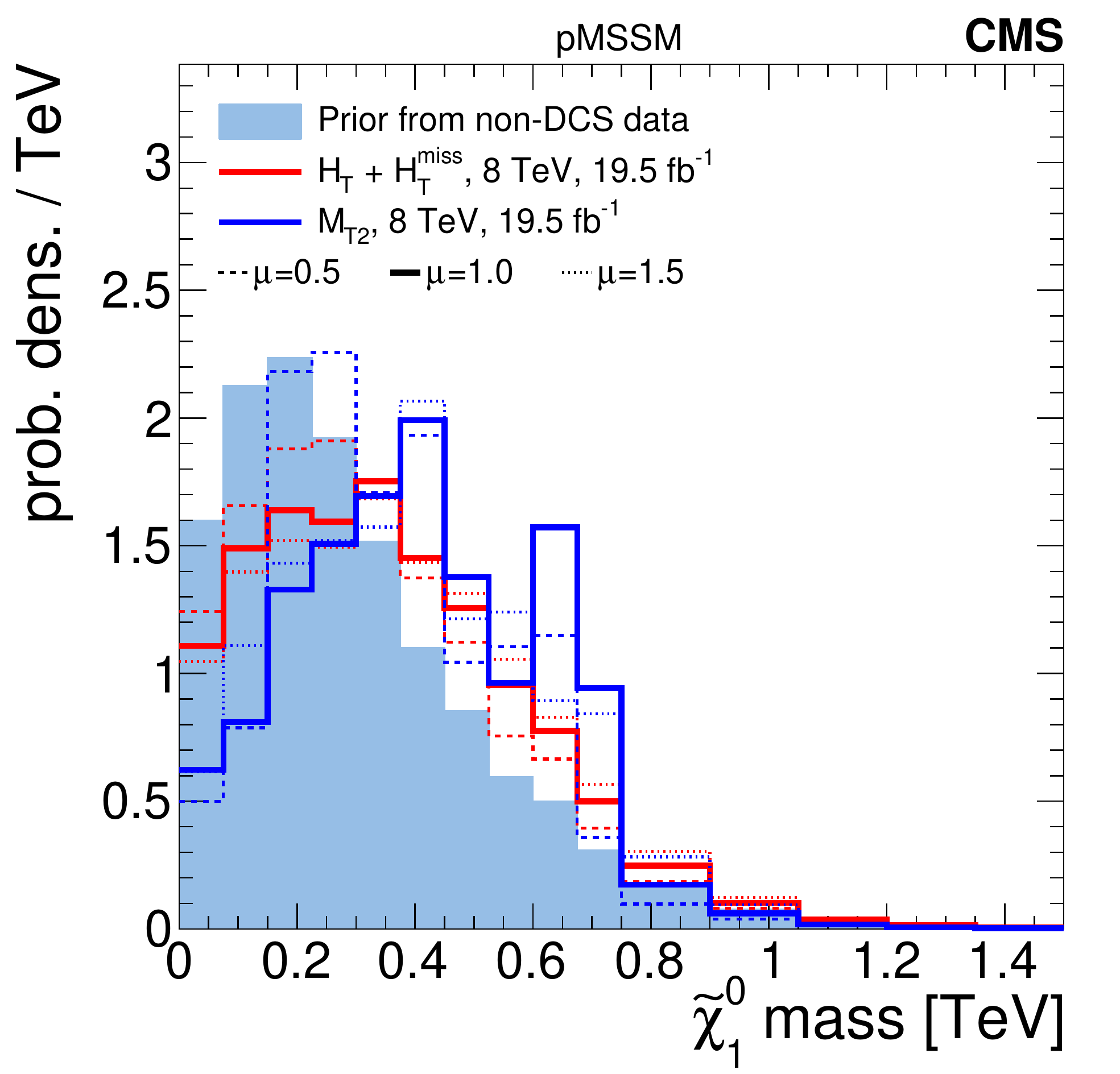}
  \includegraphics[width=0.33\textwidth]{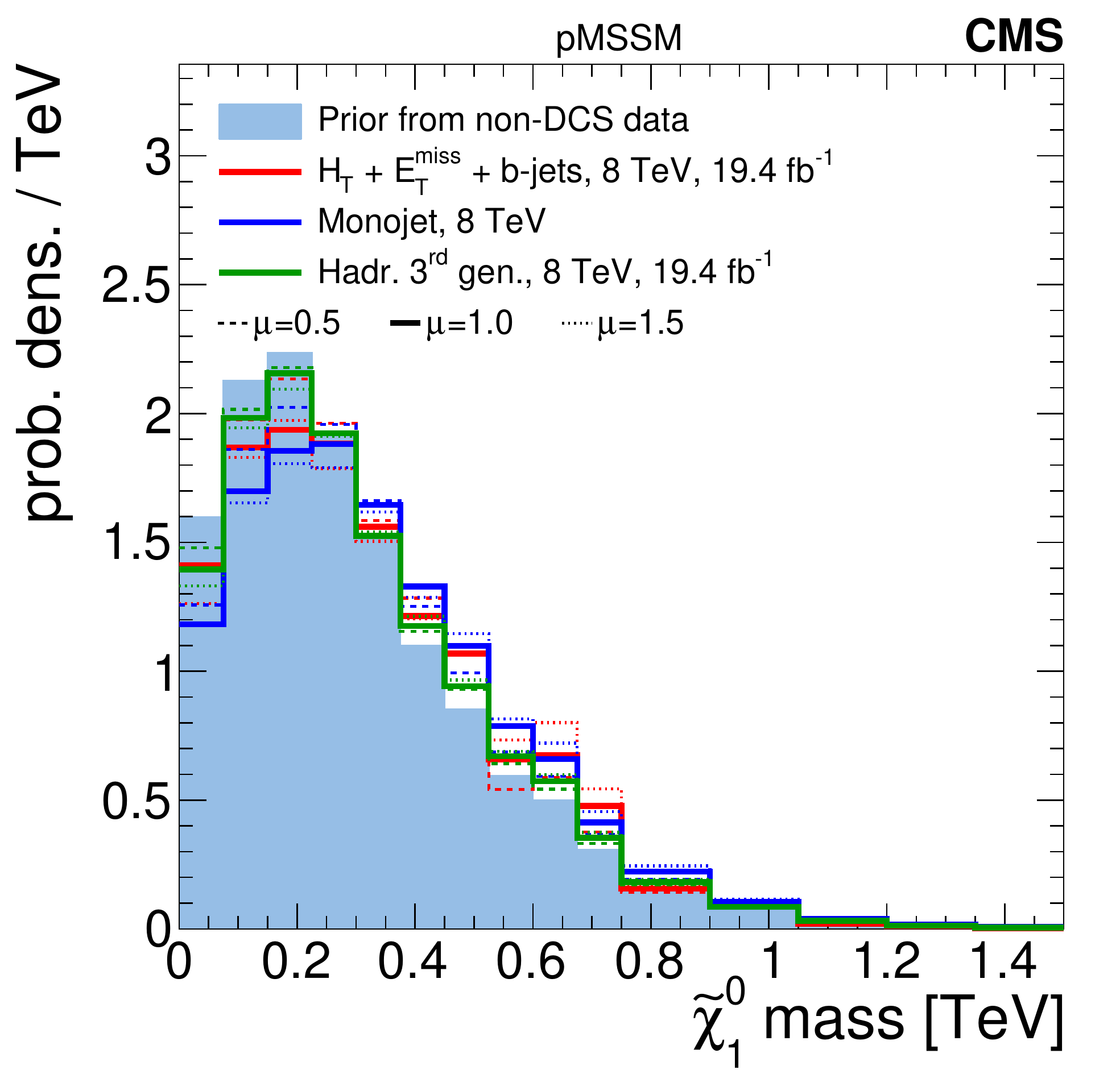}
  \includegraphics[width=0.33\textwidth]{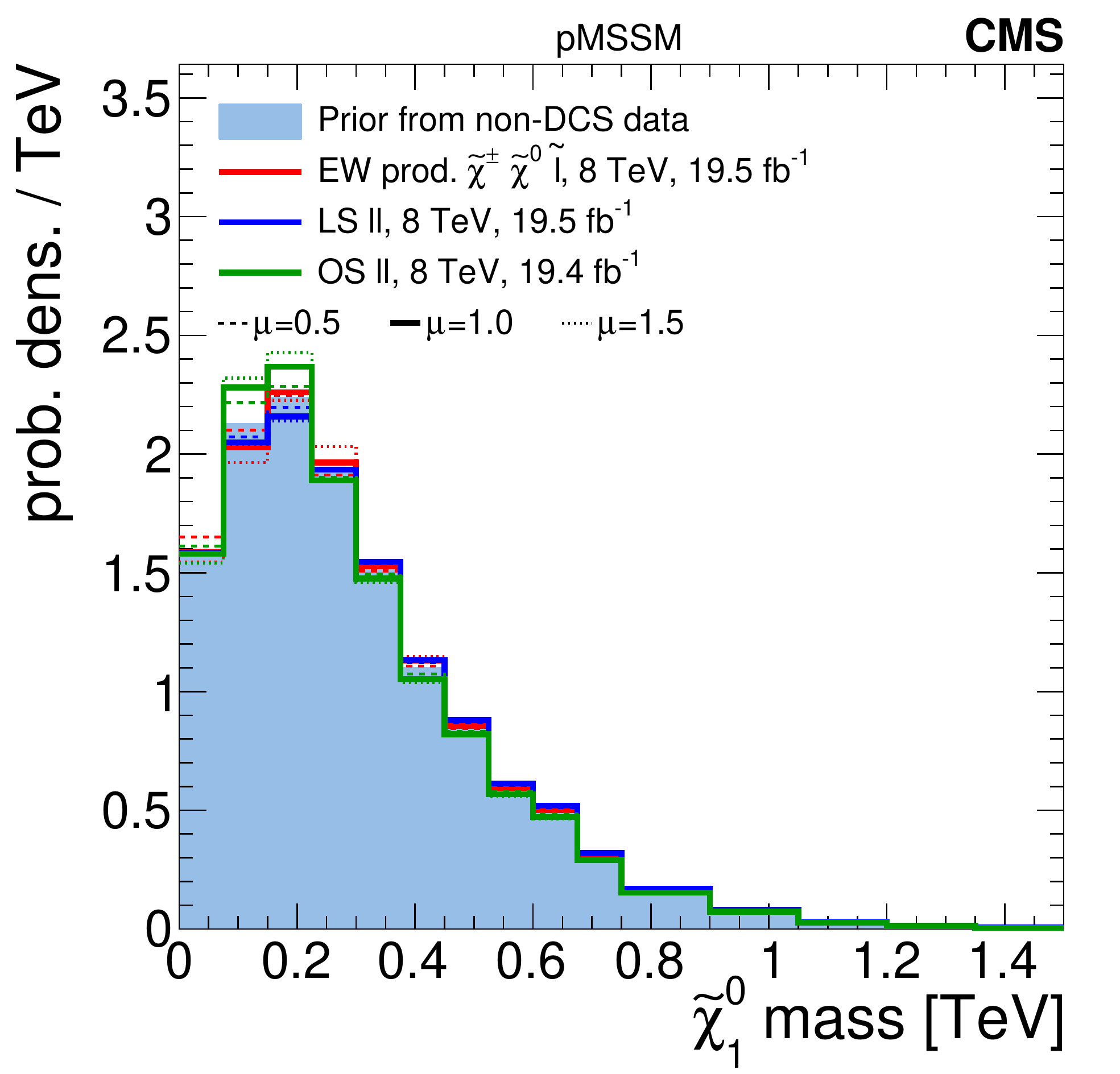}
  \includegraphics[width=0.33\textwidth]{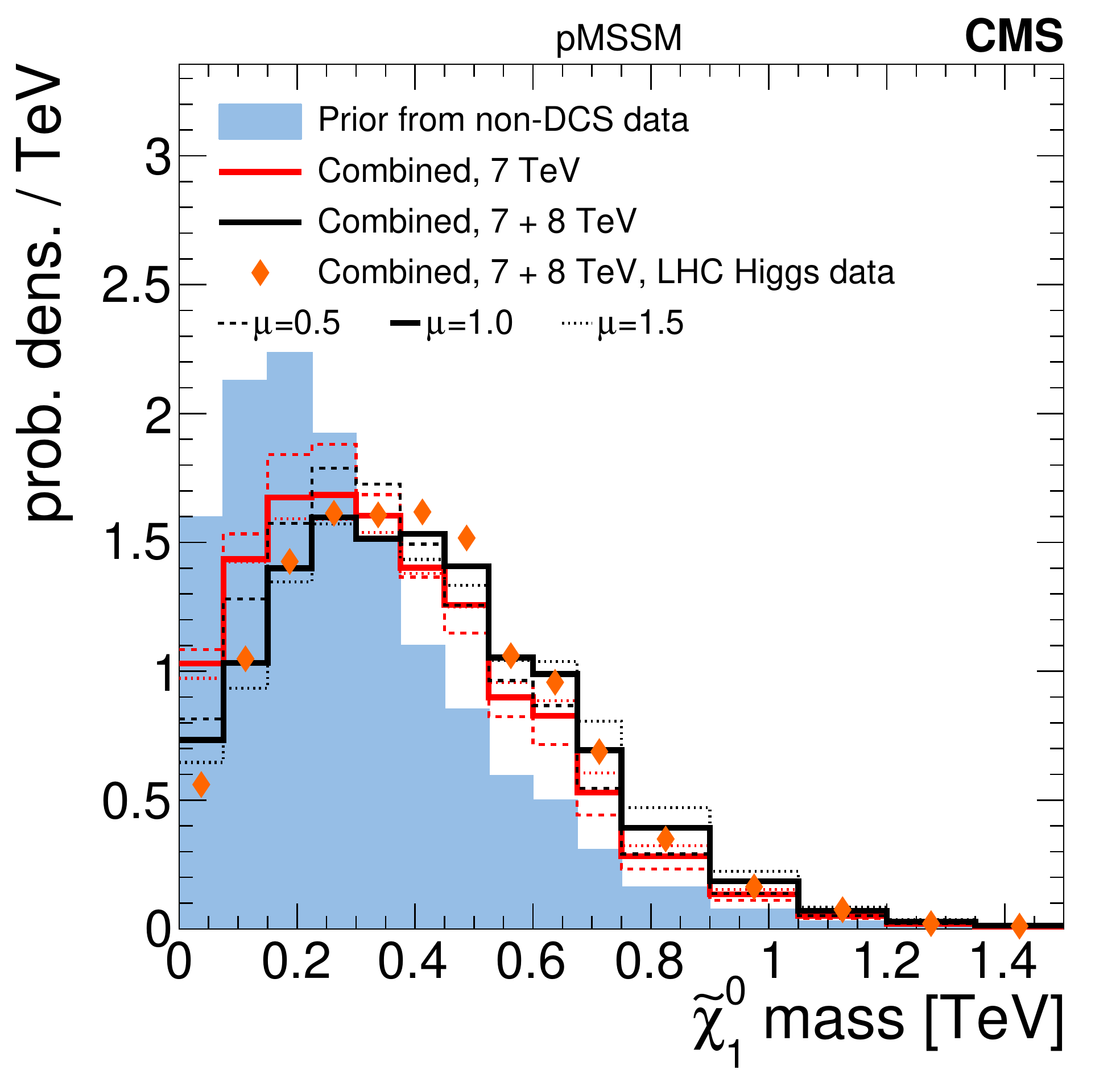}
  \includegraphics[width=0.33\textwidth]{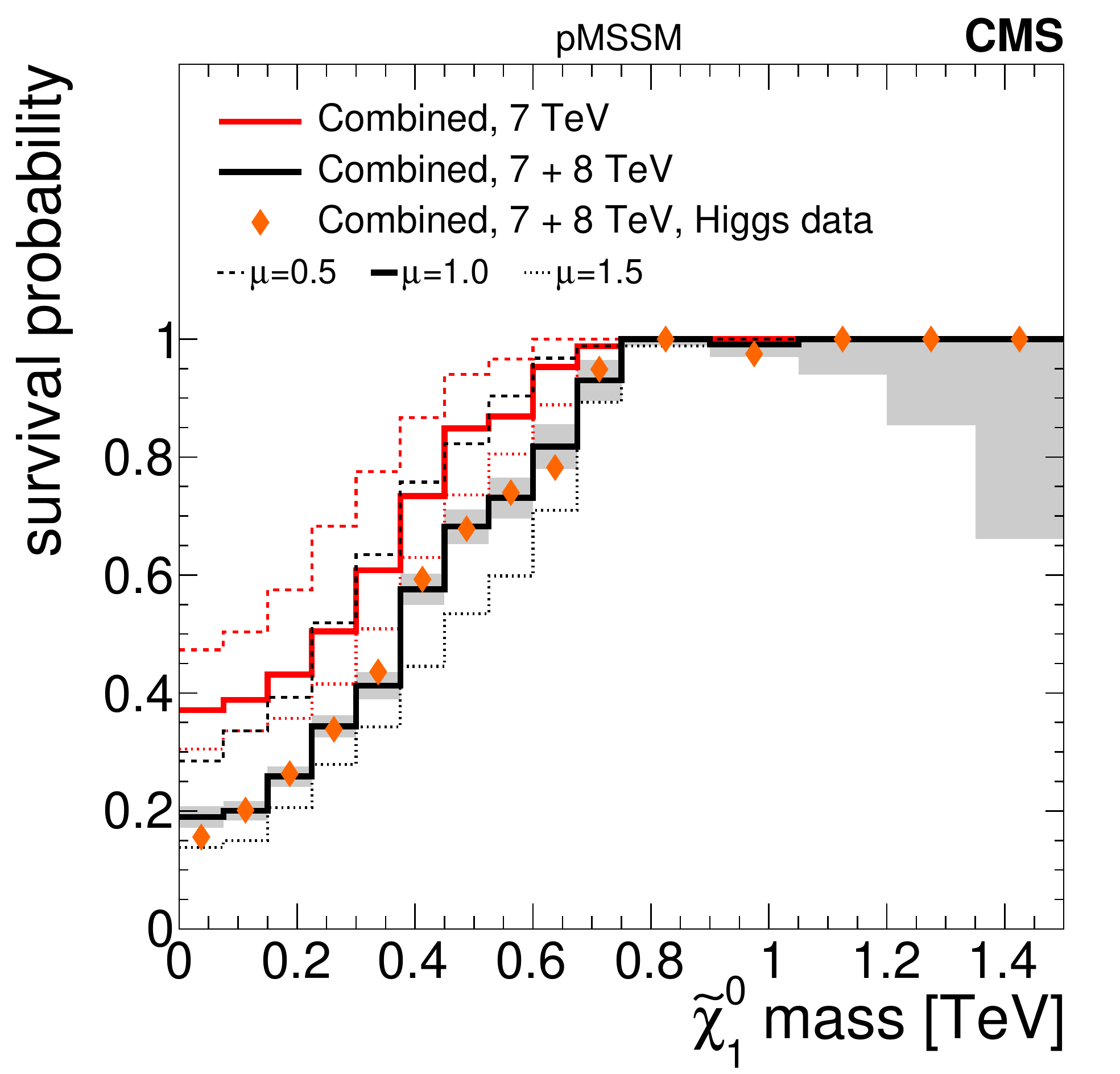}
  \includegraphics[width=0.33\textwidth]{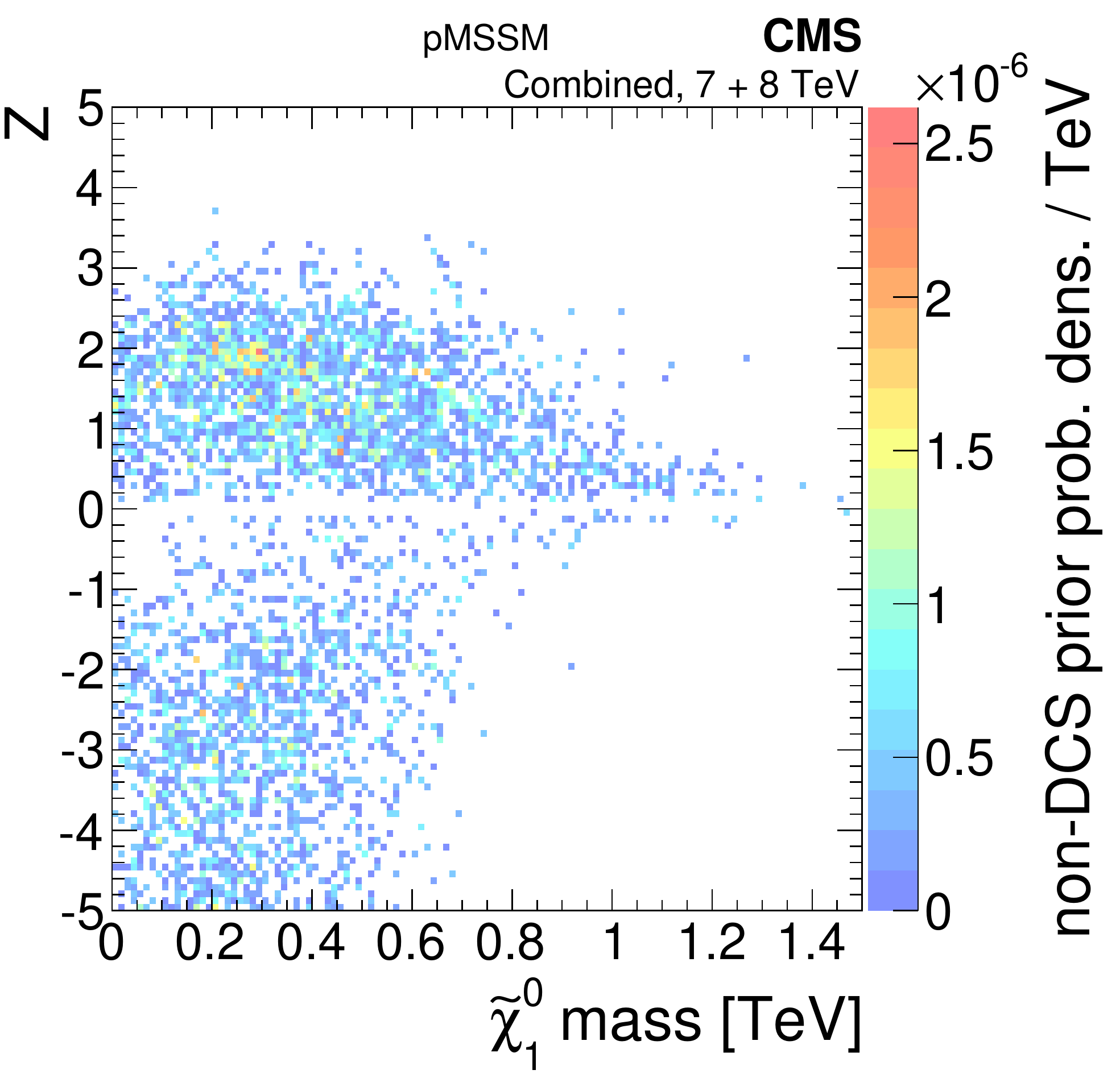}
    \caption{
        A summary of the impact of CMS searches on the probability density of the  $\PSGczDo$ mass in the pMSSM parameter space.
    The first-row and bottom-left plots compare the \preCMS prior distribution of the  $\PSGczDo$ mass to posterior distributions after data from various CMS searches, where the bottom-left plot  shows the combined effect of CMS searches and the Higgs boson results.
    The bottom-center plot shows survival probabilities as a function of the  $\PSGczDo$ mass for various combinations of CMS data and data from Higgs boson measurements.
    The bottom-right plot shows the distribution of the  $\PSGczDo$ mass versus the $Z$-significance calculated from the combination of all searches. See Fig. \ref{fig:mg} for a description of the shading.}
    \label{fig:mz1}
\end{figure*}

Turning now to the EW sector, we first show, in
Fig. \ref{fig:mz1}, the effect of the considered searches on our knowledge of the
mass of the lightest neutralino $\chiz$.  We see that the hadronic
inclusive searches disfavor low $\chiz$ masses; the hadronic
searches targeting specific topologies also have an effect, although
smaller, and the leptonic searches have a marginal impact.  The
$7{+}8\TeV$ combined distribution is very similar to the \MTtwo{}
distribution, especially in the lower mass region, indicating that this search is the most sensitive to the
$\chiz$ mass.  The main constraint
on the $\chiz$ mass arises indirectly through correlations with other sparticle masses.  Since $\chiz$ is
the LSP, its mass is constrained by the masses of
the heavier sparticles.  As CMS searches push the probability
distributions for the colored particles to higher values, more phase
space opens for $\chiz$ and the $\chiz$ distributions shift to higher
values.  The survival probability distribution shows that no $\chiz$
mass is totally excluded at the 95\% CL by CMS.  In general, the nonexcluded points with light $\chiz$ are those with heavy colored sparticles.  The fact that the survival probability decreases below a $\chiz$ mass of $\sim$700\GeV shows that CMS searches are sensitive up to this mass value.
The Higgs boson data disfavor neutralino masses below about 60\GeV, that is,
the mass range in which invisible decays
$h\to\PSGczDo\PSGczDo$ could occur; this is visible in the
first bin in Fig. \ref{fig:mz1} (bottom-left) (see Ref.~\cite{Bernon:2014vta}).

\begin{figure*}[t]
  \includegraphics[width=0.33\textwidth]{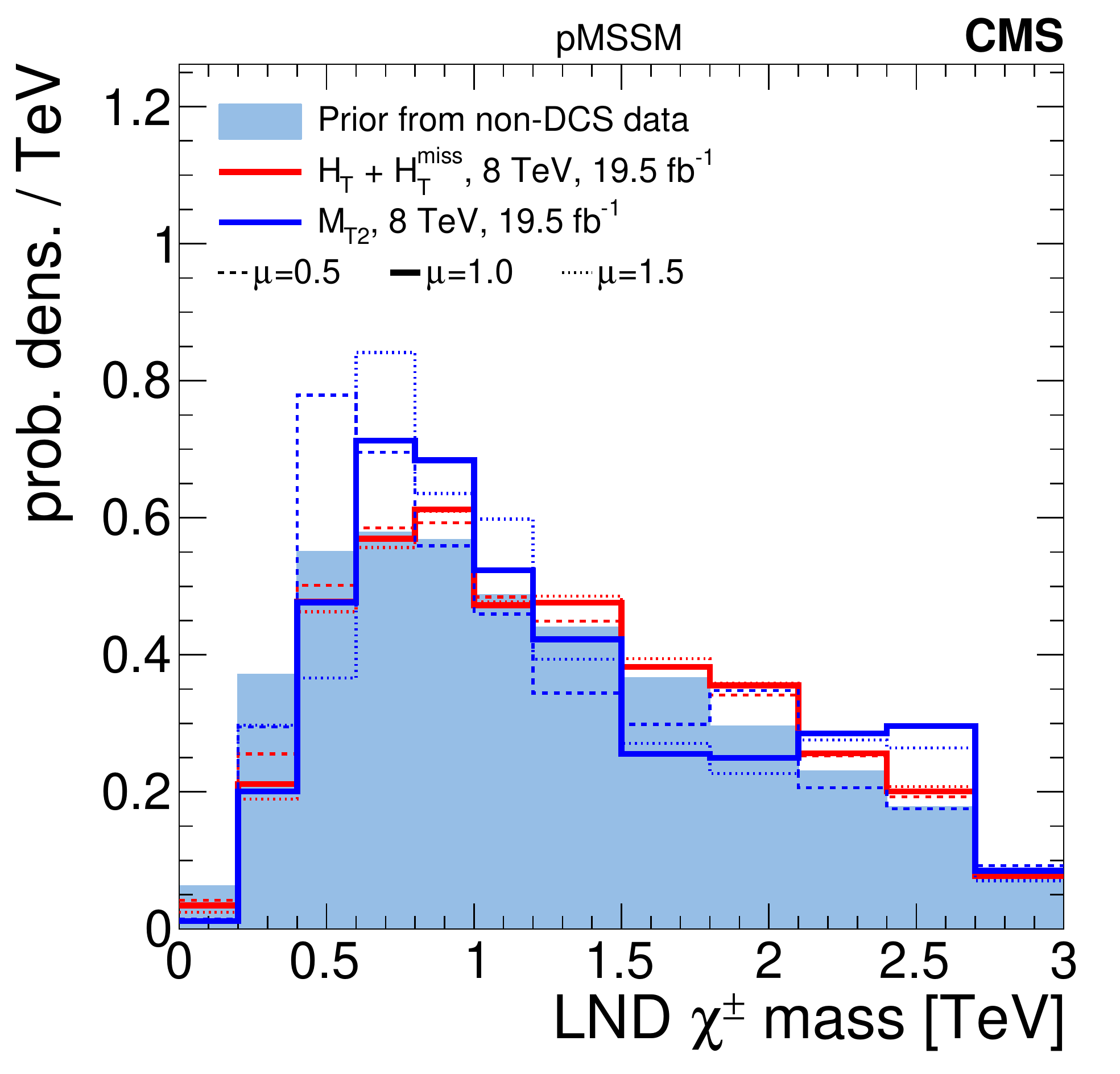}
  \includegraphics[width=0.33\textwidth]{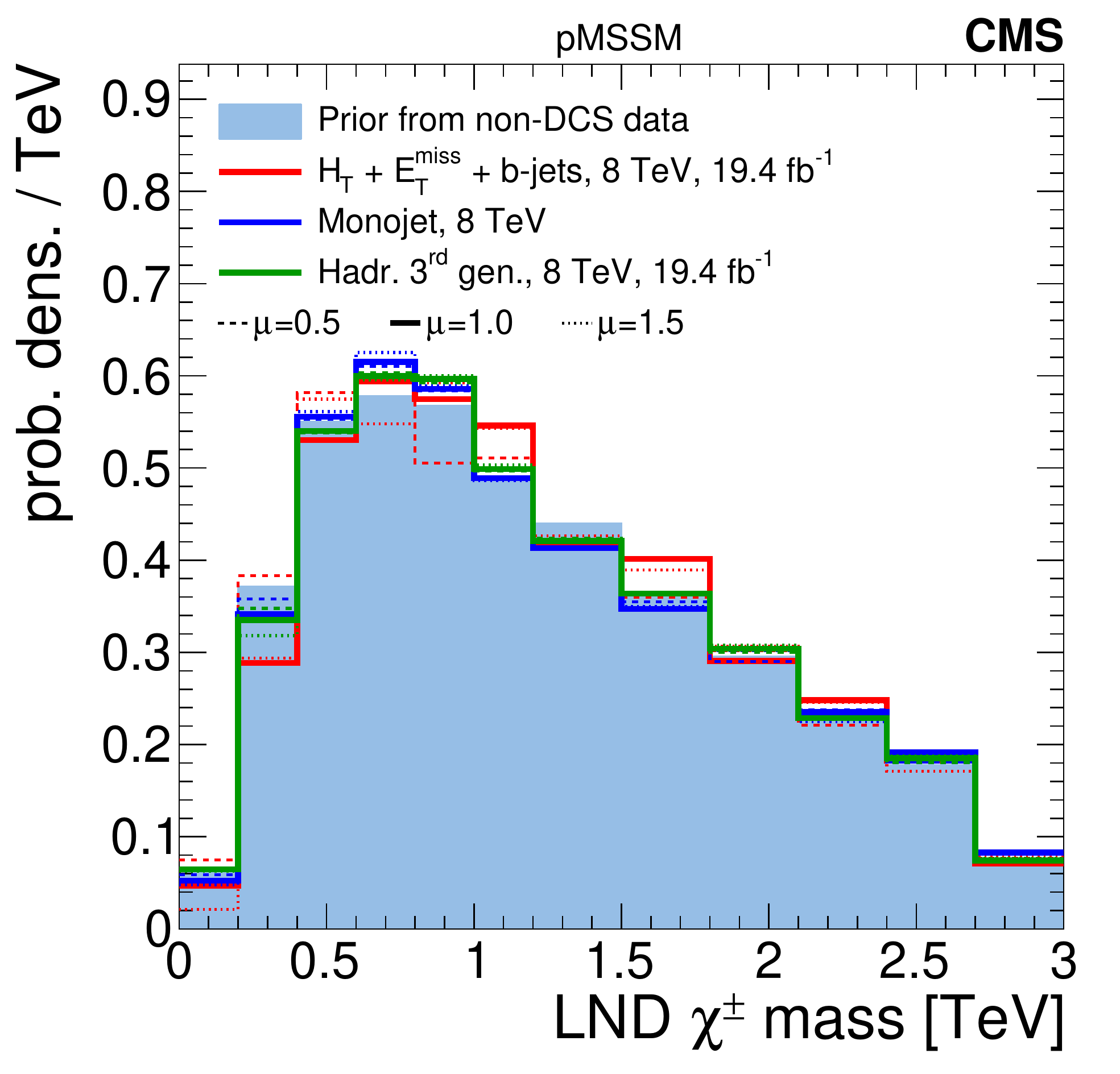}
  \includegraphics[width=0.33\textwidth]{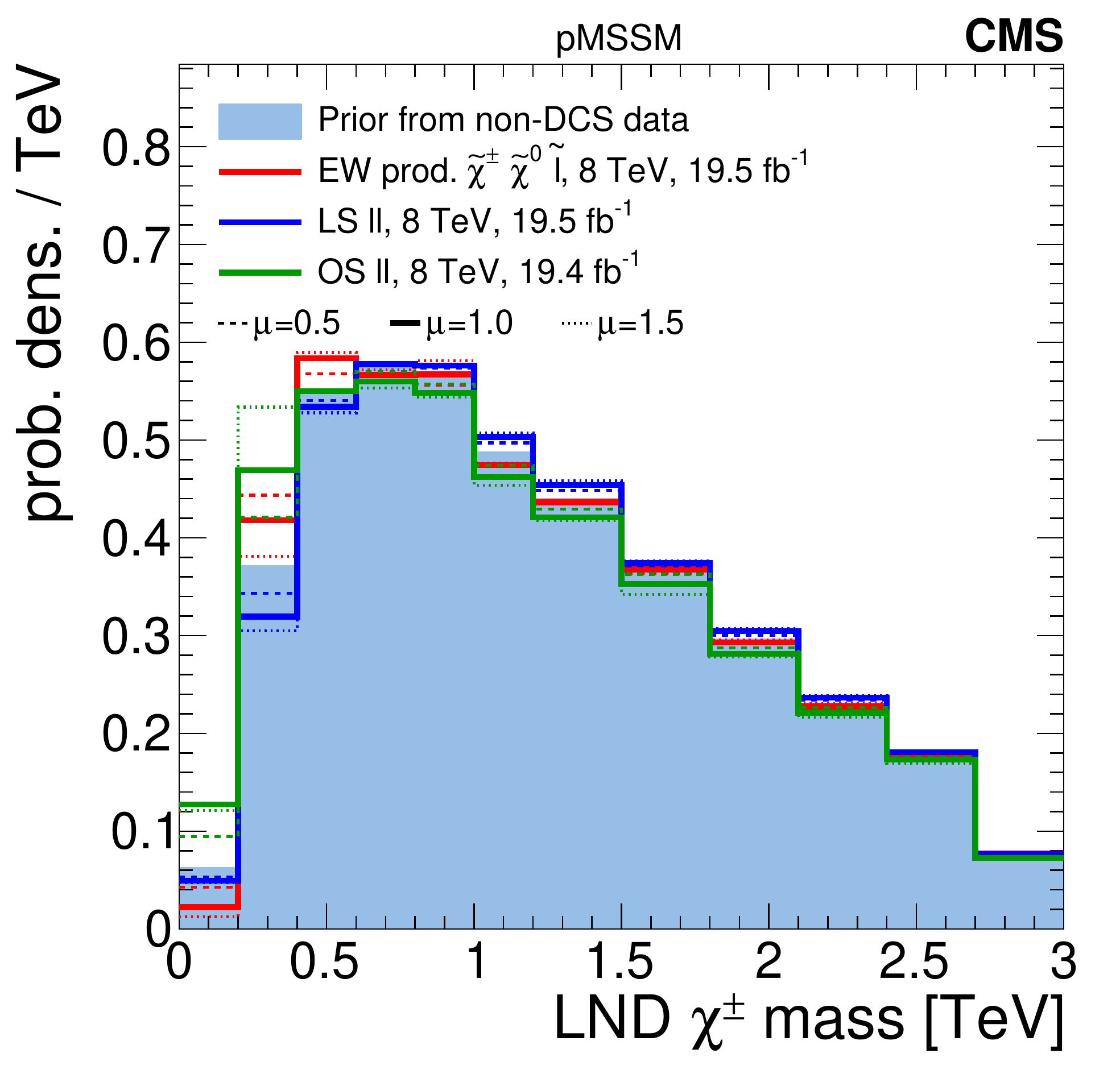}
  \includegraphics[width=0.33\textwidth]{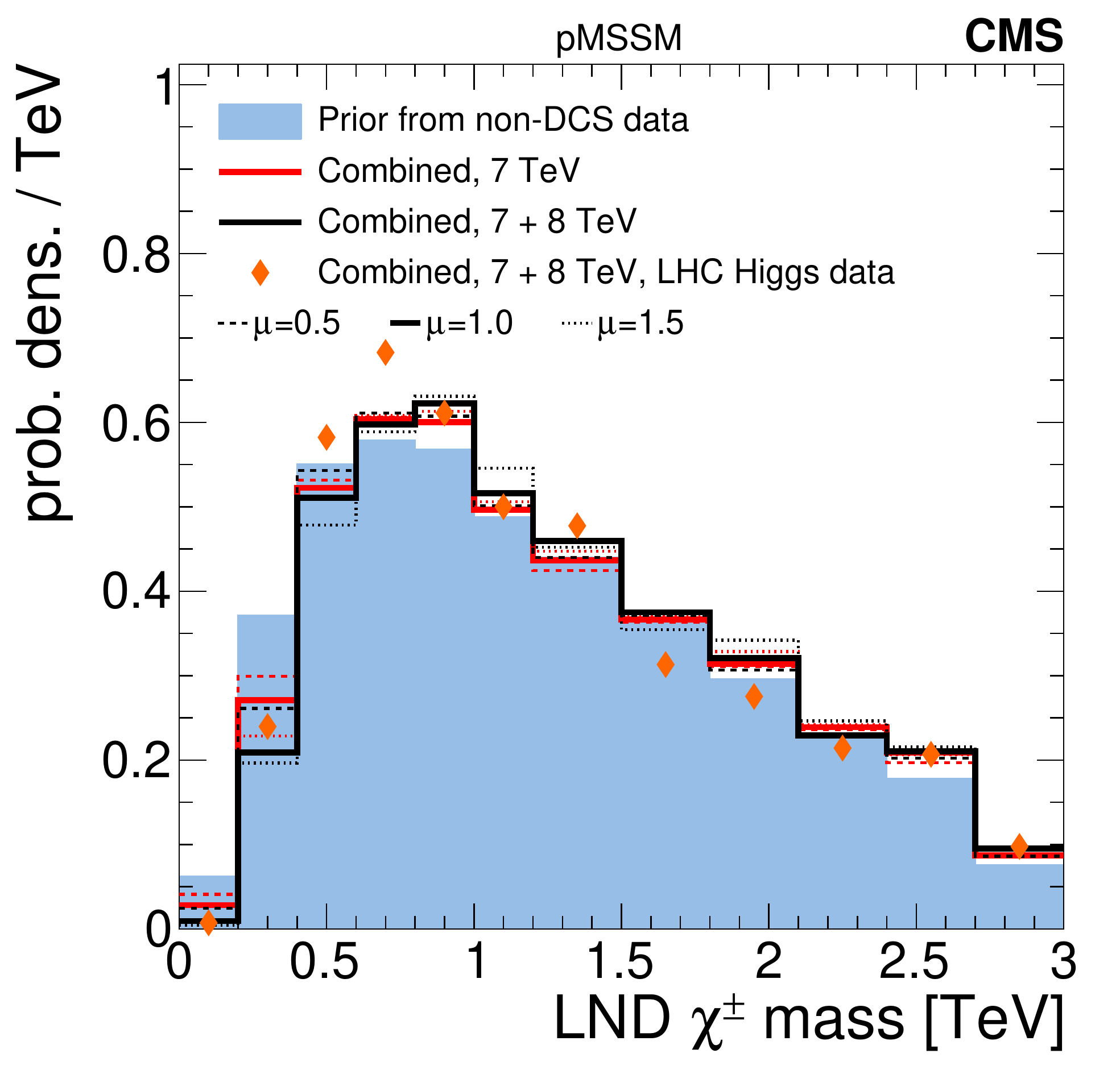}
  \includegraphics[width=0.33\textwidth]{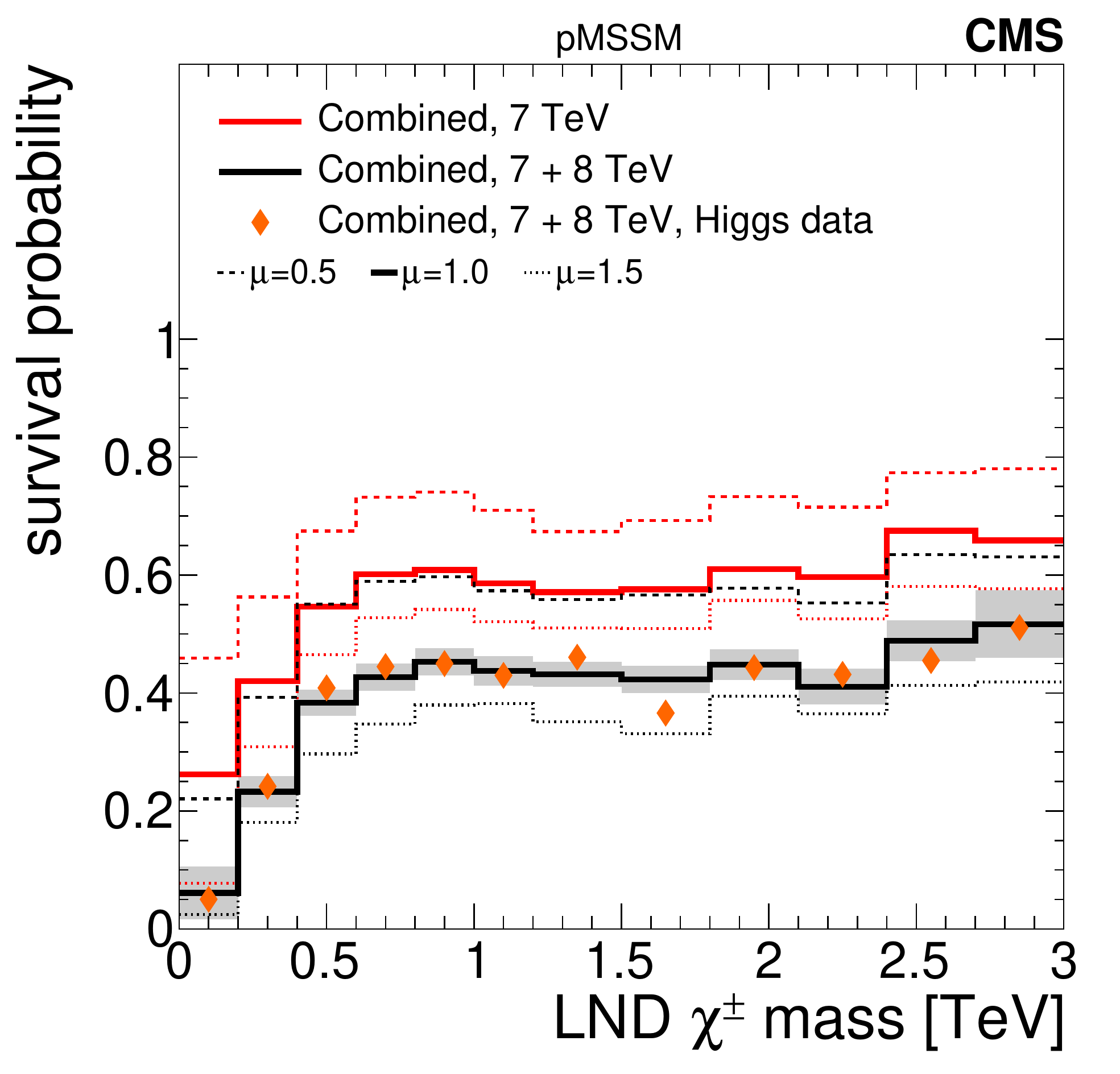}
  \includegraphics[width=0.33\textwidth]{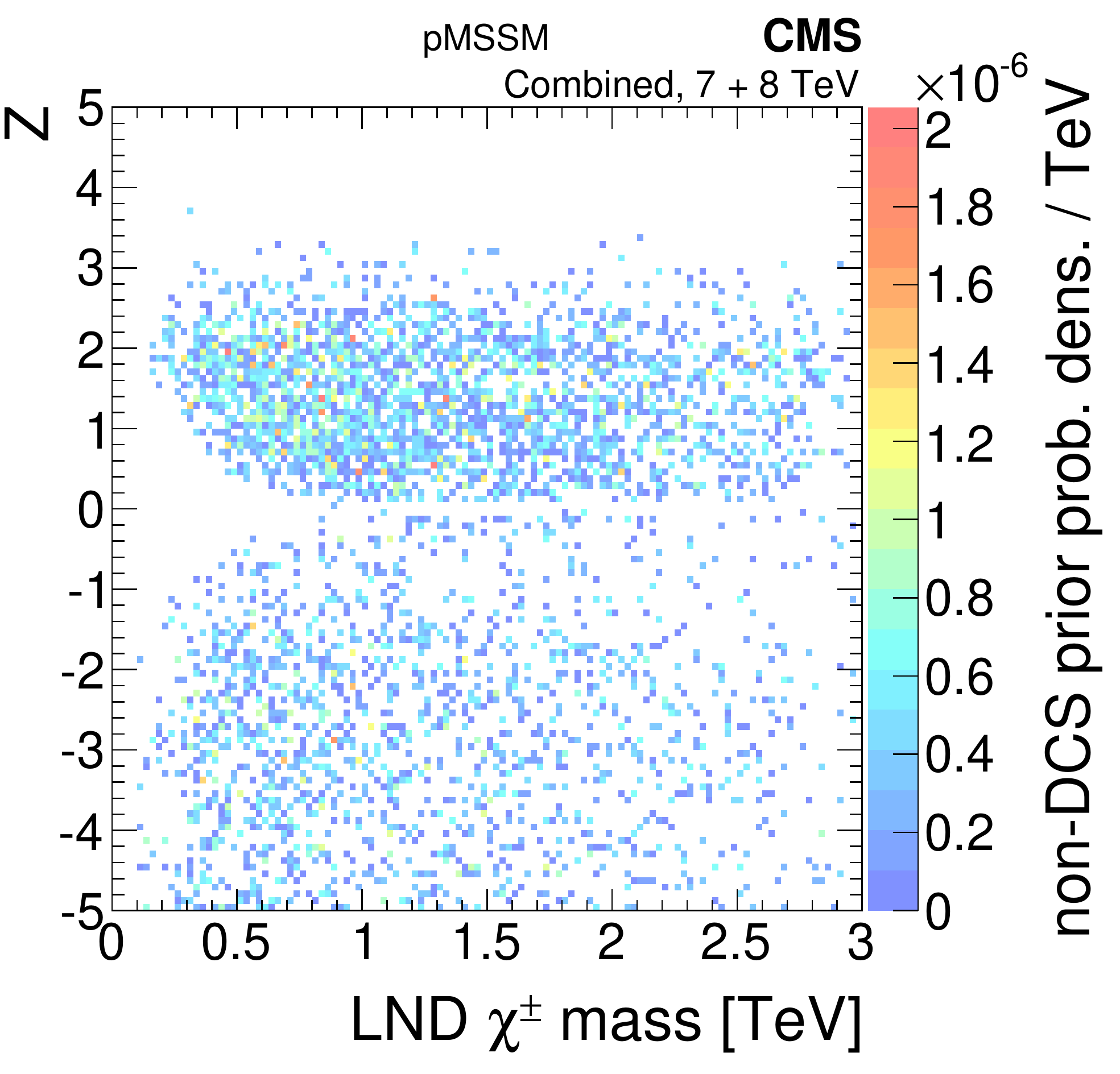}
  \caption{A summary of the impact of CMS searches on the probability density of the mass of the lightest non-degenerate (LND) chargino in the pMSSM parameter space.
    The first-row and bottom-left plots compare the \preCMS prior distribution of the LND $\widetilde{\chi}^\pm$ mass to posterior distributions after data from various CMS searches, where the bottom-left plot  shows the combined effect of CMS searches and the Higgs boson results.
    The bottom-center plot shows survival probabilities as a function of the LND $\widetilde{\chi}^\pm$ mass for various combinations of CMS data and data from Higgs boson measurements.
    The bottom-right plot shows the distribution of the LND $\widetilde{\chi}^\pm$ mass versus the $Z$-significance calculated from the combination of all searches. See Fig. \ref{fig:mg} for a description of the shading.}
 \label{fig:mLNDw}
\end{figure*}

In the MSSM, the lightest chargino becomes degenerate with the lightest neutralino for the condition
$|M_1| \geq \min(|M_2|,|\mu|)$.
Therefore, we define the lightest non-degenerate (LND) chargino as
\begin{equation}
    \mathrm{LND~}{\chi^\pm} =
    \begin{cases}
        \PSGcpm_1 & \mbox{if~~~} |M_1| < \min(|M_2|,|\mu|) \\
        \PSGcpm_2 & \mbox{if~~~} |M_1| > \min(|M_2|,|\mu|).
    \end{cases}
\end{equation}
Figure \ref{fig:mLNDw} summarizes what information has been gained about the mass of the
LND chargino.
Again, the impact of the CMS searches is found to be rather limited and no chargino mass
can be reliably excluded.
It is worth noticing the impact of the leptonic searches.
In Fig. \ref{fig:mLNDw} (top-right), the distributions differ from the \preCMS distribution,
while these searches have negligible impact on most of the other  SUSY observables and parameters considered in this study. We also note that the survival probability is lowest in the first bin where the LND $\chi^{\pm}$ mass is between 0 and 200\GeV, but a small percentage of points still survive.

\begin{figure*}[t]
  \includegraphics[width=0.33\textwidth]{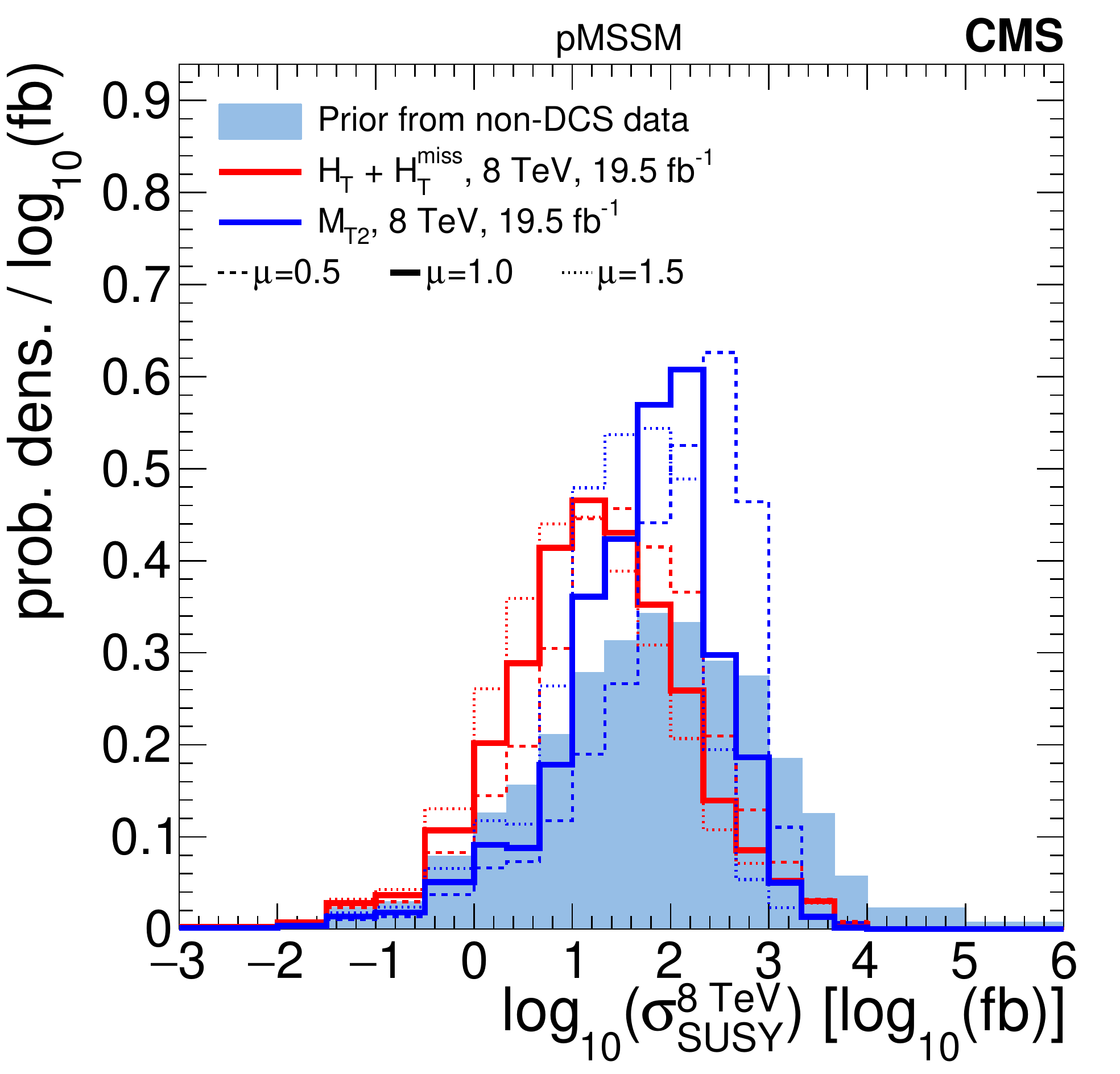}
  \includegraphics[width=0.33\textwidth]{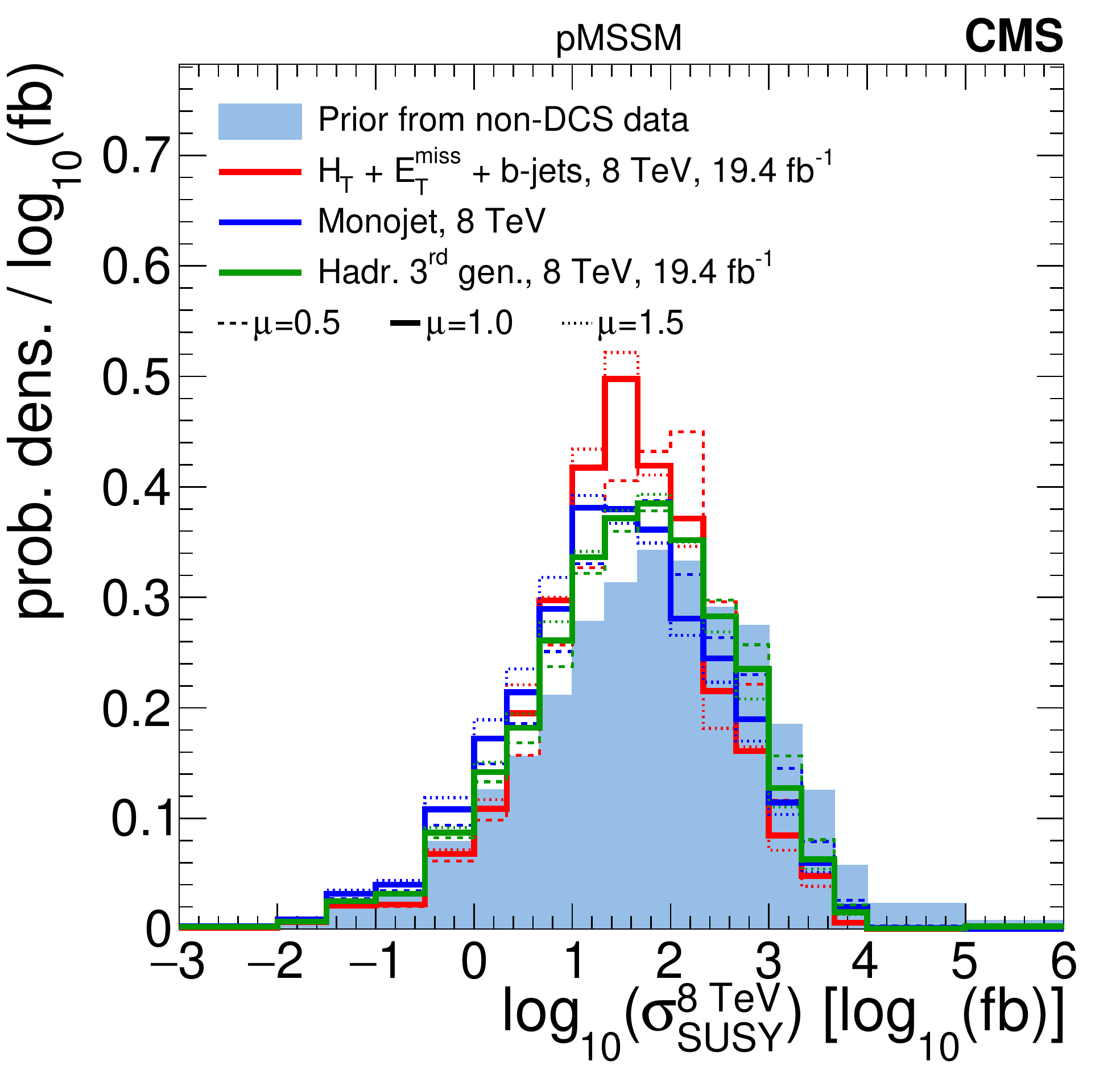}
  \includegraphics[width=0.33\textwidth]{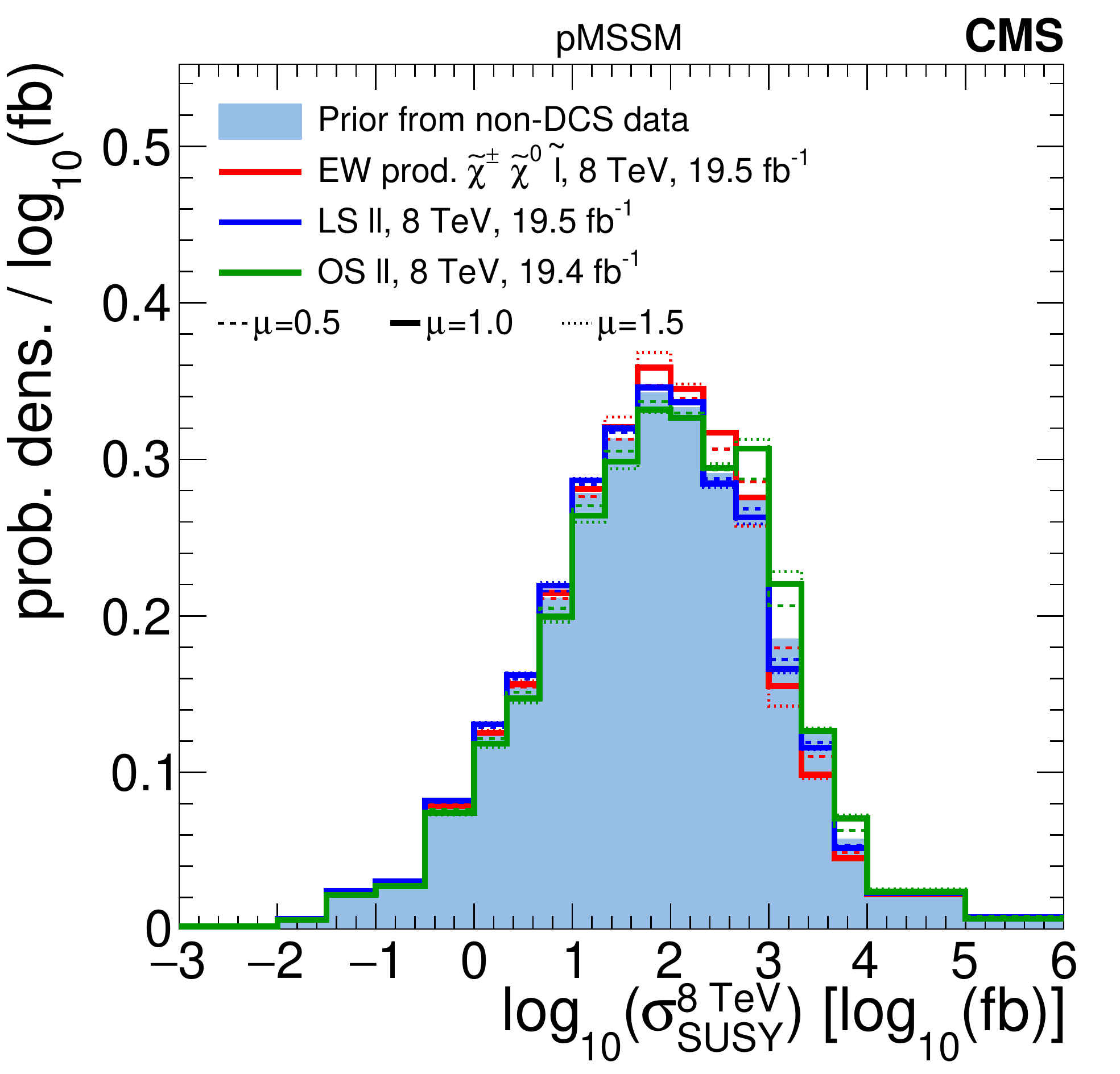}
  \includegraphics[width=0.33\textwidth]{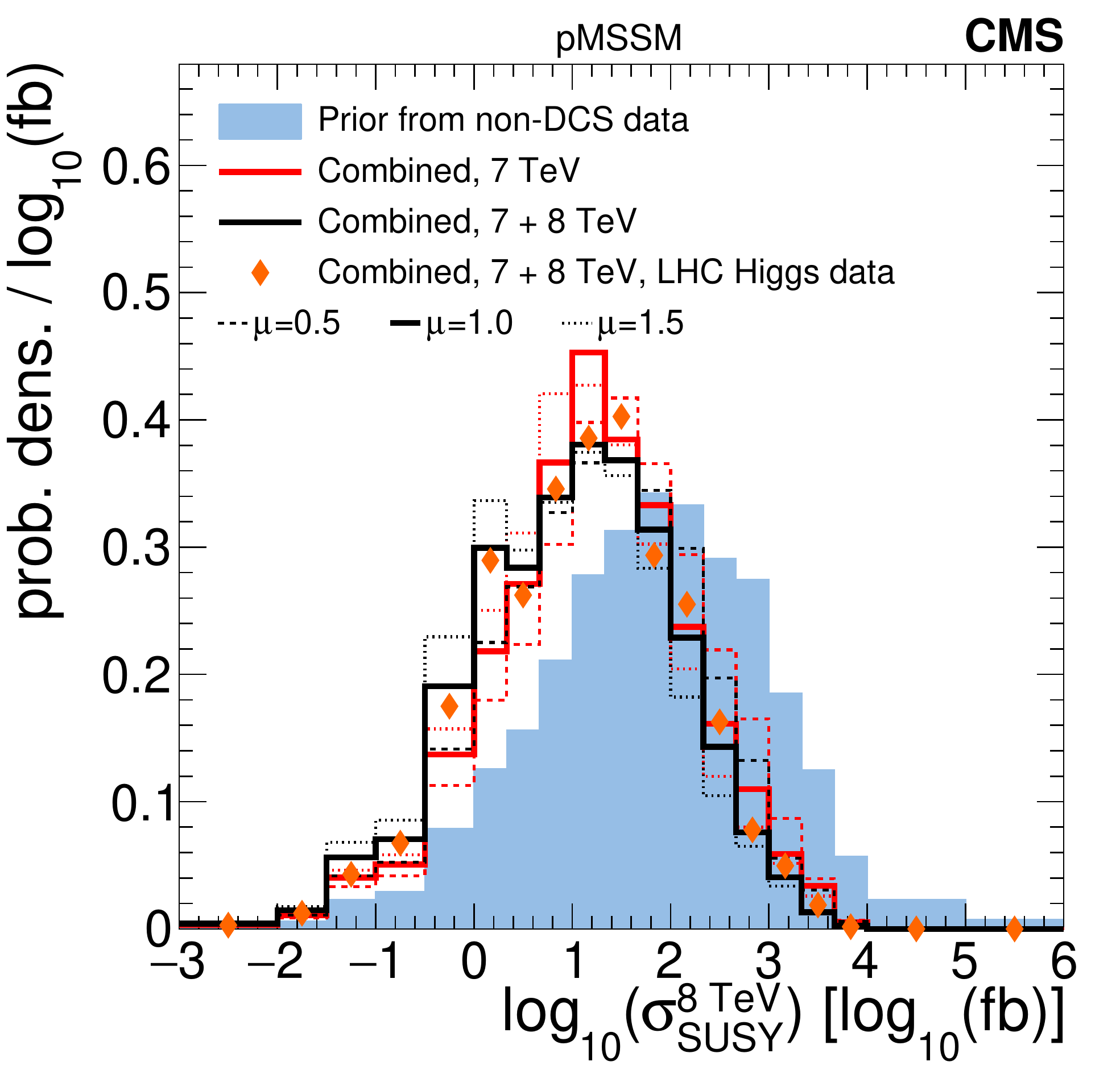}
  \includegraphics[width=0.33\textwidth]{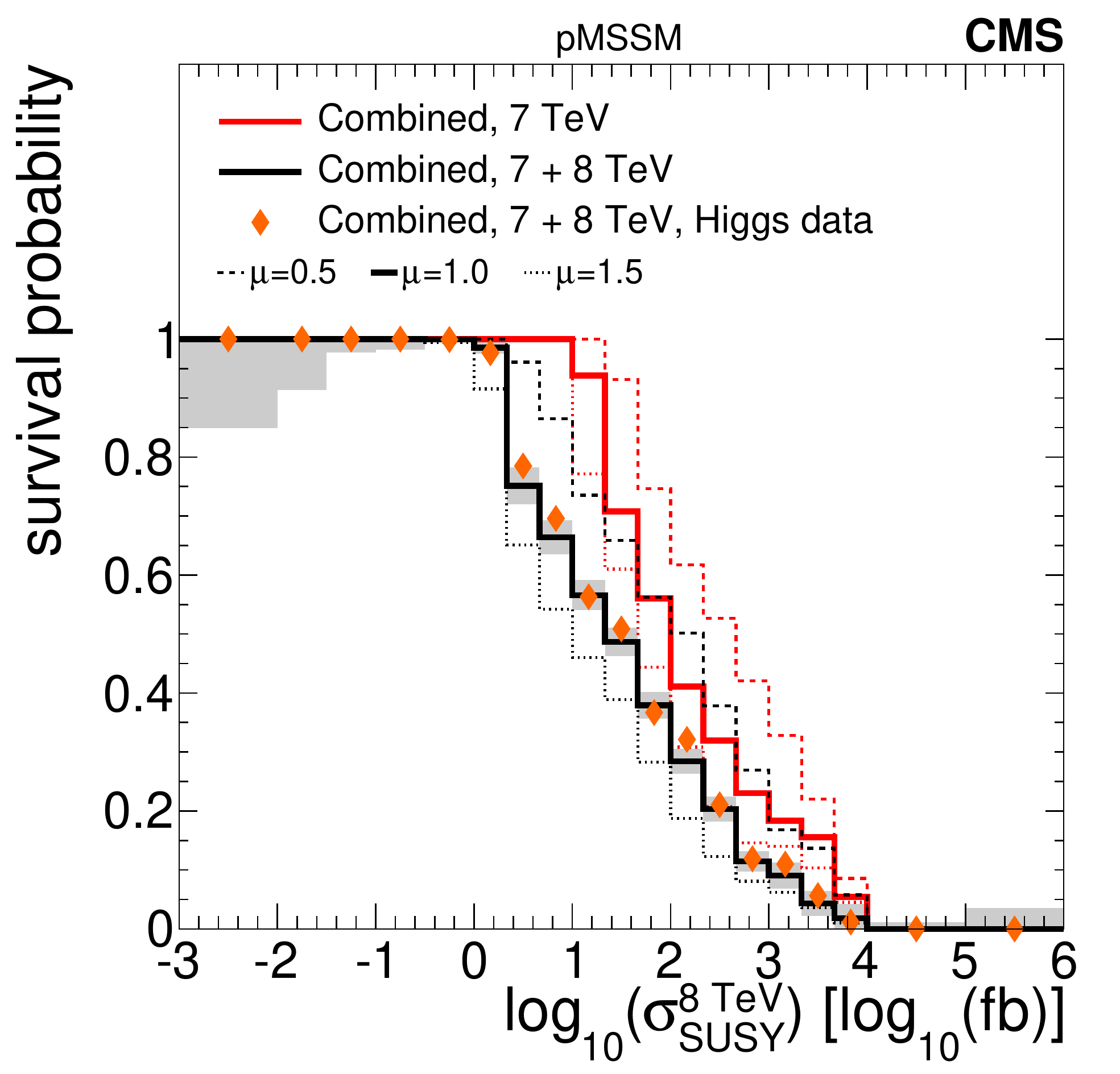}
  \includegraphics[width=0.33\textwidth]{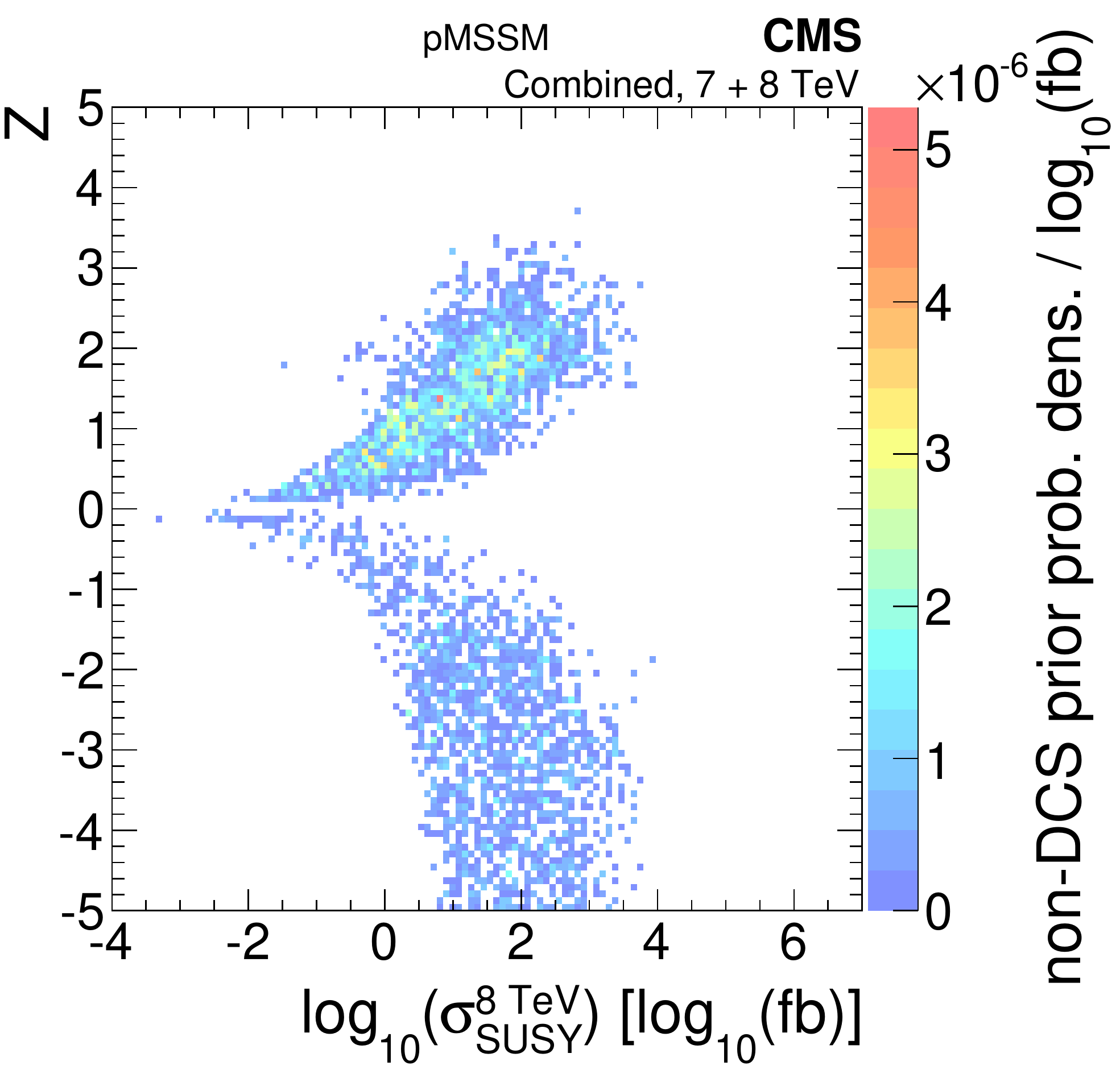}
    \caption{A summary of the impact of CMS searches on the probability density of the  logarithm of the cross
        section for inclusive sparticle production in 8\TeV pp
        collisions, $\log_{10}(\sigma^{8\TeV}_\mathrm{SUSY})$, in the pMSSM parameter space.
    The first-row and bottom-left plots compare the \preCMS prior distribution of the $\log_{10}(\sigma^{8\TeV}_\mathrm{SUSY})$ to posterior distributions after data from various CMS searches, where the bottom-left plot  shows the combined effect of CMS searches and the Higgs boson results.
    The bottom-center plot shows survival probabilities as a function of the $\log_{10}(\sigma^{8\TeV}_\mathrm{SUSY})$ for various combinations of CMS data and data from Higgs boson measurements.
    The bottom-right plot shows the distribution of the $\log_{10}(\sigma^{8\TeV}_\mathrm{SUSY})$ versus the $Z$-significance calculated from the combination of all searches. See Fig. \ref{fig:mg} for a description of the shading. In the bottom-left plot, the apparent enhancement of the left tail of the posterior density with respect to the prior is due to the suppression of the right tail and an overall renormalization.}
    \label{fig:xsect}
\end{figure*}

A more generic view is possible by looking at the overall CMS impact on the inclusive SUSY production cross section for 8\TeV, which is shown in Fig. \ref{fig:xsect}.
The most probable total sparticle cross section in \preCMS prior is
approximately 100\unit{fb}; the low tail of this distribution is shaped by the upper limits on the masses of sparticles in the prior. The effect of the CMS SUSY searches is to reduce this
value by an order of magnitude.  The inclusive \HT{}$+$\MHT{} search
has the largest individual contribution to this because of its ability to address a great diversity of final states comprising different sparticle compositions.  The survival probability distribution confirms that CMS is sensitive to SUSY scenarios with total cross sections as low as 1\unit{fb}.

\begin{figure*}[p]
\centering
\includegraphics[width=0.31\textwidth]{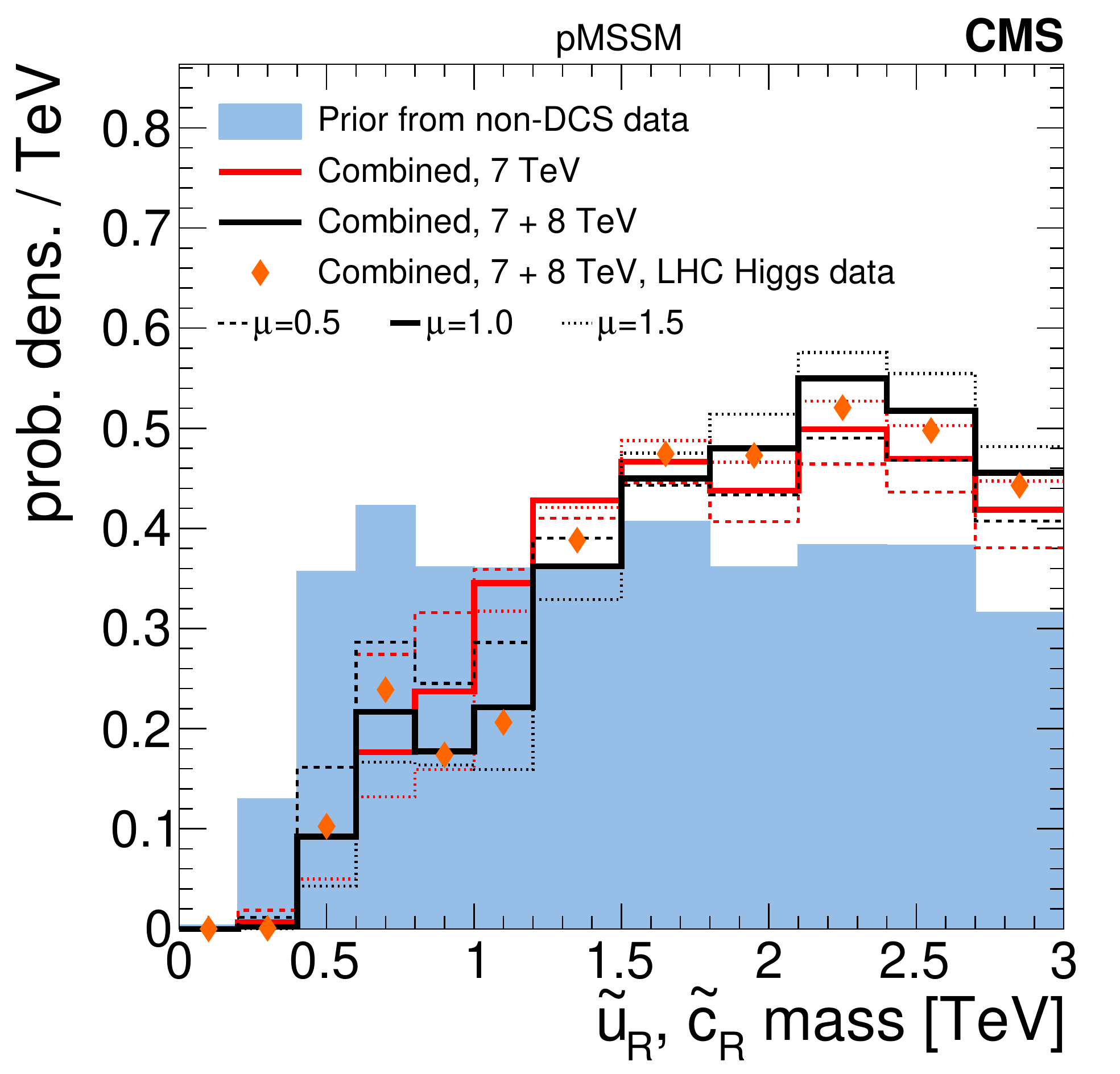}
\includegraphics[width=0.31\textwidth]{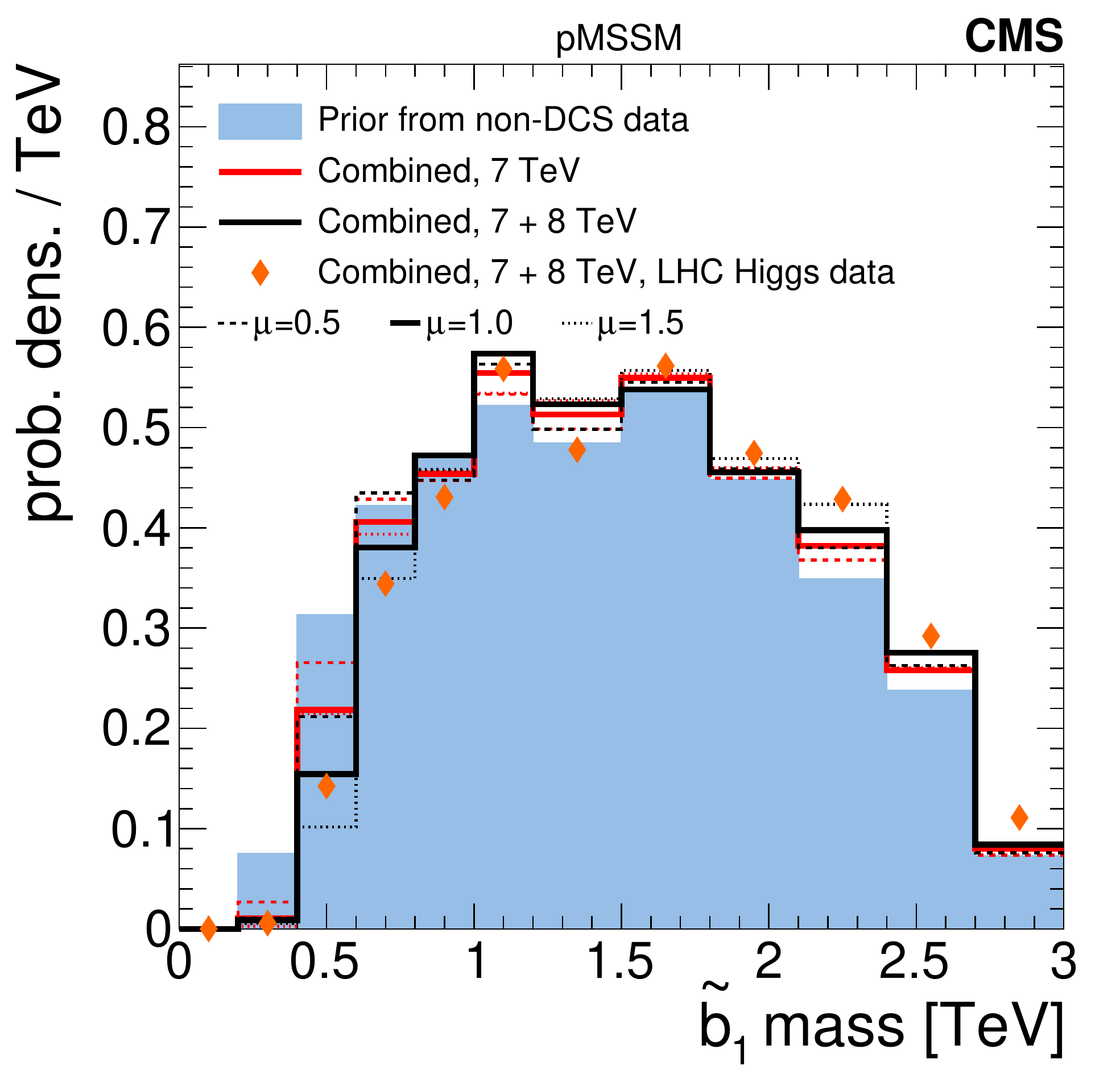}
\includegraphics[width=0.31\textwidth]{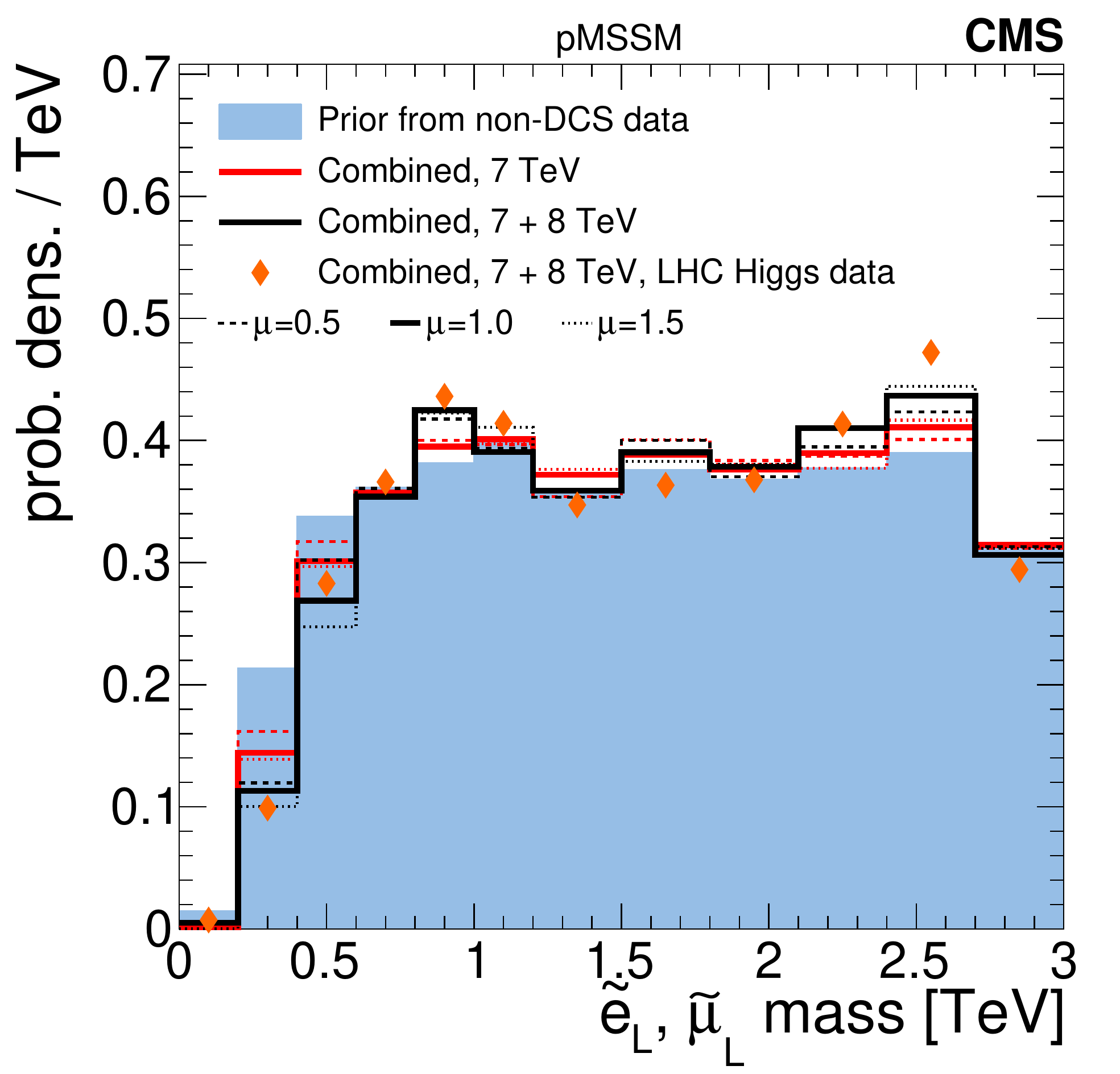}
\includegraphics[width=0.31\textwidth]{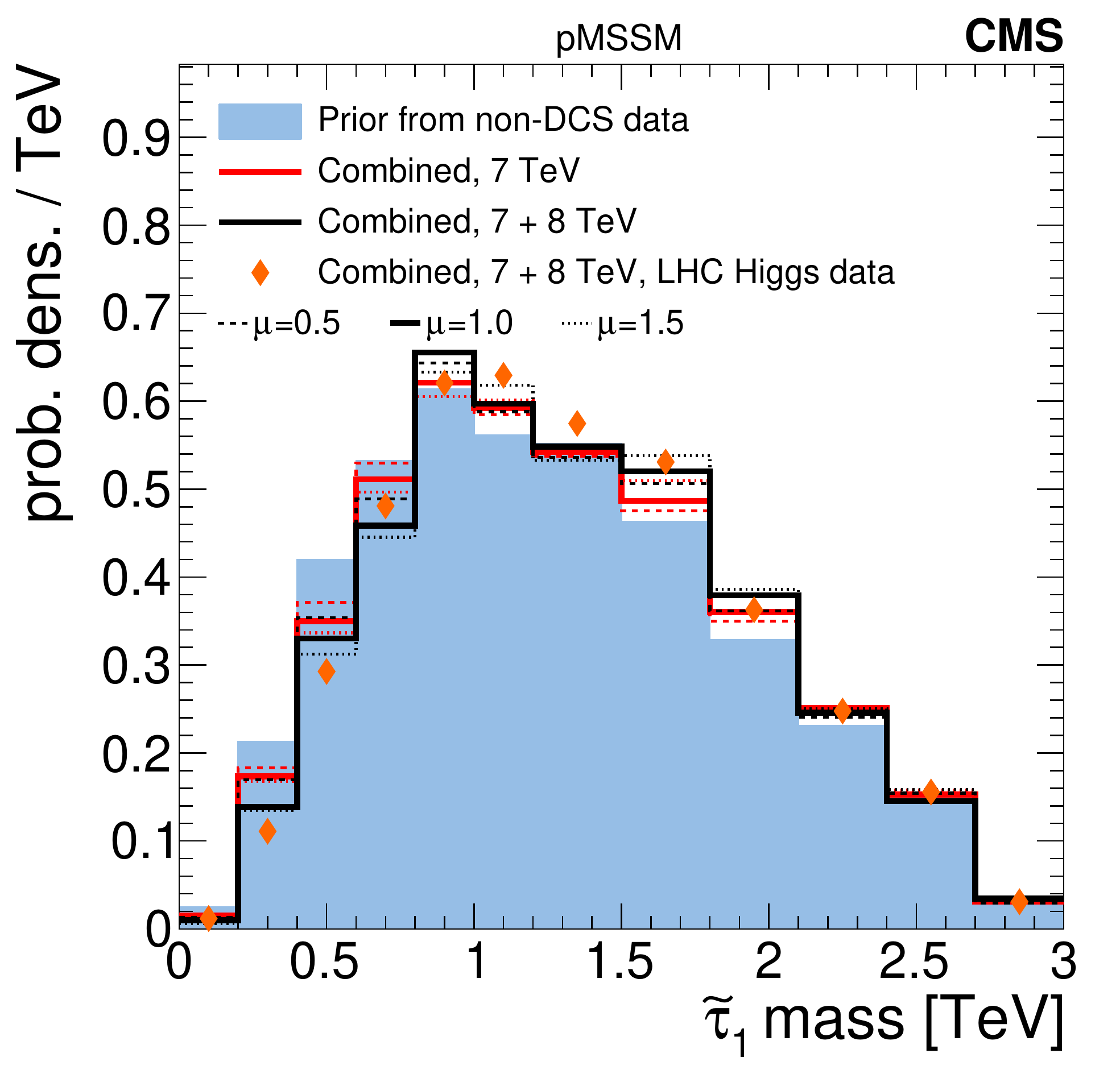}
\includegraphics[width=0.31\textwidth]{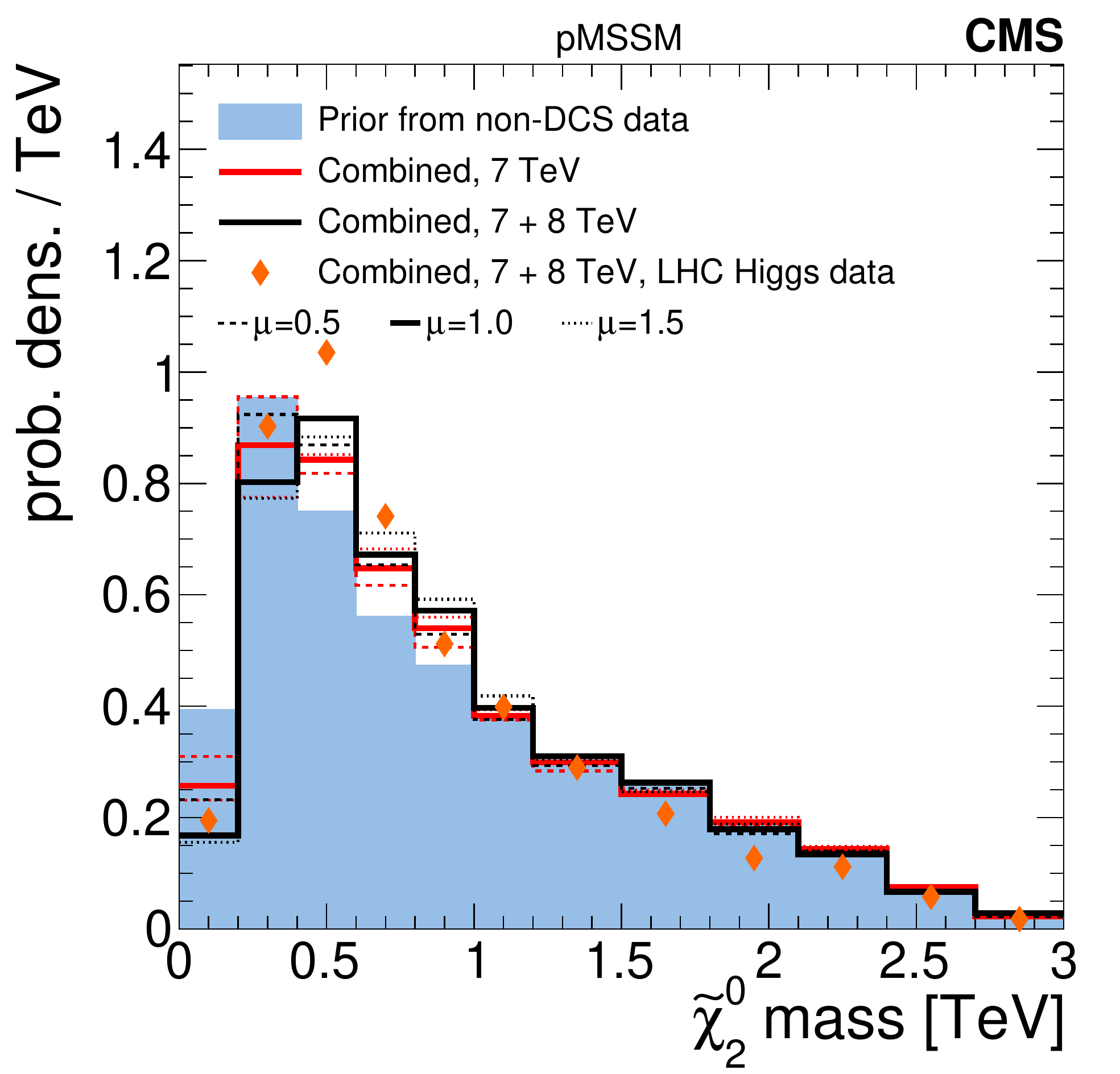}
\includegraphics[width=0.31\textwidth]{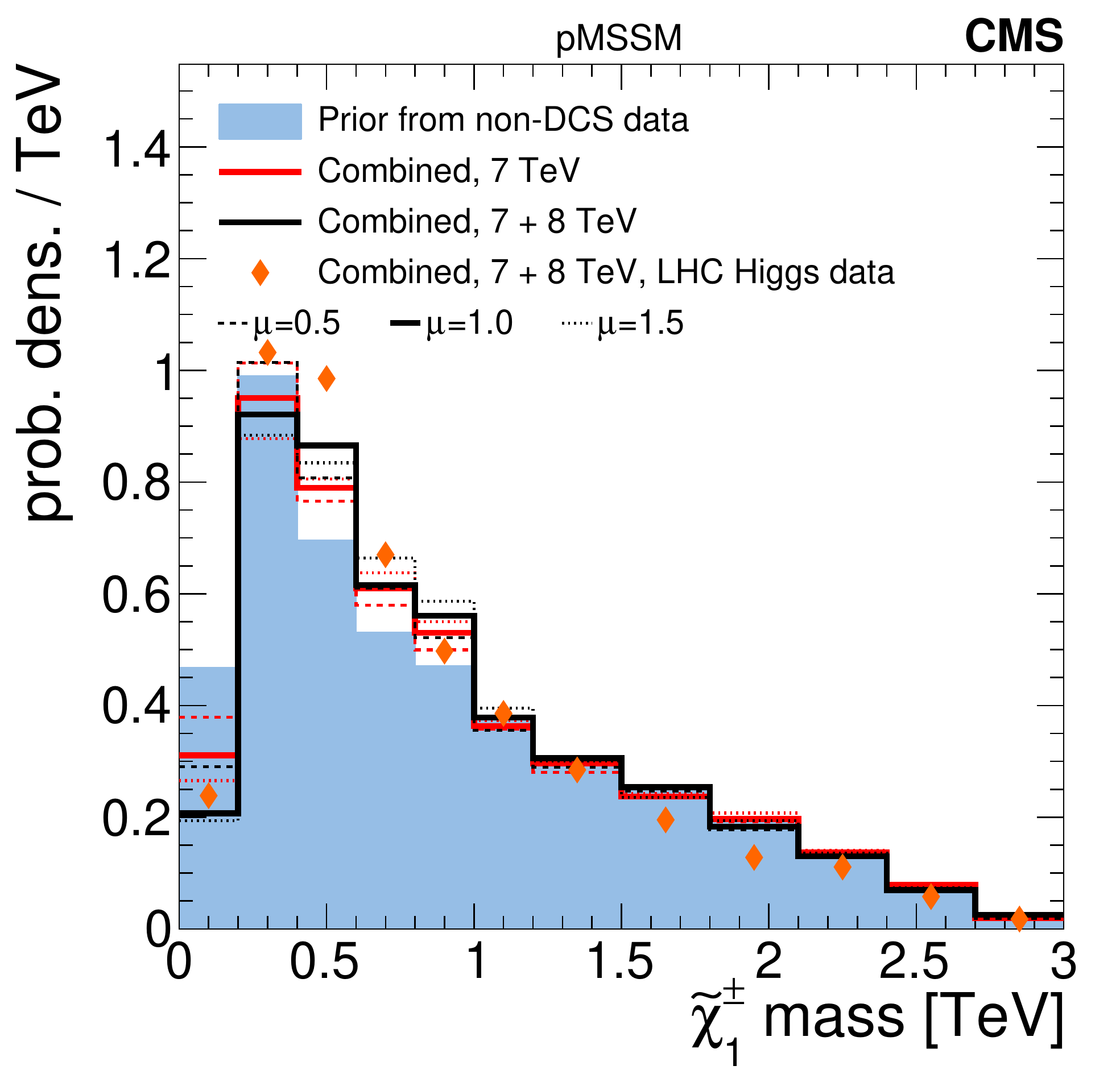}
\includegraphics[width=0.31\textwidth]{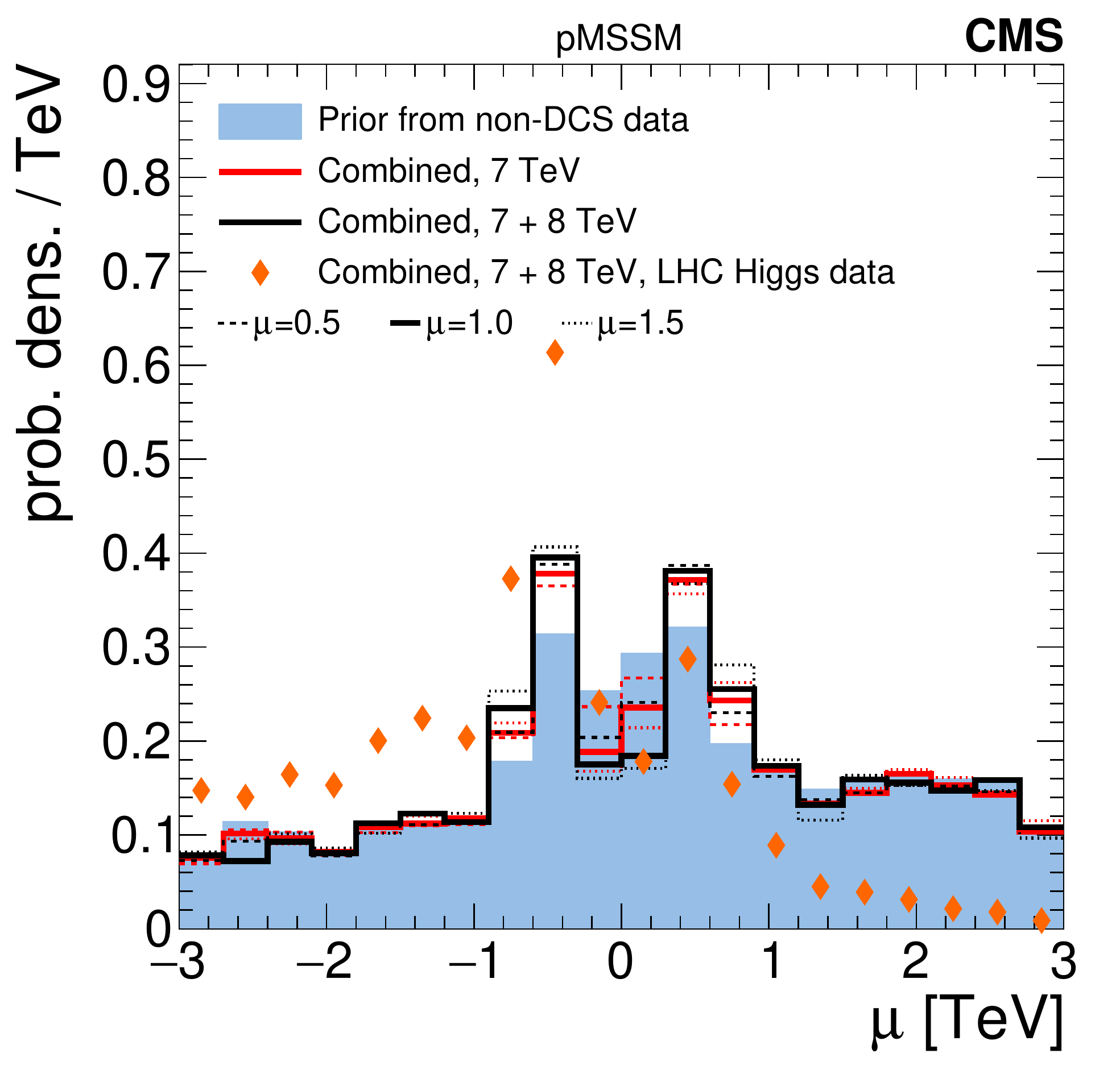}
\includegraphics[width=0.31\textwidth]{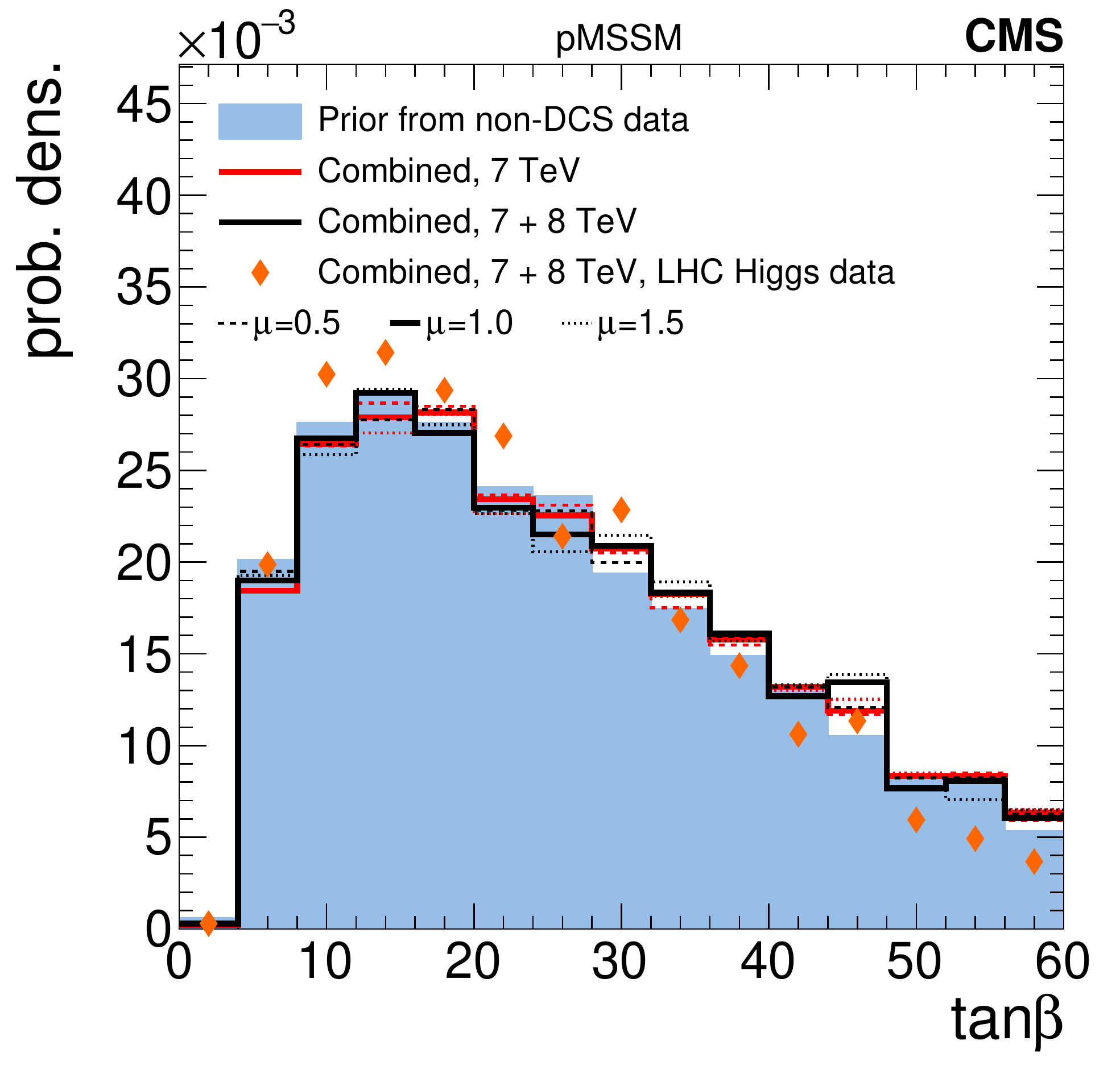}
\includegraphics[width=0.31\textwidth]{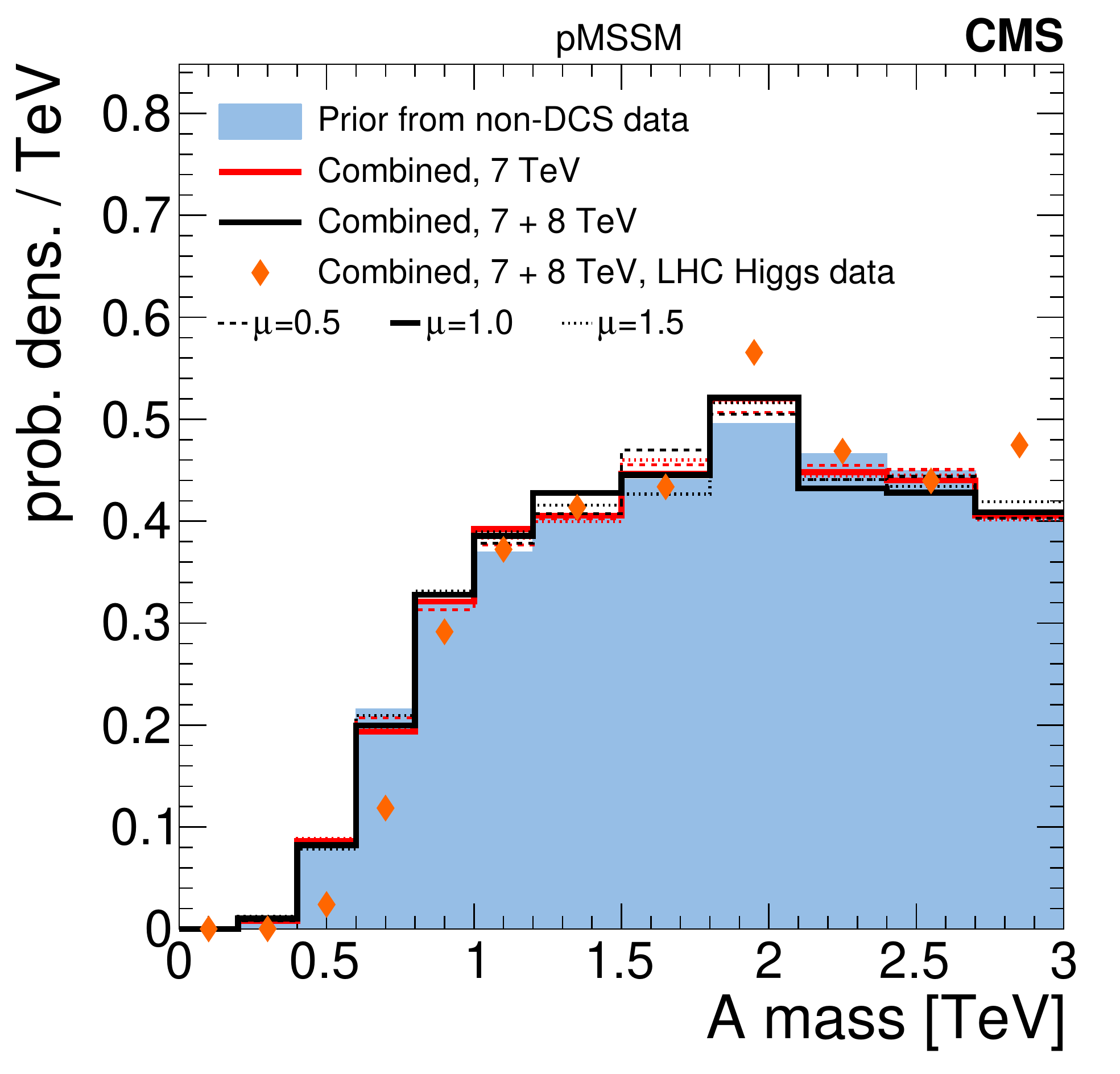}
\caption{Comparison of prior and posterior distributions after several combinations of data from the CMS searches for the
$\suR,\scR$ mass, $\sbone$ mass, $\seL,\smuL$ mass, $\stauOne$ mass, $\PSgxz{}_2$ mass, $\chipm$ mass, the higgsino mass parameter $\mu$, \tanb, and A mass.}
\label{fig:more1D}
\end{figure*}

In Fig. \ref{fig:more1D}, the  \preCMS and post-CMS distributions are
compared after 7 and $7{+}8\TeV$ data for several other important
observables.  We first note that the impact of the CMS data on the
first and second generation right-handed up squarks is lower than on the
corresponding left-handed up squarks (Fig. \ref{fig:mq}). This is because left-handed up squarks in
the MSSM form doublets with mass-degenerate left-handed down squarks, while
the right-handed up and down squarks are singlets and their
masses are unrelated.  Therefore, for the left-handed up squarks, the CMS
sensitivity for a given mass is increased by the left-handed down squarks,
which have the same mass.  We also observe a mild impact on the bottom
squark mass, where CMS disfavors masses below 400\GeV.  The CMS
searches also have some sensitivity to the selectron and stau masses,
which comes from the leptonic searches.  The impact on $\PSgxz{}_2$ and
$\chipm$ masses is larger, mostly due to the dedicated EW
analyses. The CMS SUSY searches have no impact on the masses of the light and heavy pseudoscalar Higgs
bosons. The preference of the Higgs data for negative values of the higgsino mass parameter $\mu$ comes primarily from the fact that the measured signal
strength normalized to its SM value for $\mathrm{V}\Ph\to\bbbar$ (where V is a W or a Z boson) is currently slightly below one. In a SUSY model, this requires that radiative corrections reduce the bottom Yukawa coupling, thereby creating a preference for $\mu<0$~\cite{Dumont:2013npa}. The $\tan\beta$ distribution is largely unaffected by both the CMS SUSY searches and the current Higgs boson data evaluated via \textsc{Lilith} 1.01.

\begin{figure*}[htbp]
  \centering
  \includegraphics[width=0.31\textwidth]{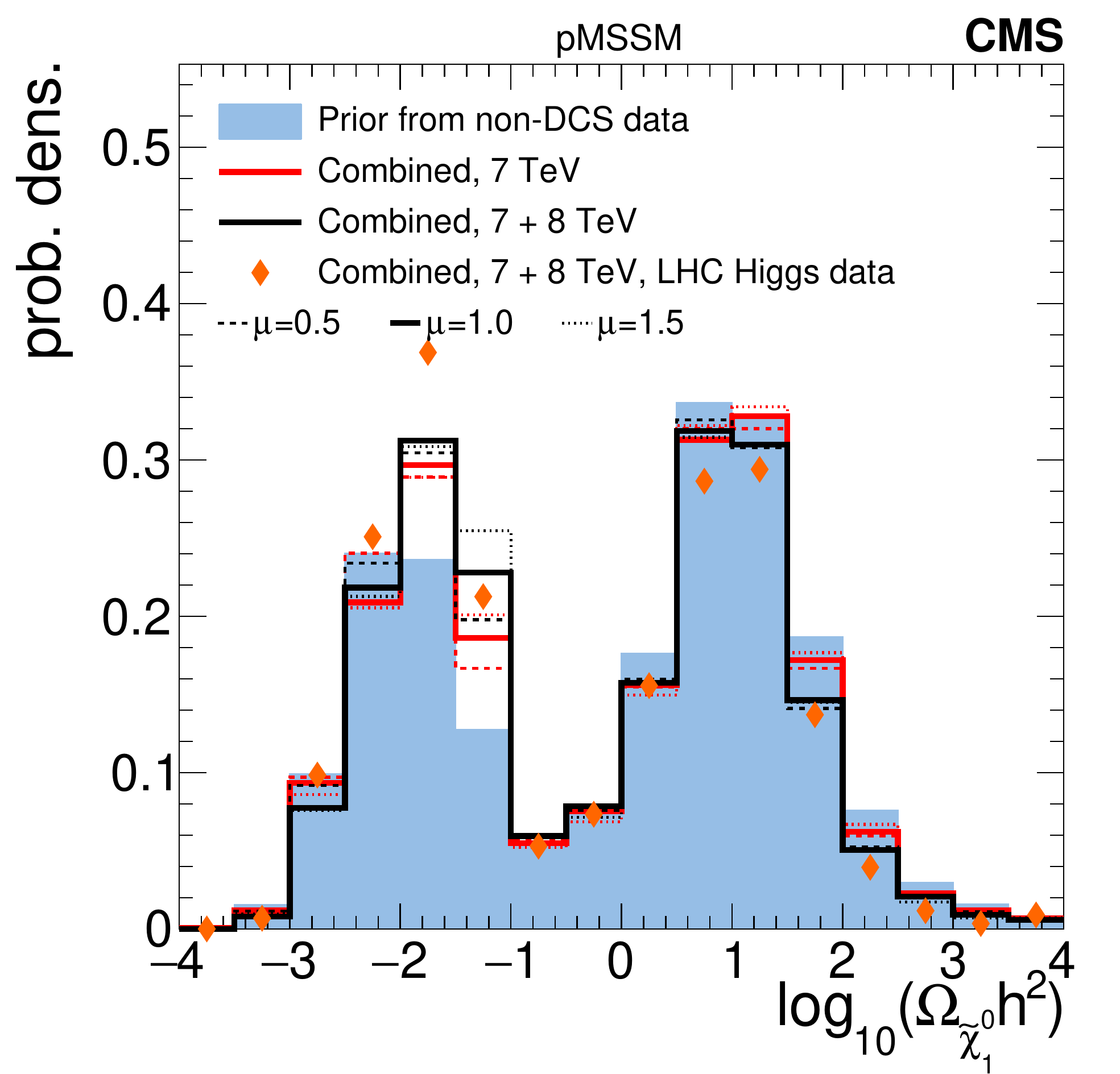}
  \includegraphics[width=0.31\textwidth]{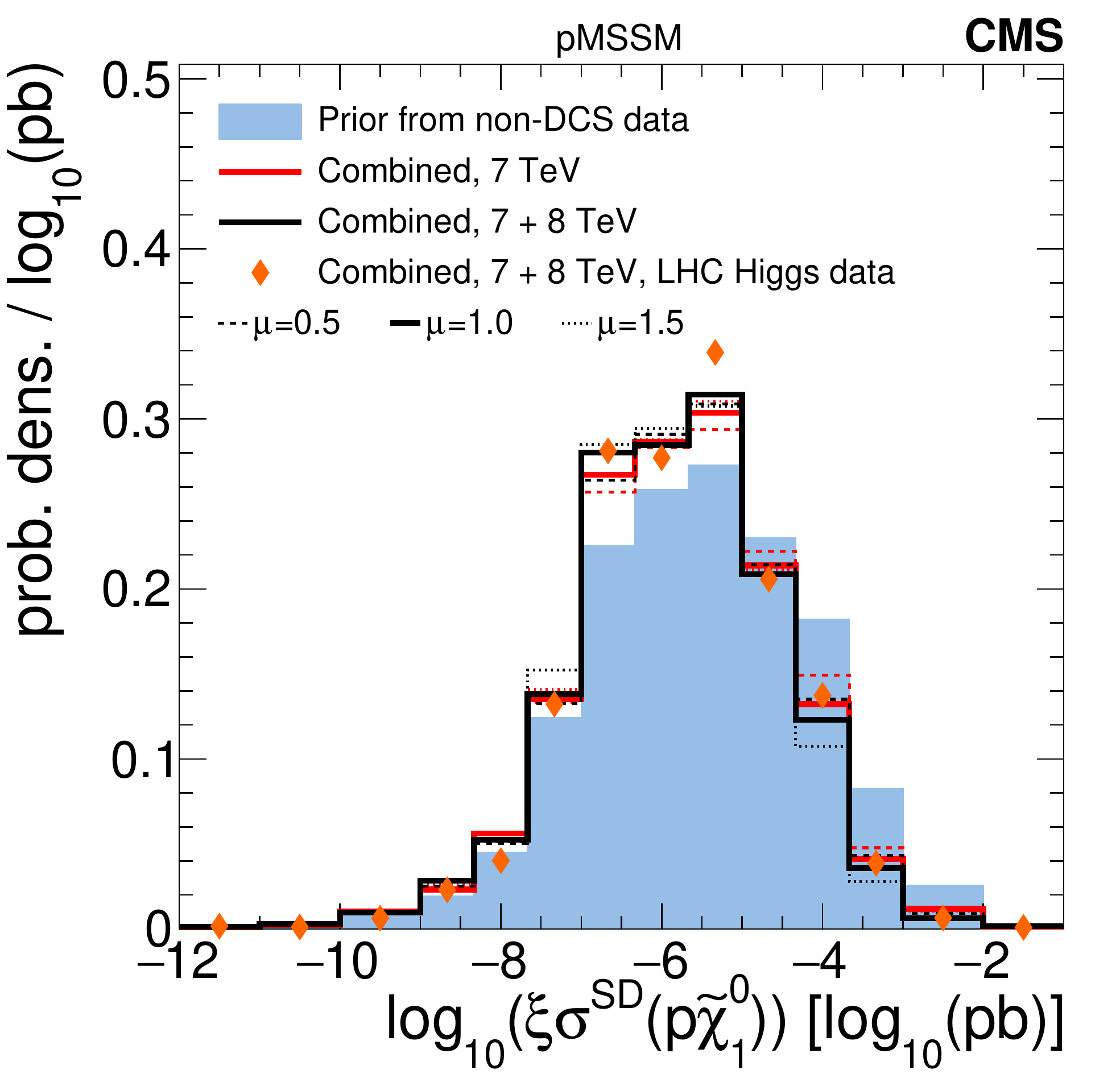}
  \includegraphics[width=0.31\textwidth]{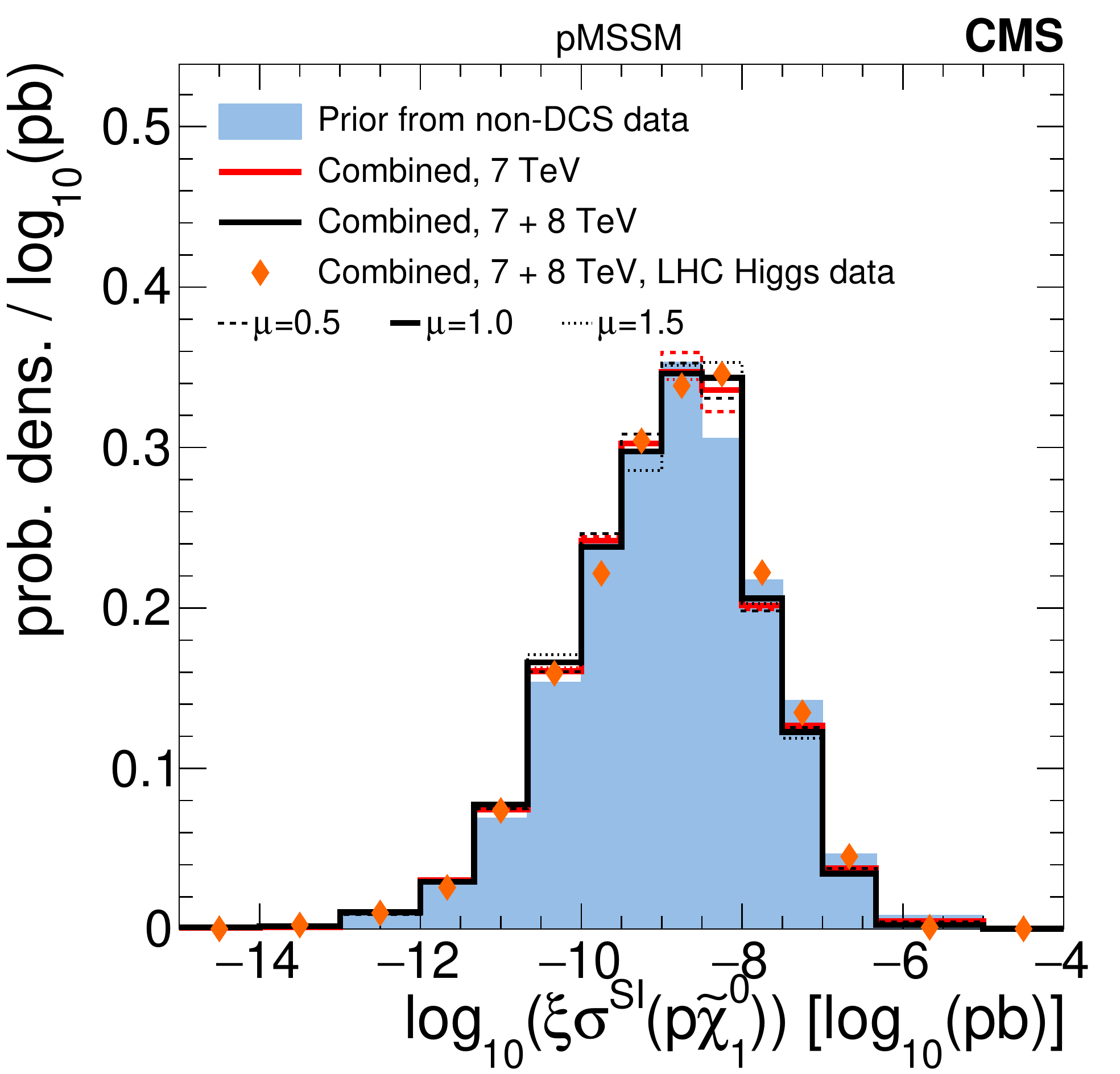}
    \caption{Comparison of prior and posterior distributions after several
  combinations of data from the CMS searches for $\Omega_{\PSGcz_{1}}$, $\xi\sigma^{\text{SD}}(p\PSGczDo)$, and $\xi\sigma^{\text{SI}}(p\PSGczDo)$.}
    \label{fig:more1D_dm}
\end{figure*}

We also investigate the impact of the considered searches on some observables
related to dark matter. Figure~\ref{fig:more1D_dm}  shows distributions of the
dark matter relic density, the spin-dependent (SD) direct detection cross section,
and spin-independent (SI) direct detection cross section. In Fig.~\ref{fig:more1D_dm} (left),
 the relic density is seen to take on a bimodal probability density.
The lower peak corresponds primarily to model points with bino-like LSPs,
and the upper peak is mainly due to points with wino- and higgsino-like LSPs. The combined CMS searches lead to a noticeable enhancement of the lower peak. In Fig.~\ref{fig:more1D_dm} (center) and (right), minor differences are seen between the prior and posterior densities for the direct detection cross section.

\subsection{Correlations among pMSSM parameters}
A virtue of high-dimensional models like the pMSSM is that they
enable the examination of correlations among parameters not
possible in the context of more constrained models.

\begin{figure*}
  \centering
  \includegraphics[width=0.32\textwidth]{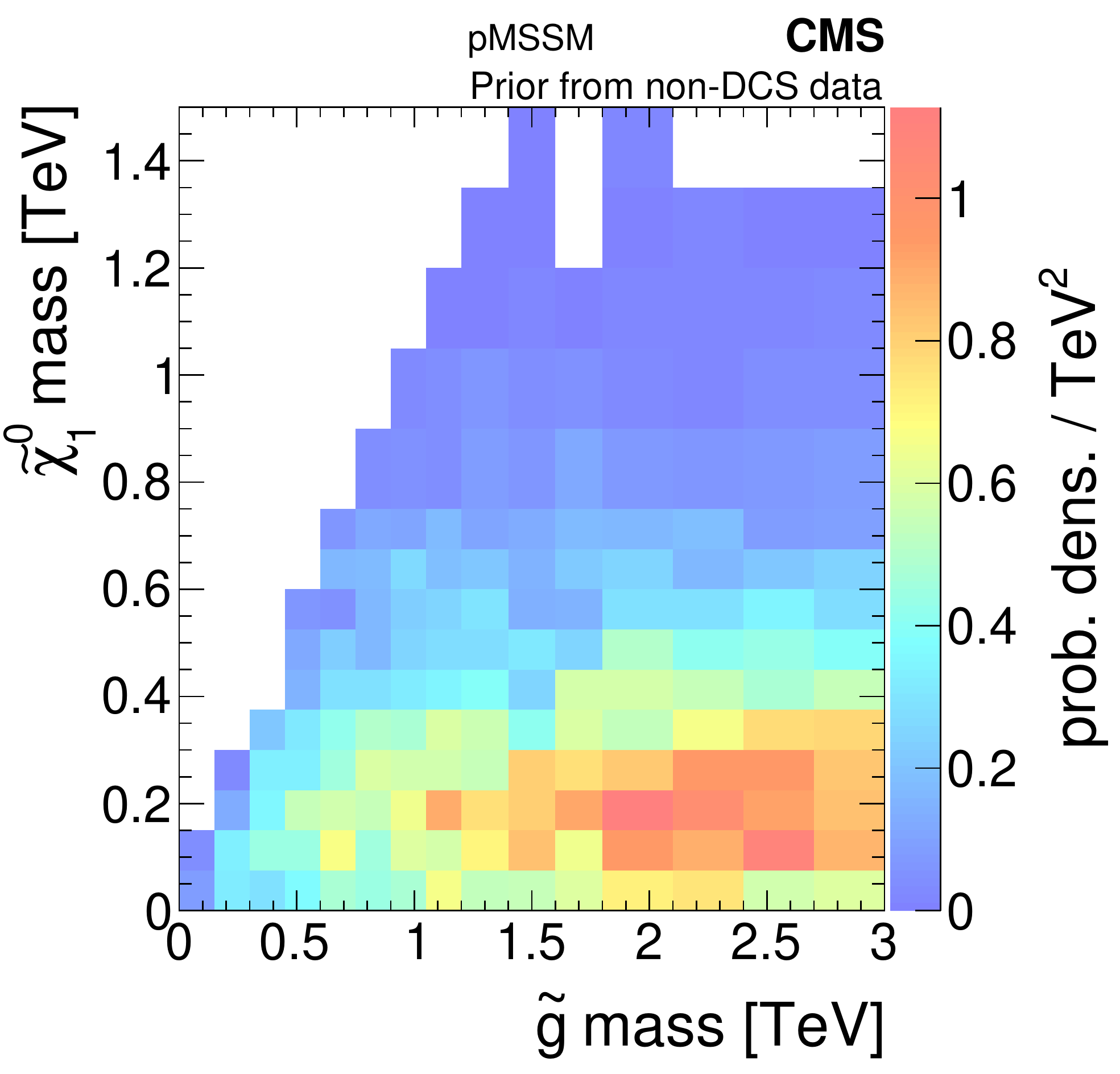}
  \includegraphics[width=0.32\textwidth]{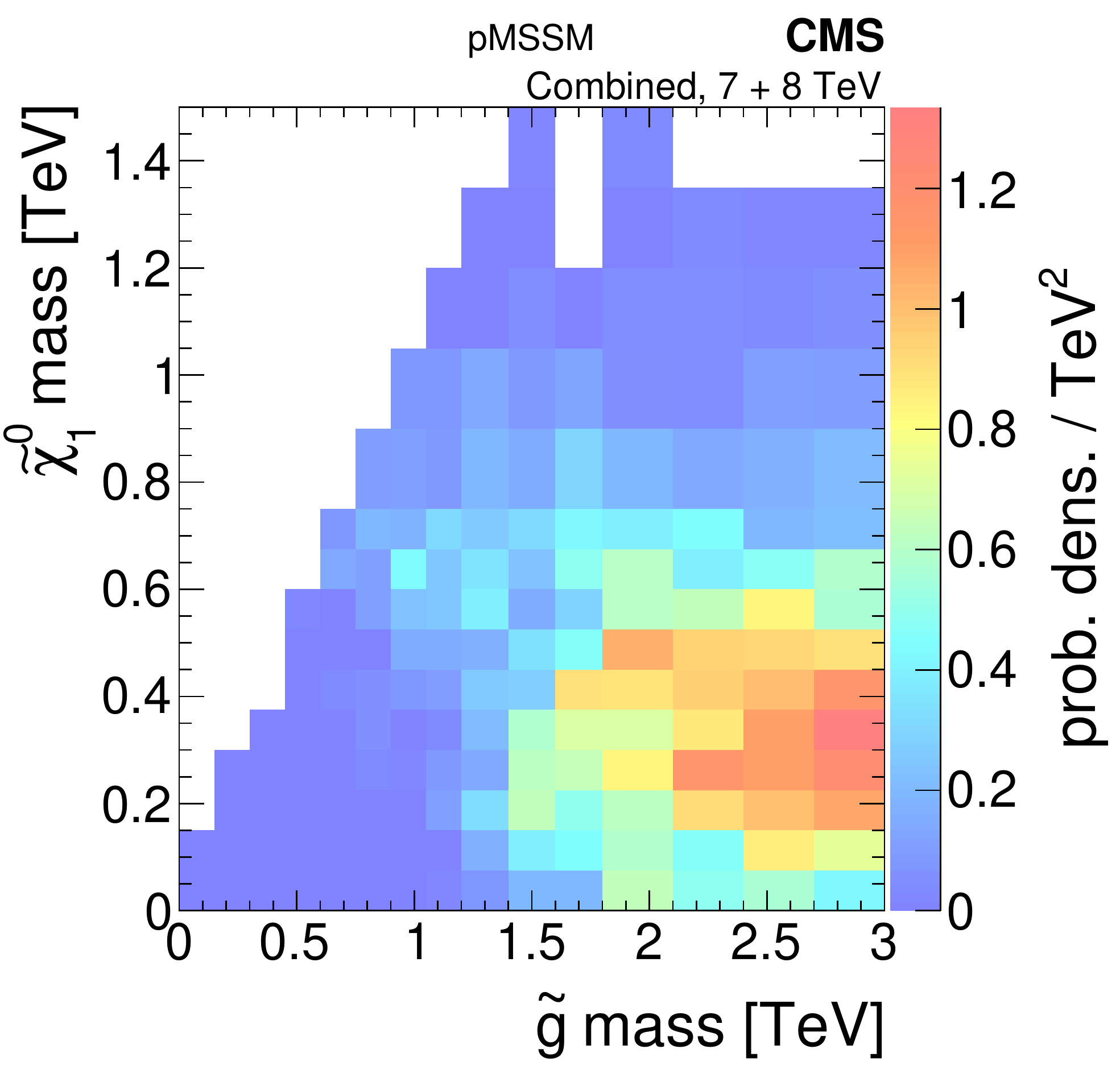}
  \includegraphics[width=0.32\textwidth]{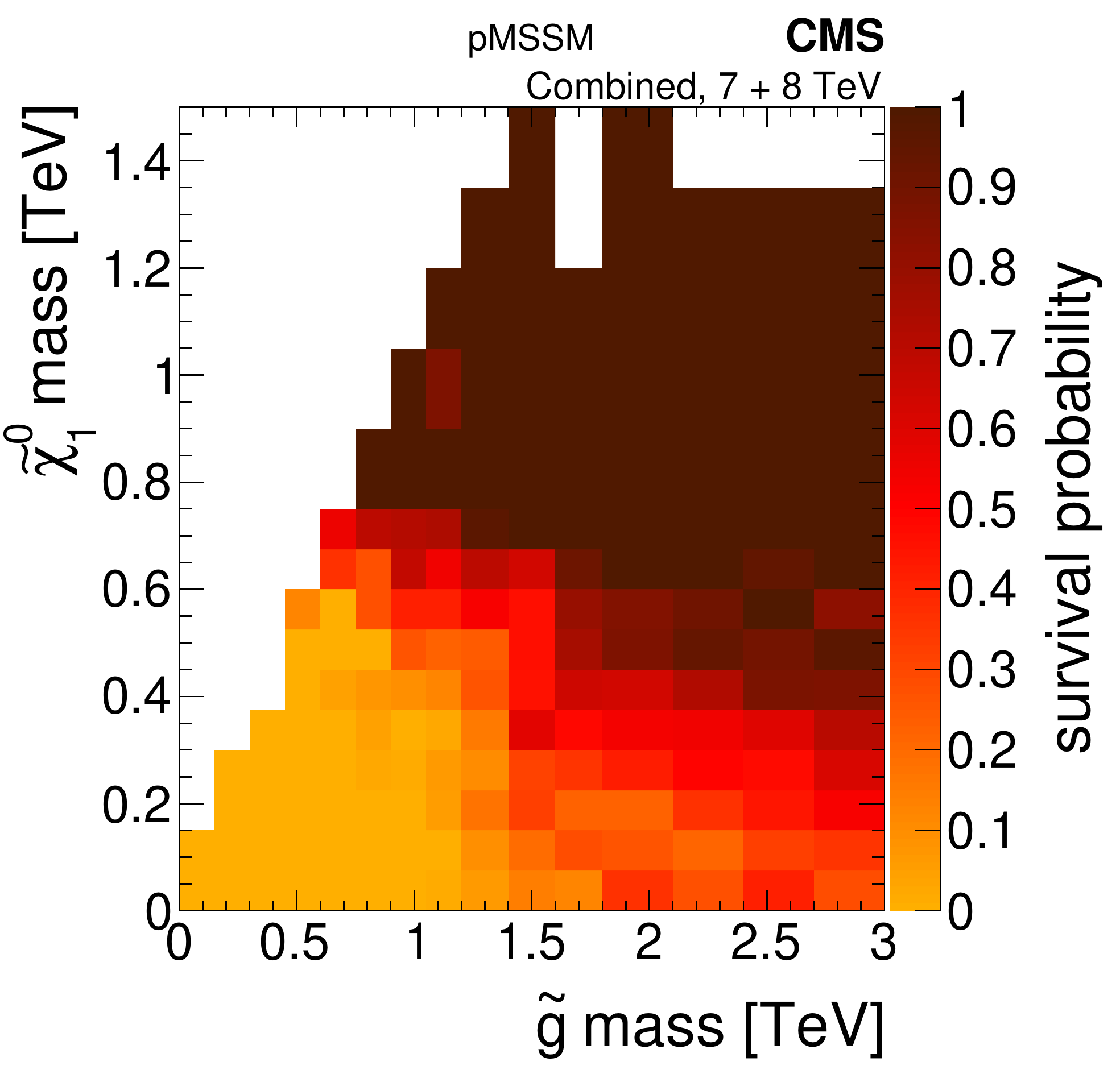}

  \includegraphics[width=0.32\textwidth]{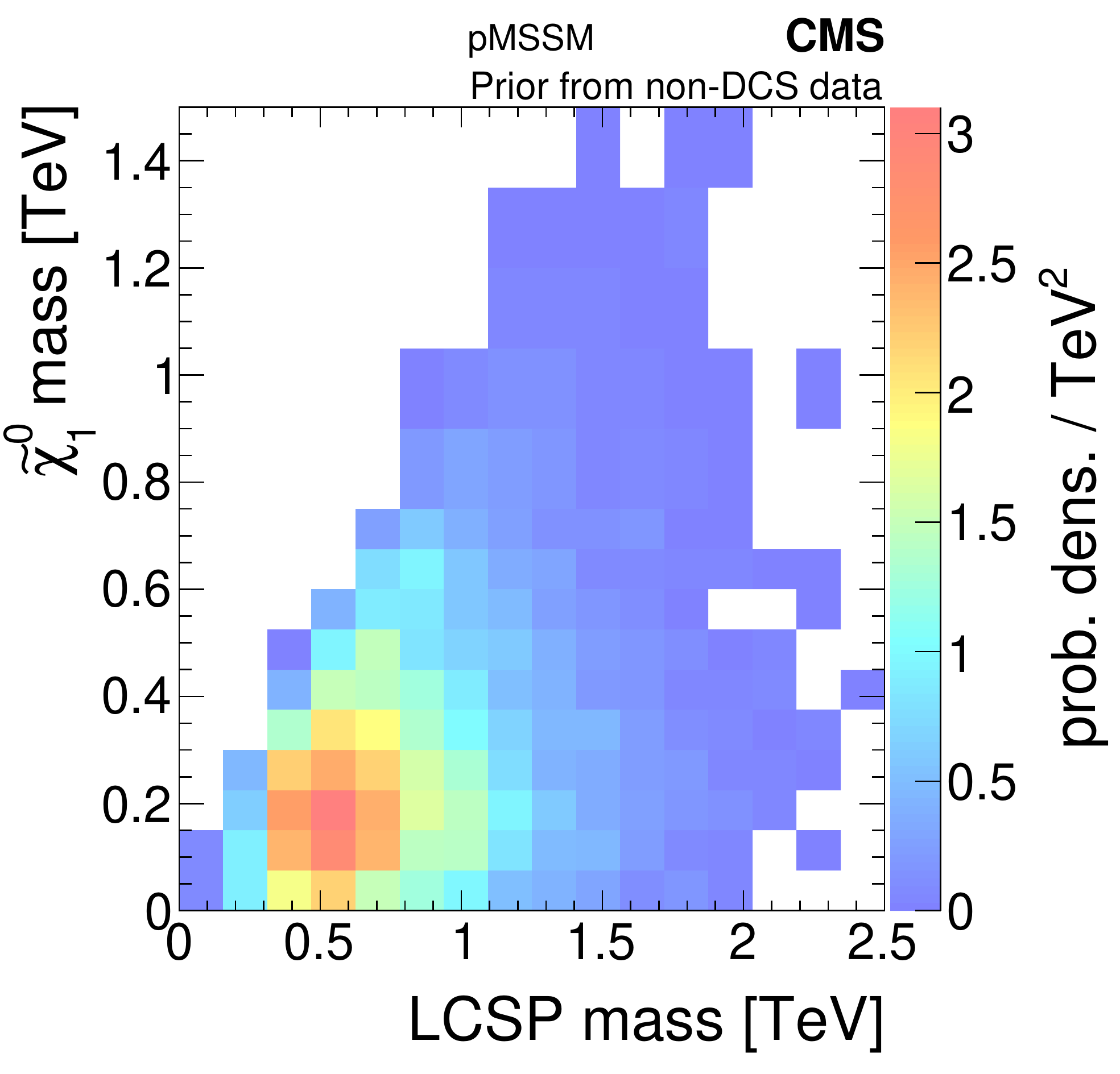}
  \includegraphics[width=0.32\textwidth]{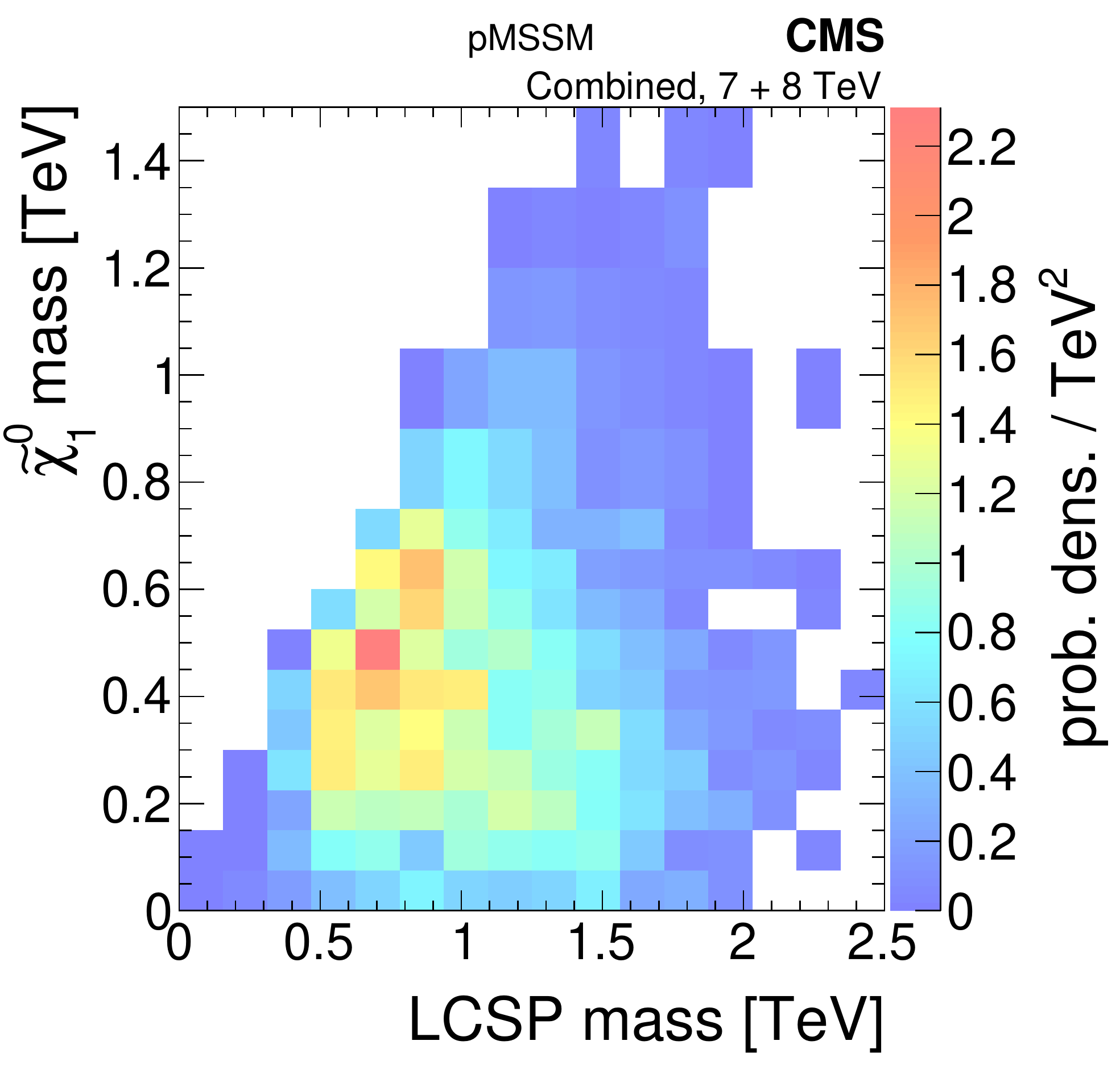}
  \includegraphics[width=0.32\textwidth]{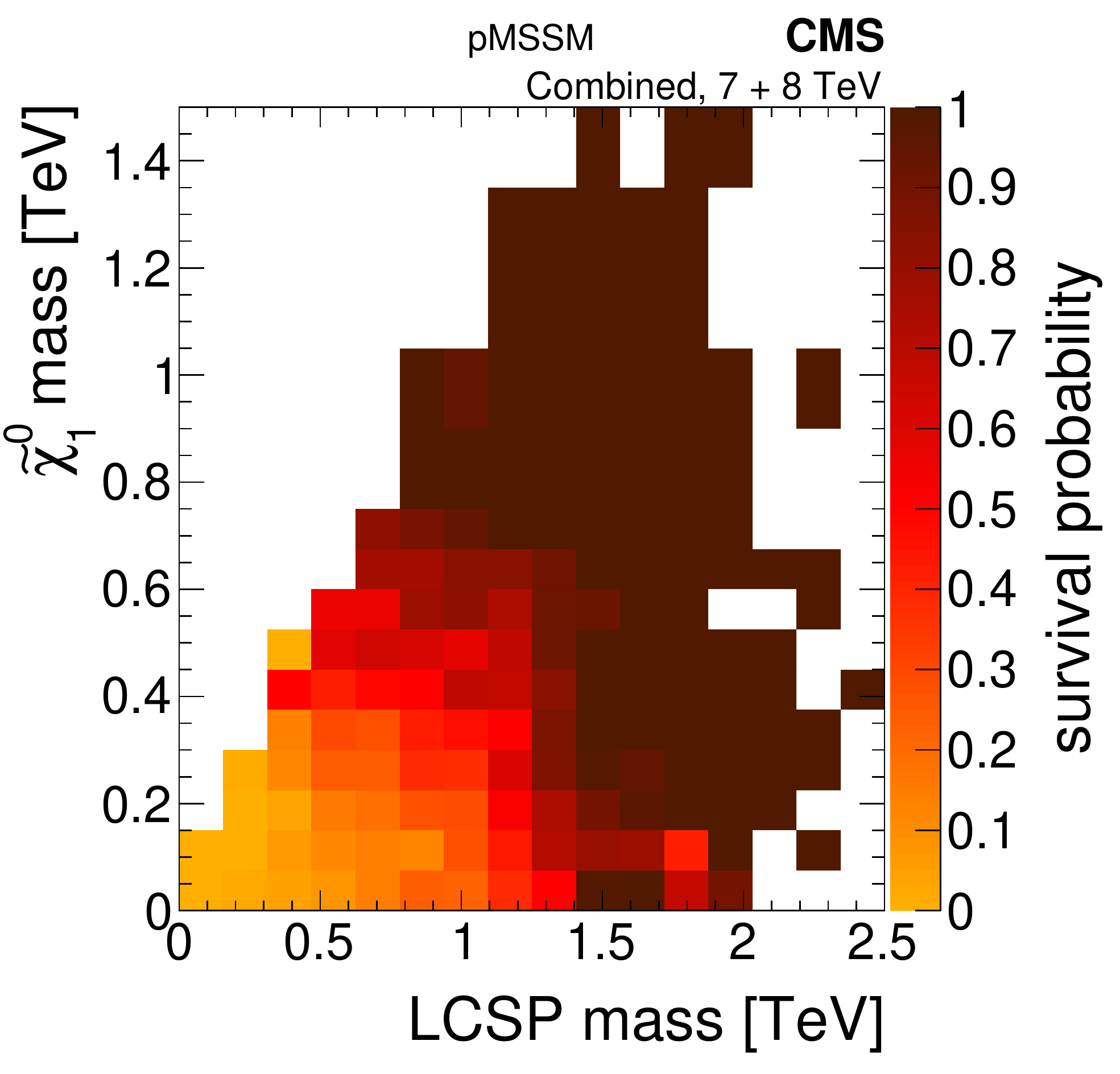}

  \includegraphics[width=0.32\textwidth]{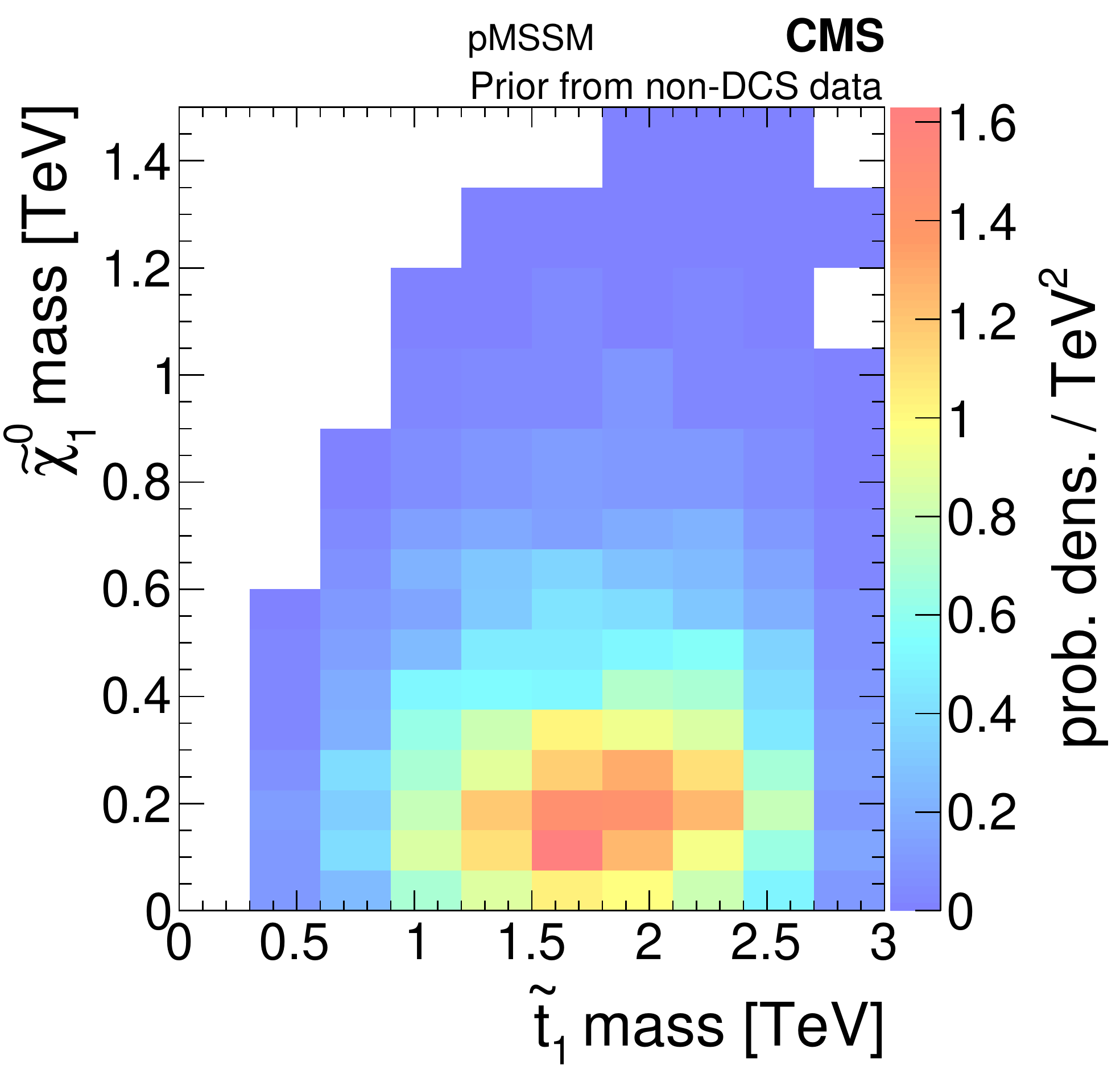}
  \includegraphics[width=0.32\textwidth]{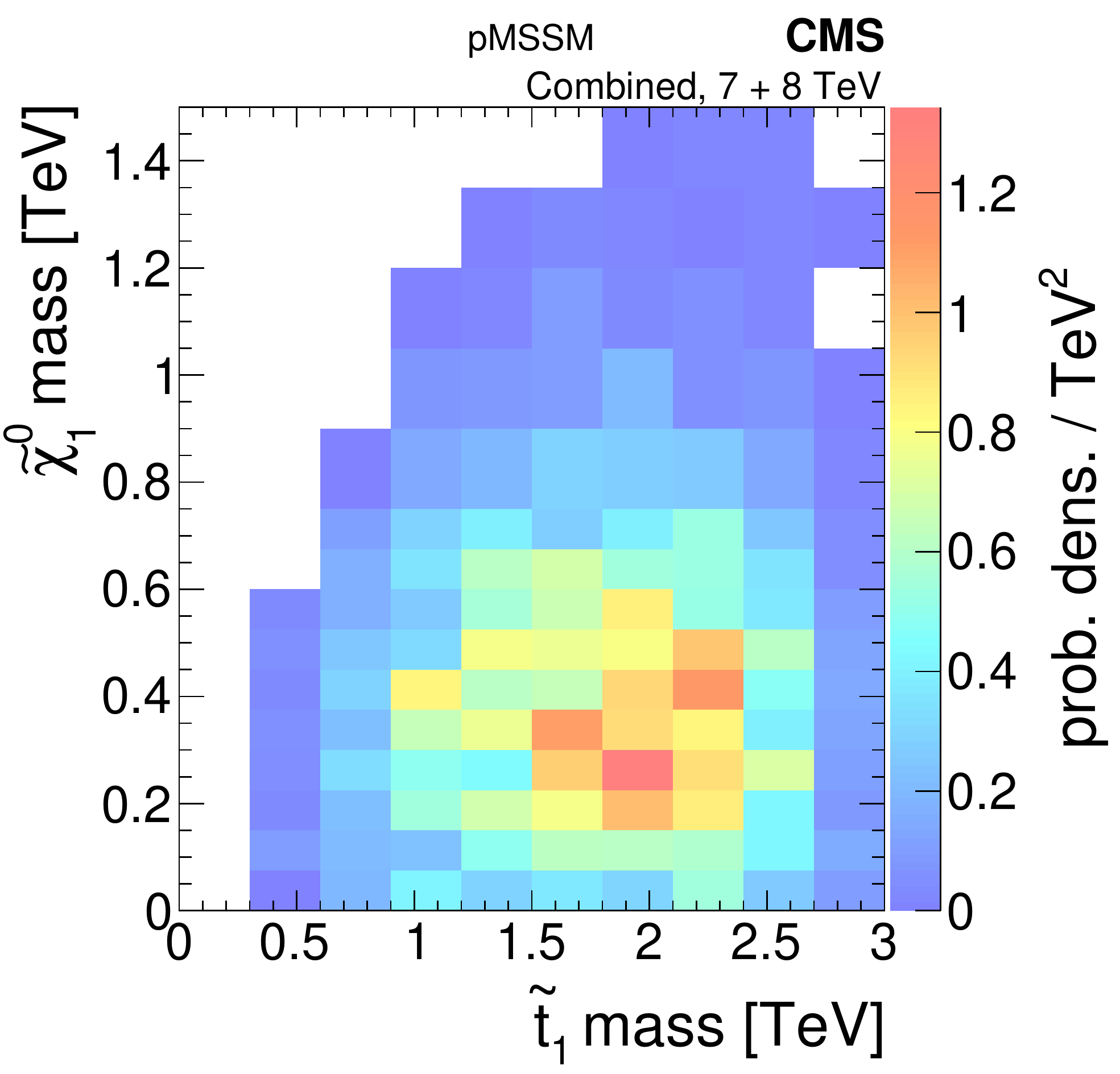}
  \includegraphics[width=0.32\textwidth]{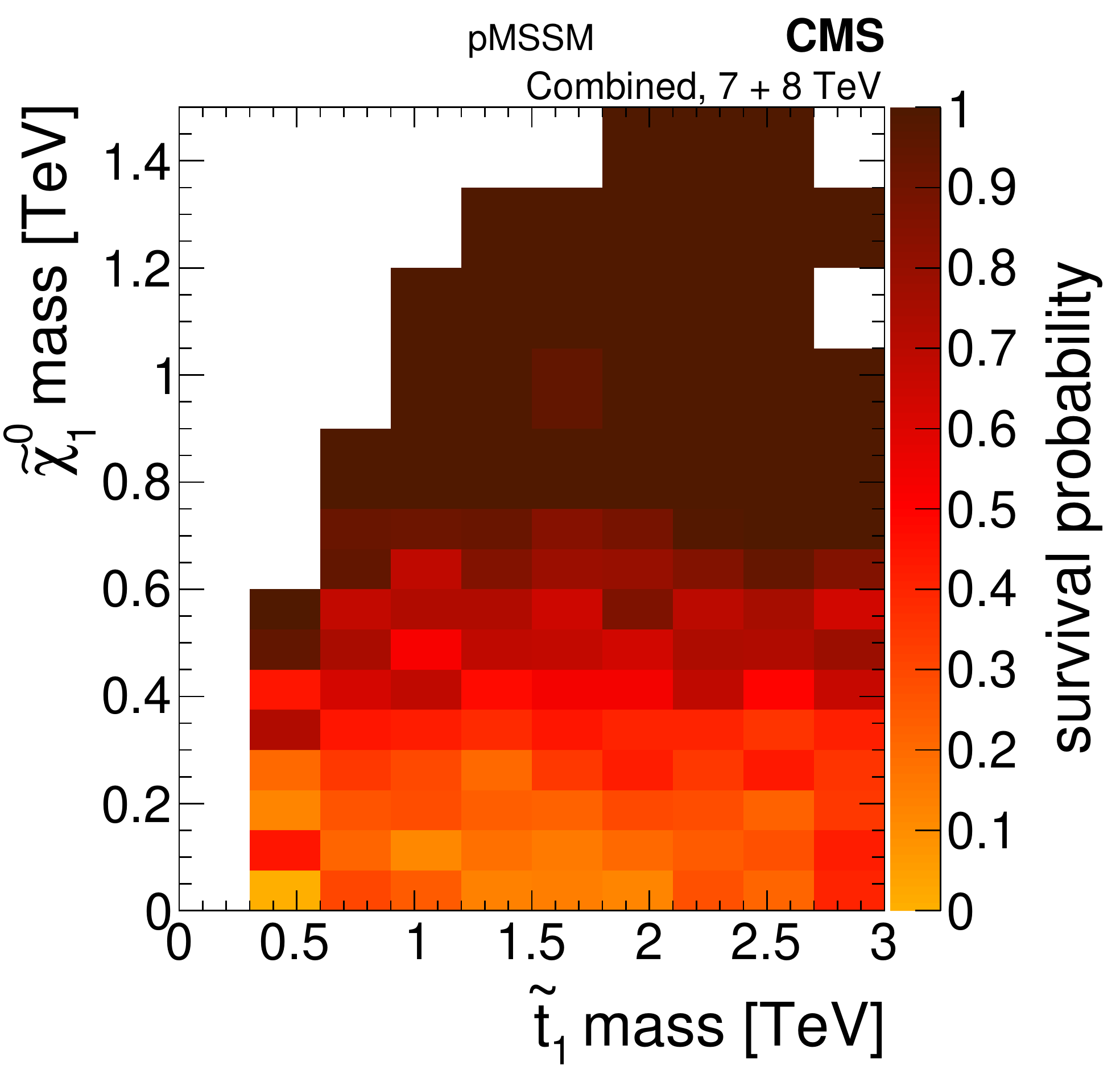}

  \includegraphics[width=0.32\textwidth]{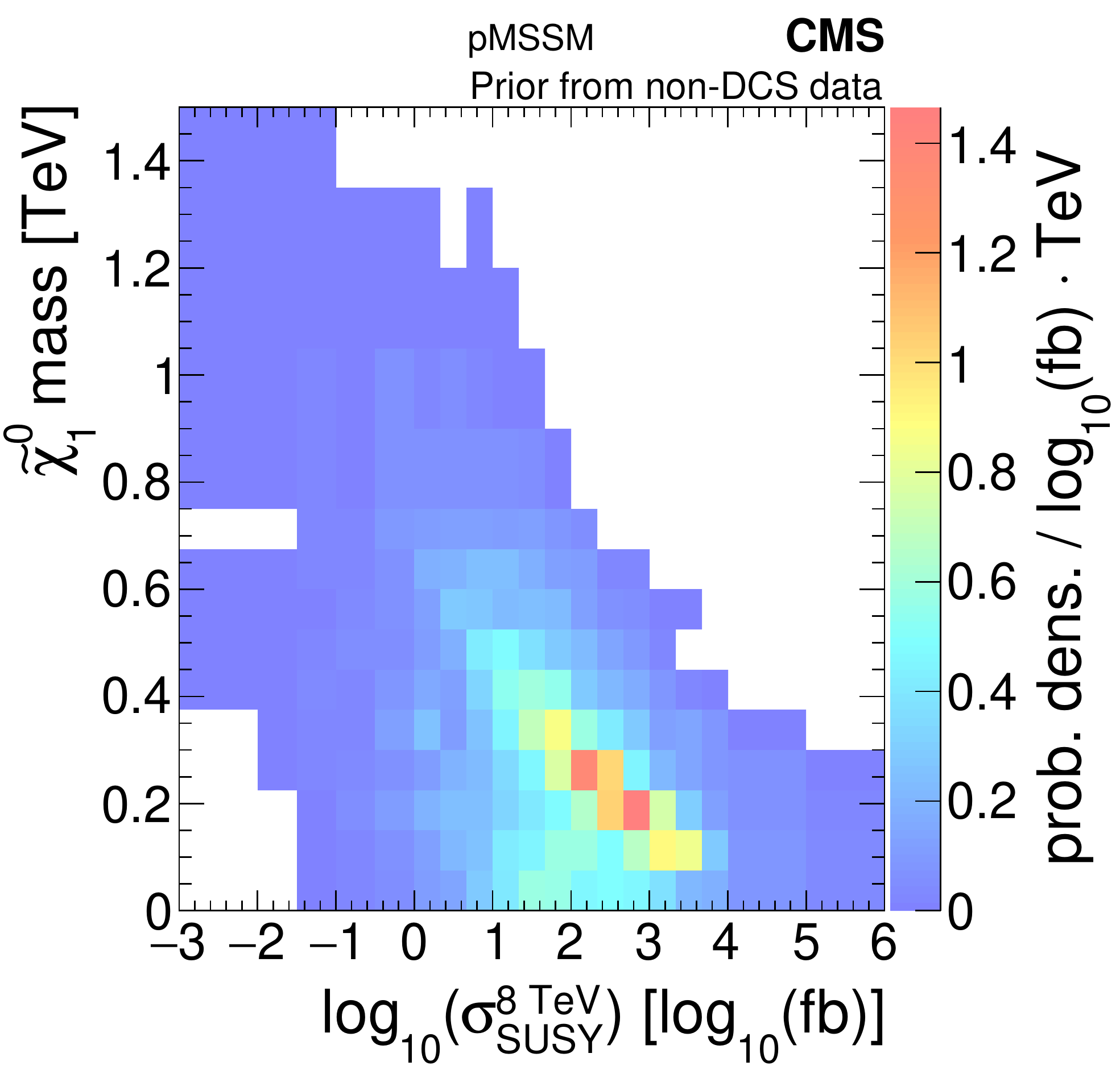}
  \includegraphics[width=0.32\textwidth]{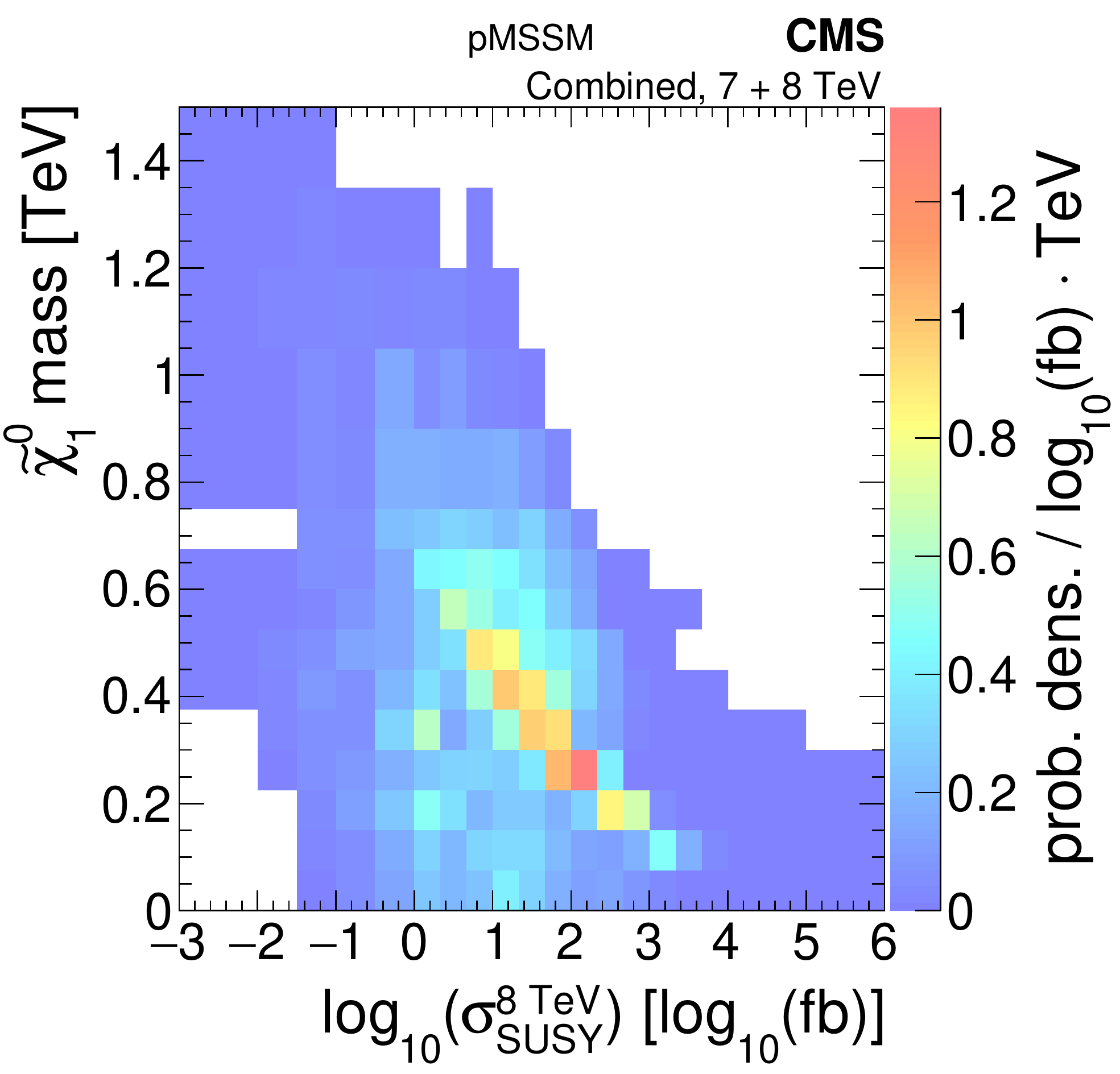}
  \includegraphics[width=0.32\textwidth]{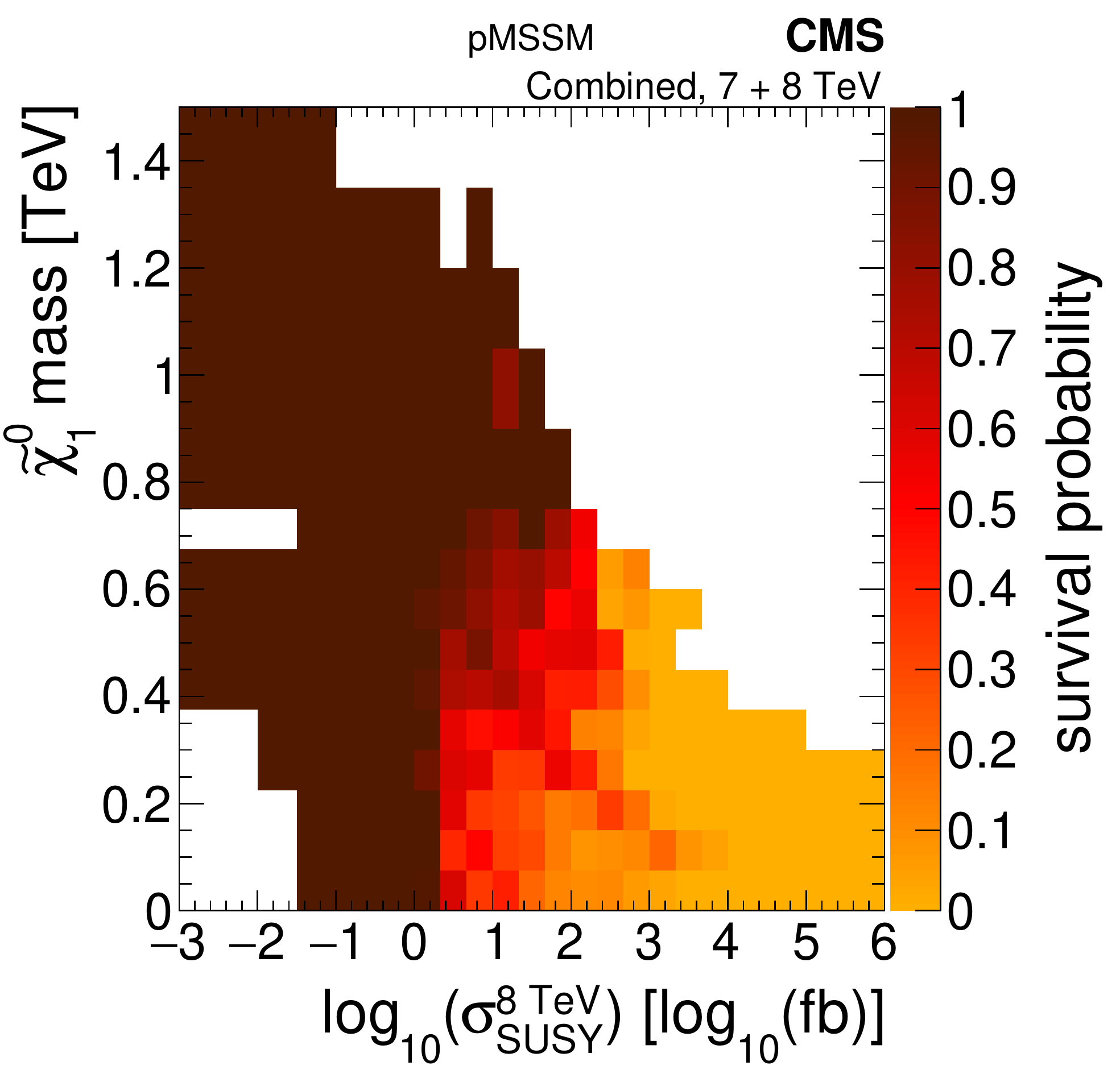}

  \caption{Marginalized \preCMS distributions (first column),
    compared with posterior distributions (second column)
    and survival probabilities (third column) after inclusion of the considered CMS searches,
    are shown for the
    $\PSGczDo$ mass versus gluino mass (first row), the LCSP mass
    (second row), the top squark mass (third row),  and the
    logarithm of the cross section for inclusive sparticle production at 8\TeV (bottom row).
  }
  \label{fig:twoD}
\end{figure*}

Figure \ref{fig:twoD} compares marginalized distributions in two dimensions of \preCMS (left) to post-CMS distributions (middle), and also shows the post-CMS to \preCMS survival probability (right) for several observable pairs.  The first two rows of distributions show that the CMS impact on our knowledge of
the $\chiz$ mass is strongly correlated with the gluino or the LCSP mass.  Since $\chiz$ is the LSP,  light colored particles imply a light $\chiz$.
Consequently, the disfavoring of light colored sparticles implies the disfavoring of a light $\chiz$.  In the last row, it is seen that the $\chiz$ mass is correlated most strongly with the cross section and that light $\chiz$ LSPs are indeed disfavored for the reason just given.
We note, however, that scenarios with $\chiz$ masses around 100\GeV
can still survive even though they have cross sections above 1\,pb. These and other high cross section model points are discussed in Section \ref{sec:unexplored}.
In the third row, we show the probability distributions and survival
probability for $\chiz$ versus $\stl$ mass.  Here we see that, although the
post-CMS probabilities shift towards higher values, the survival
probabilities never really go down to zero.  Although current SMS
scenarios exclude large parts of the $\stl$-$\chiz$ plane, we see that
pMSSM scenarios with relatively low $\stl$ masses (500\GeV) are
not significantly disfavored by the CMS searches considered.  We note that the searches for top squark production considered here focus primarily on the decay channel $\tilde{\text{t}}_1\to \text{t}\PSGczDo$, and it may be that a greater impact would be observed if the searches targeting leptonic channels were incorporated in this study.

Studies were performed to assess how the conclusions would change if a
different choice of initial prior had been made. A log-uniform prior ($p_0(\theta)$ in Eq. \ref{eq:prior}) is found to yield posterior densities very similar to those from the nominal uniform prior. The most significant exception is that
the densities for the masses of the $\chiz$ and  $\chipm$ are shifted 10\textendash20\%
toward higher values with respect to the densities derived from the
uniform prior. It is found that the marginalized
likelihood distributions are consistent with the profile likelihoods, suggesting that a
frequentist analysis based on the profile likelihoods would yield similar conclusions.

\section{Nonexcluded regions in the pMSSM parameter space}

\label{sec:unexplored}
Of the 7200 pMSSM points considered in this study, about 3700 cannot be
excluded by CMS analyses based on their $Z$-significance (Fig.~\ref{fig:Z} (bottom right)), although more
than half of these nonexcluded points have a total cross section greater than 10\unit{fb} at $\sqrt{s}$ = 8\TeV.
It is of interest to characterize this nonexcluded
subspace in order to shed light on why the CMS analyses are not
sensitive to these points, which can help guide the design of future analyses. To this end, we decompose the nonexcluded subspace into the dominant physical processes and
follow with an idealized analysis of final state
observables.

\begin{figure*}[htbp]
\centering
\hspace*{\fill}
\includegraphics[width=0.20\textwidth]{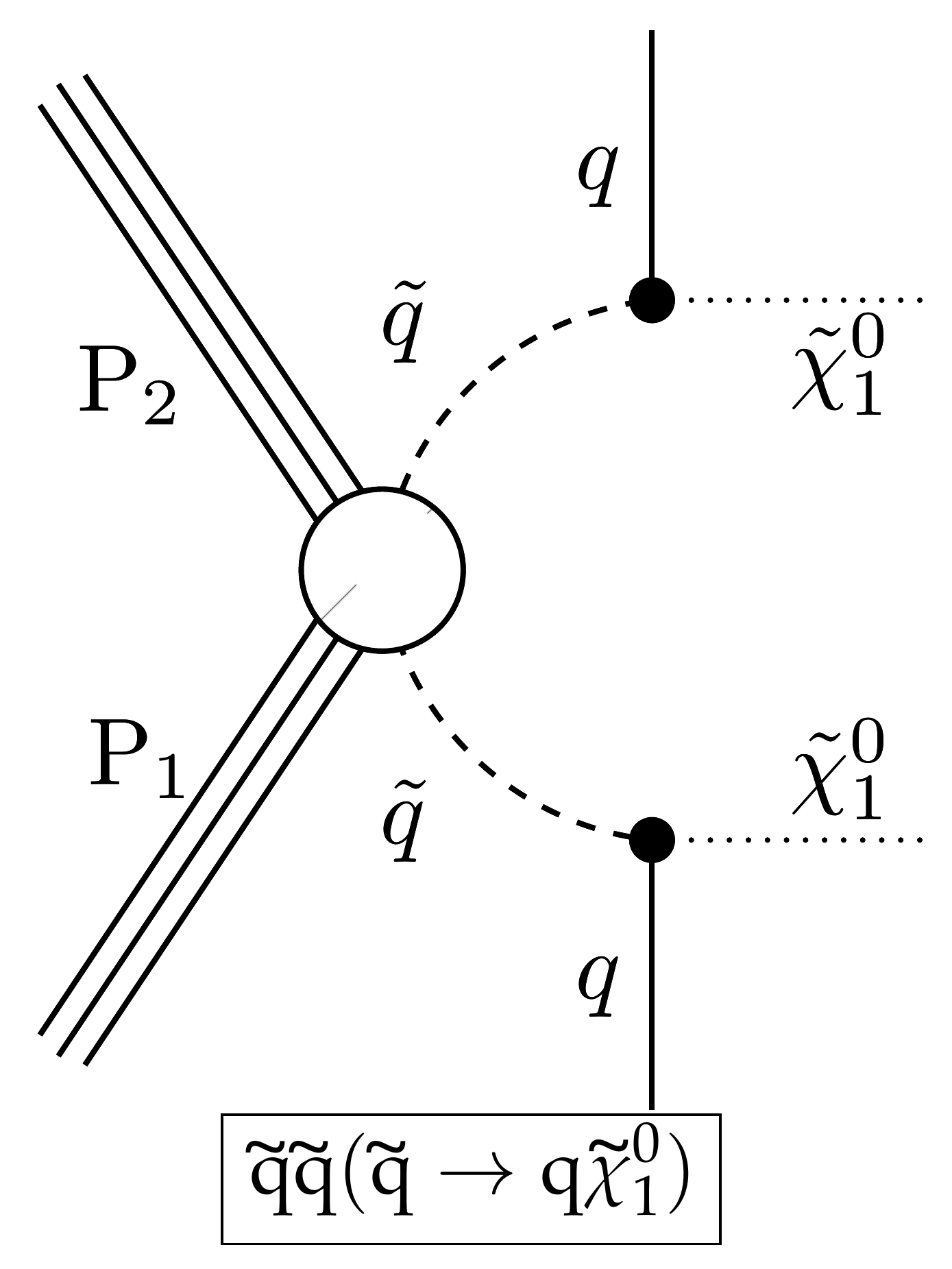}\hspace*{\fill}
\includegraphics[width=0.20\textwidth]{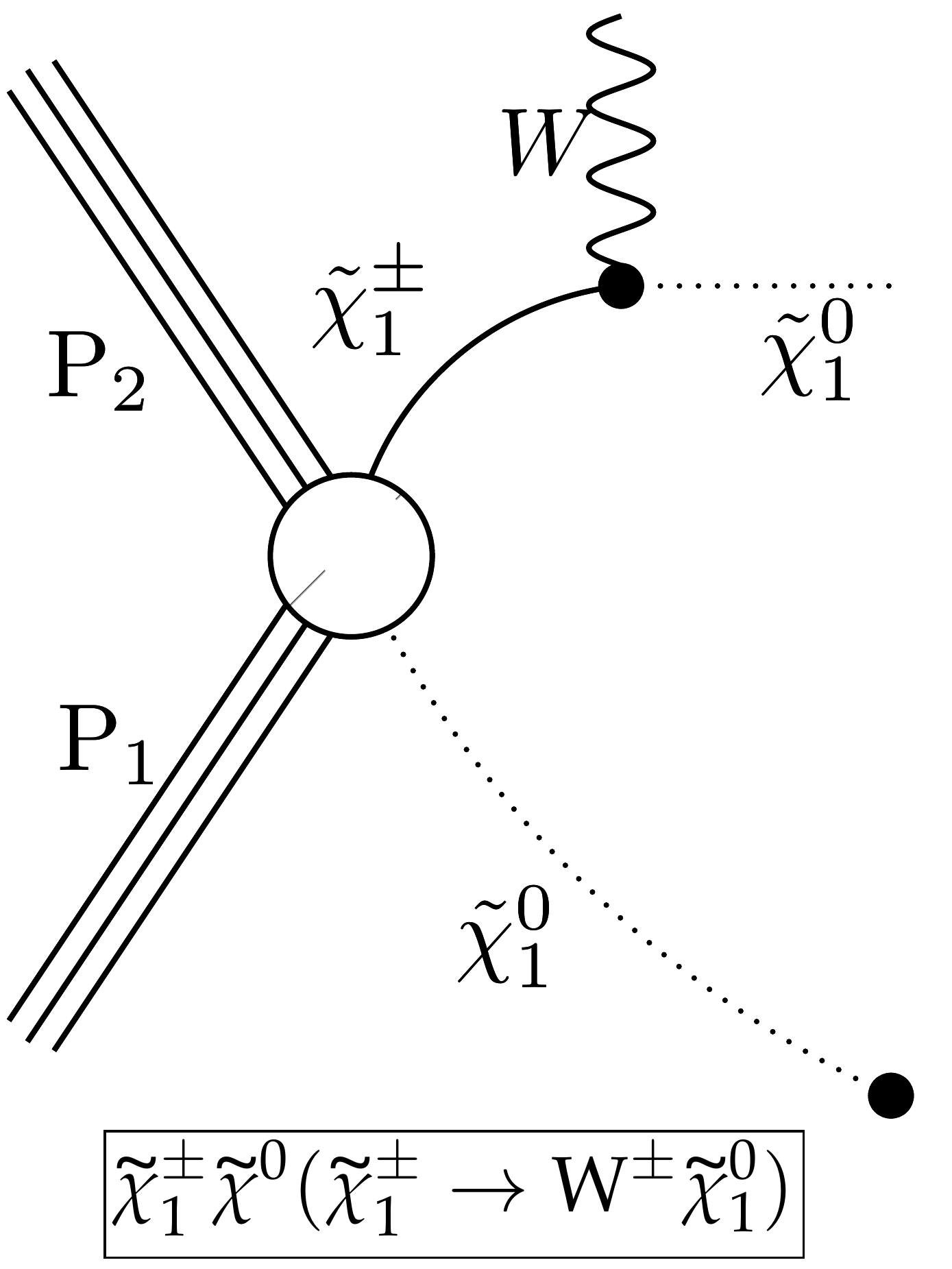}\hspace*{\fill}
\includegraphics[width=0.20\textwidth]{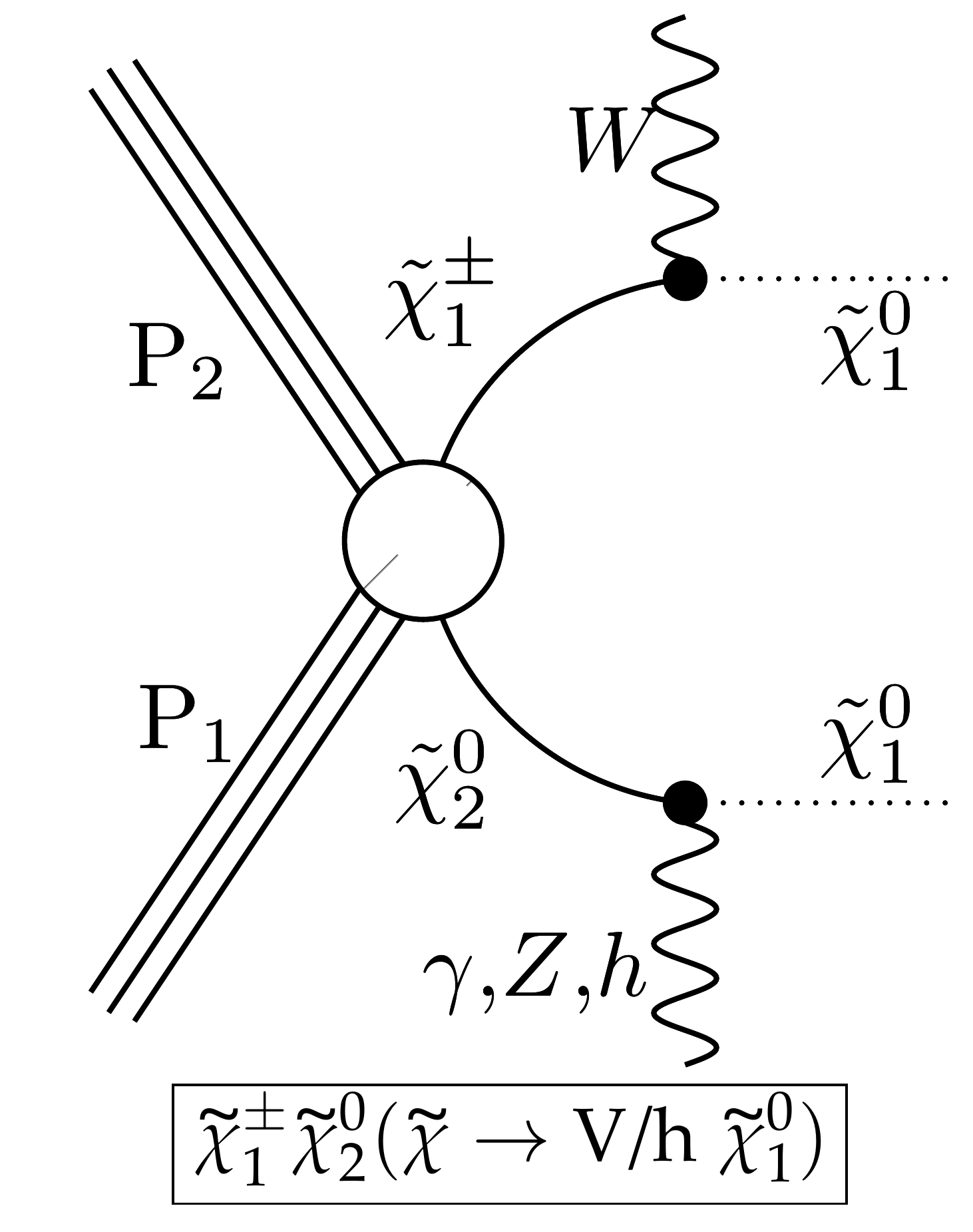}\hspace*{\fill}
\includegraphics[width=0.20\textwidth]{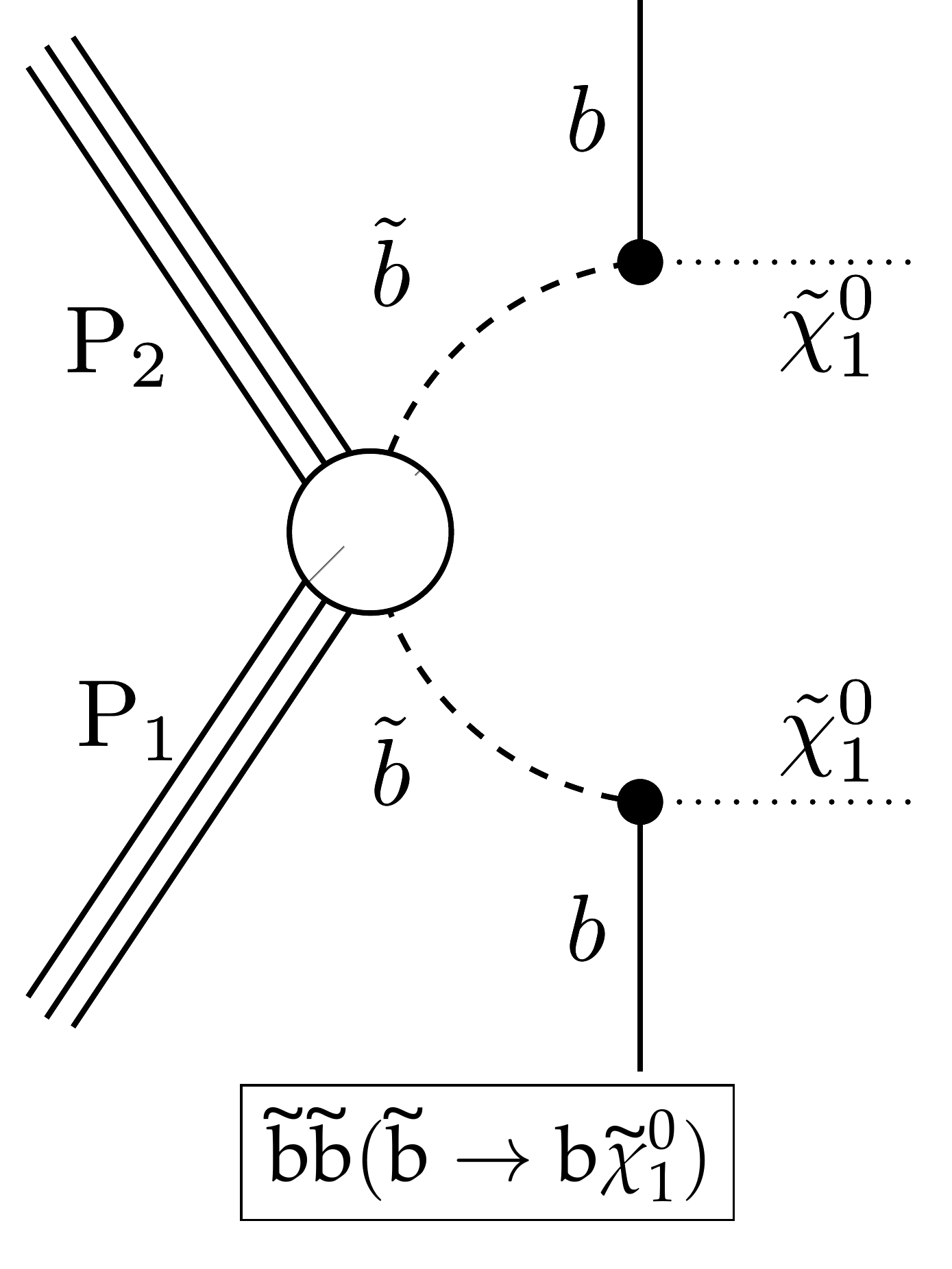}\hspace*{\fill}
\\[2em]
\hspace*{\fill}
\includegraphics[width=0.20\textwidth]{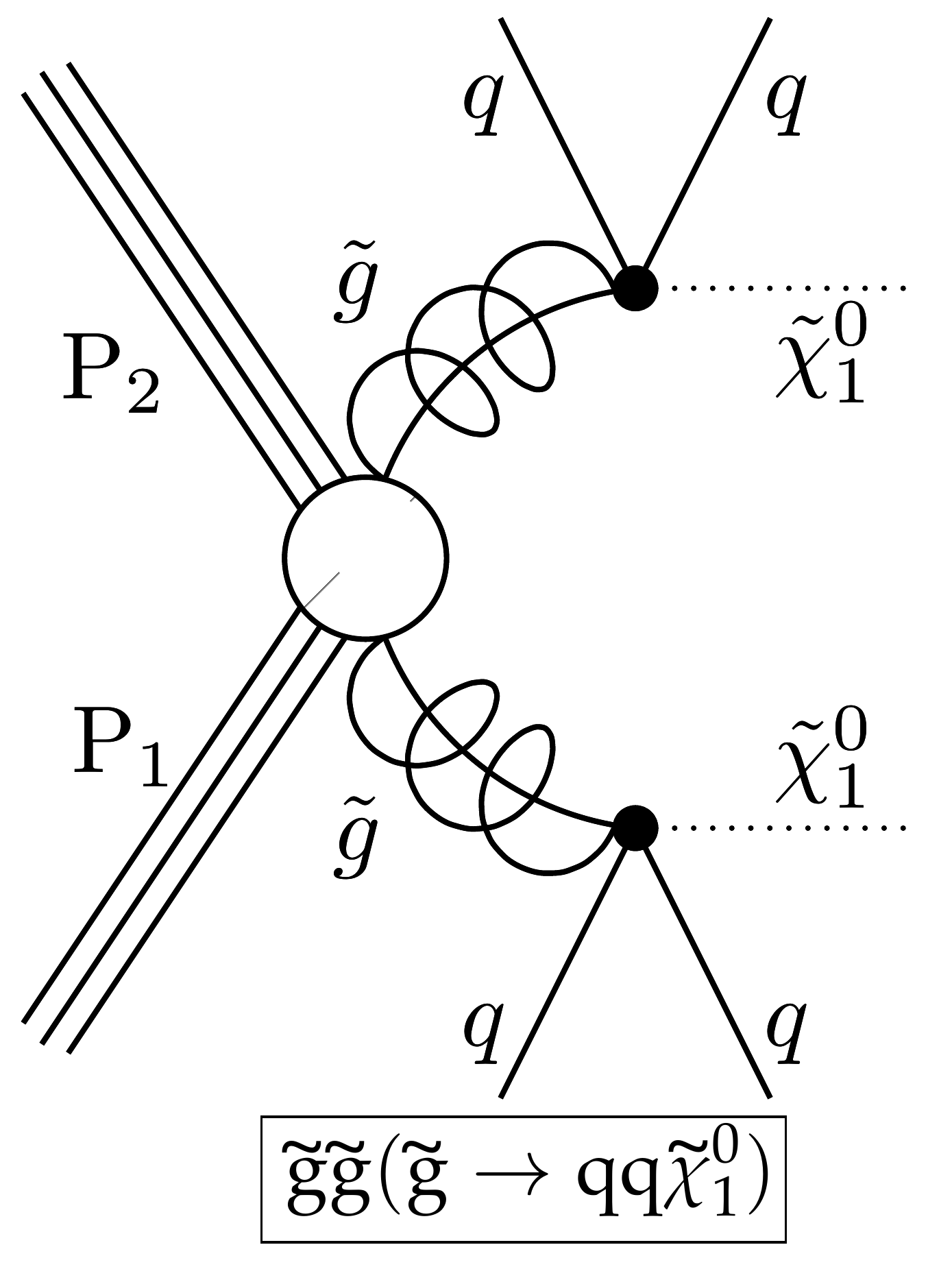}\hspace*{\fill}
\includegraphics[width=0.20\textwidth]{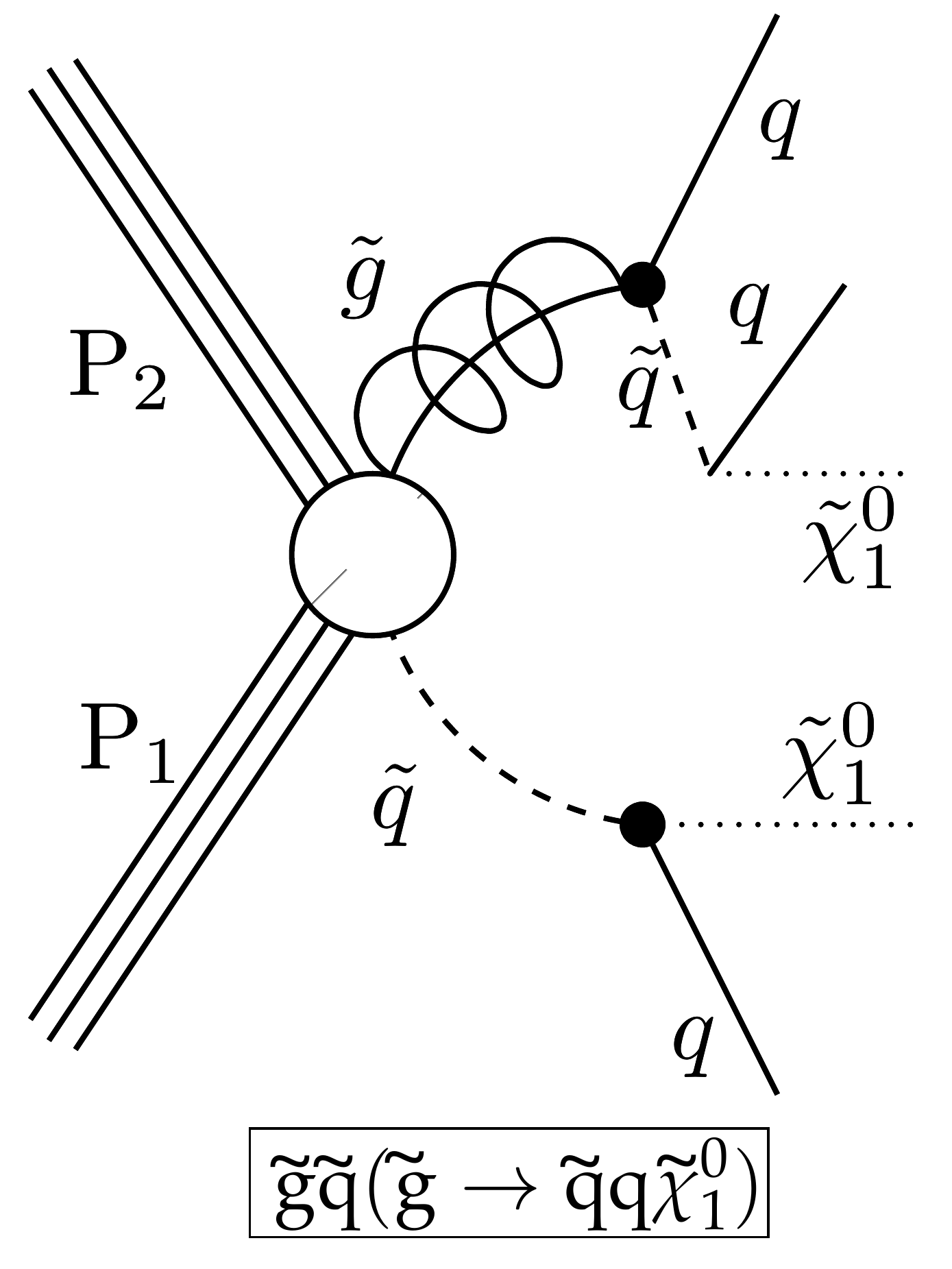}\hspace*{\fill}
\includegraphics[width=0.20\textwidth]{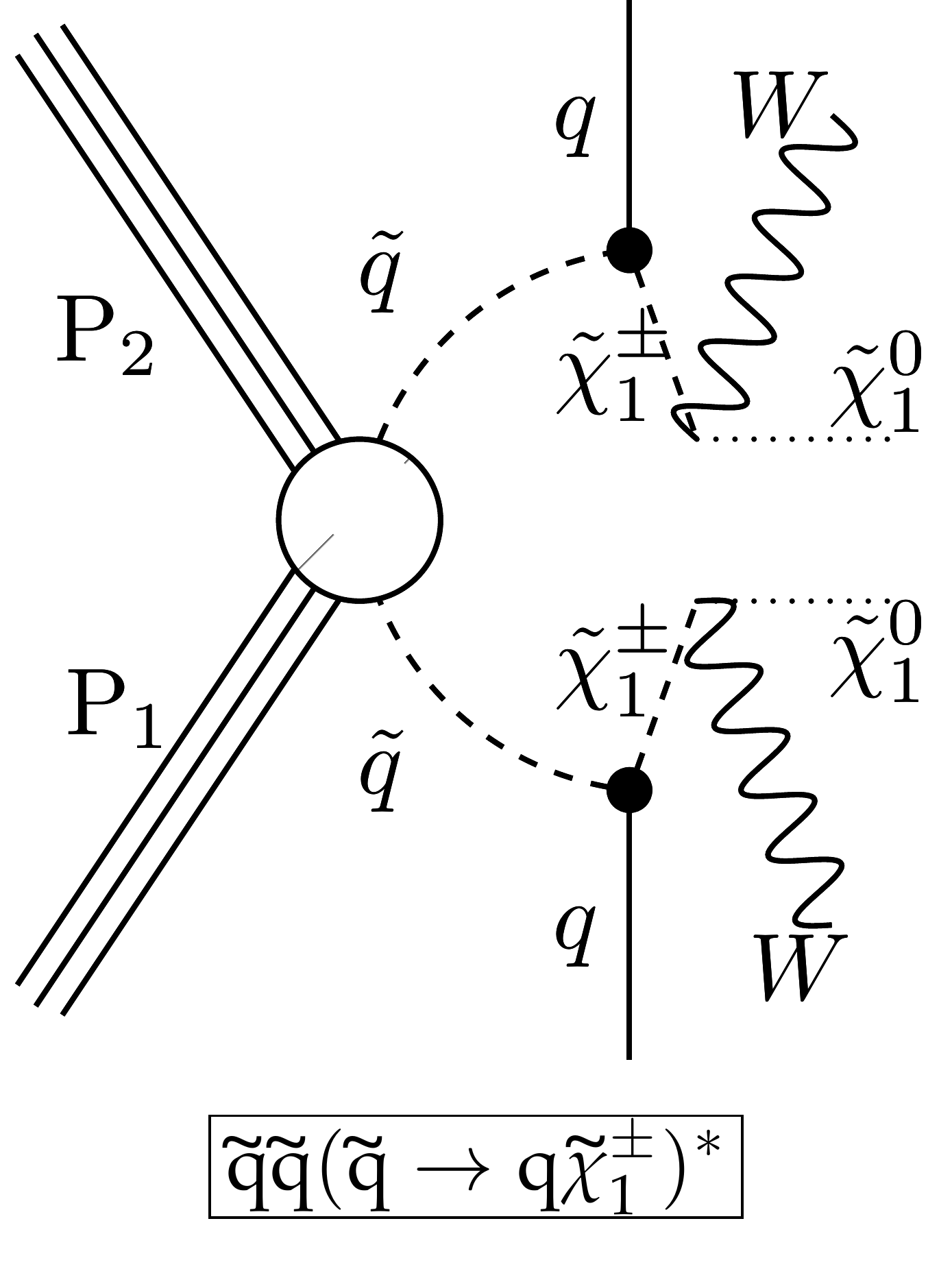}\hspace*{\fill}
\includegraphics[width=0.20\textwidth]{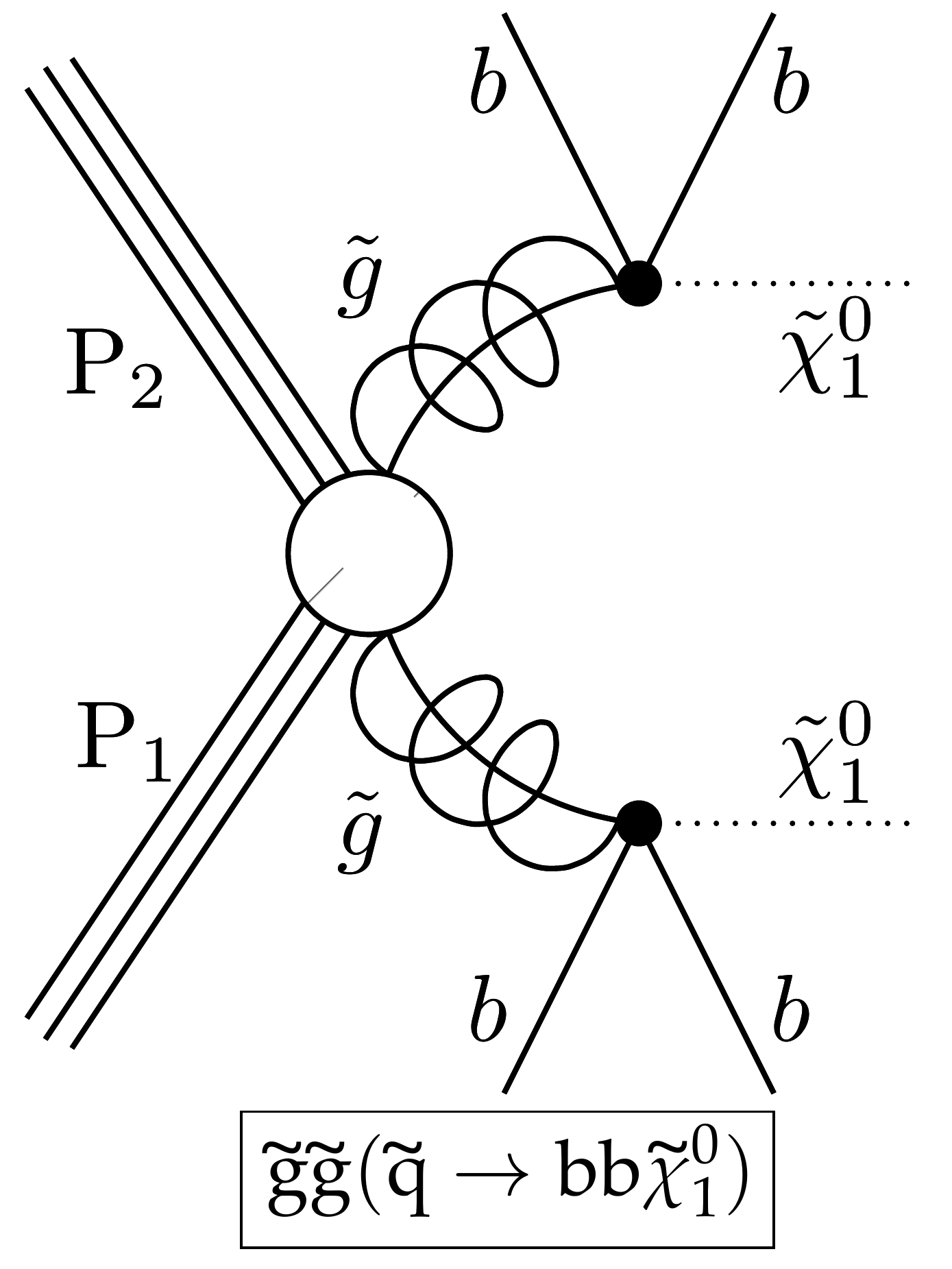}\hspace*{\fill}
\\[2em]
\hspace*{\fill}
\includegraphics[width=0.20\textwidth]{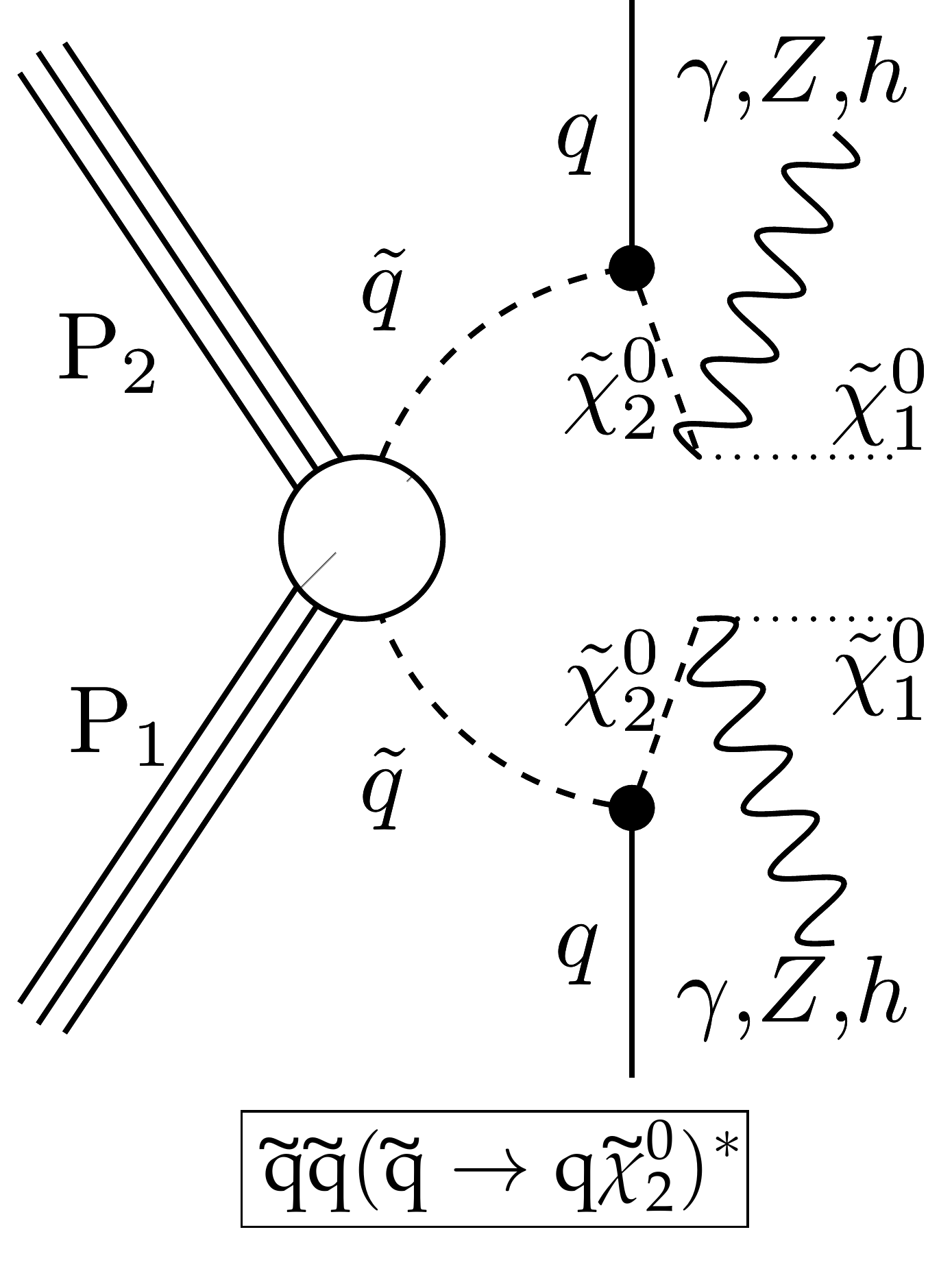}\hspace*{\fill}
\includegraphics[width=0.20\textwidth]{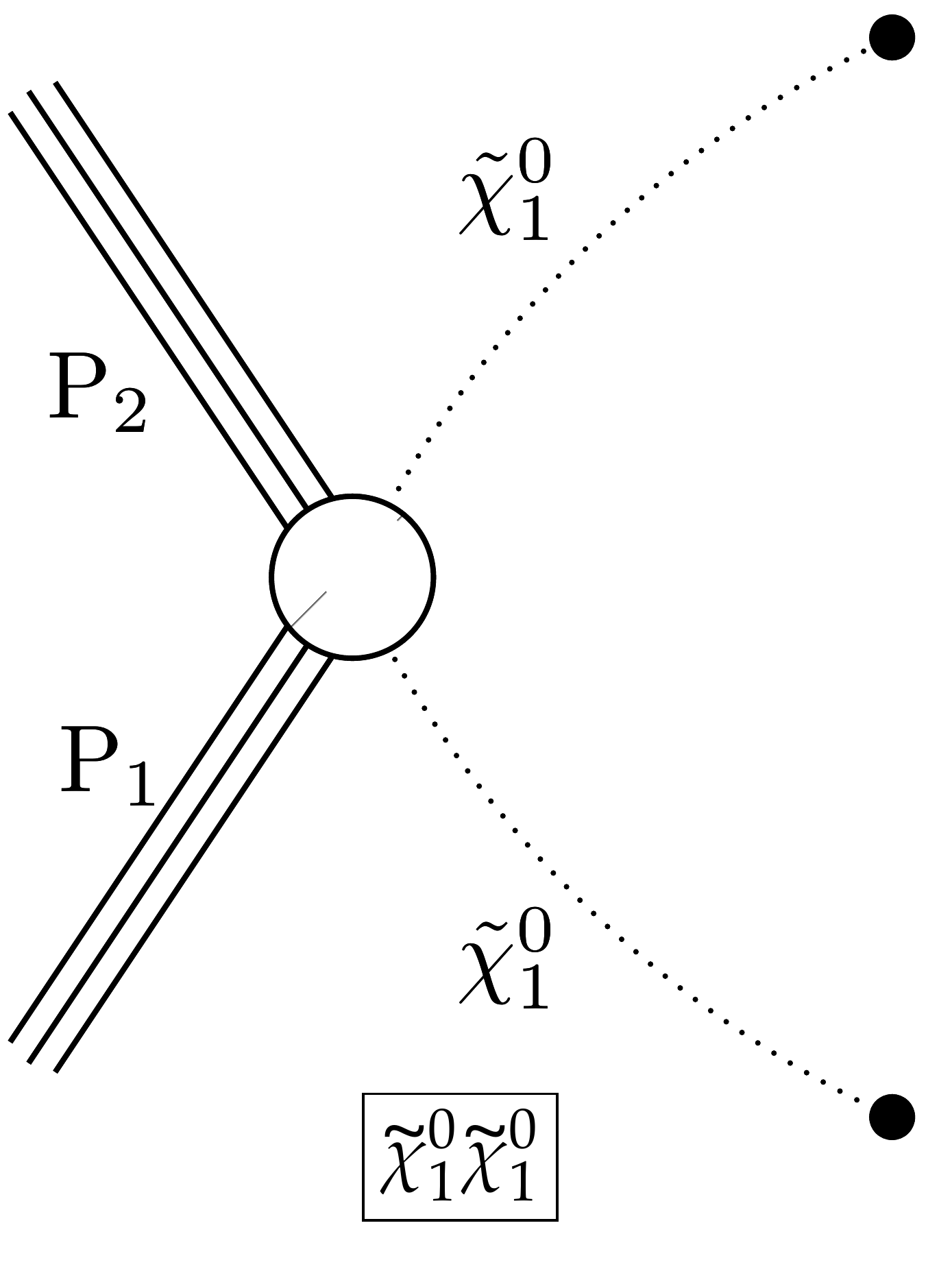}\hspace*{\fill}
\includegraphics[width=0.20\textwidth]{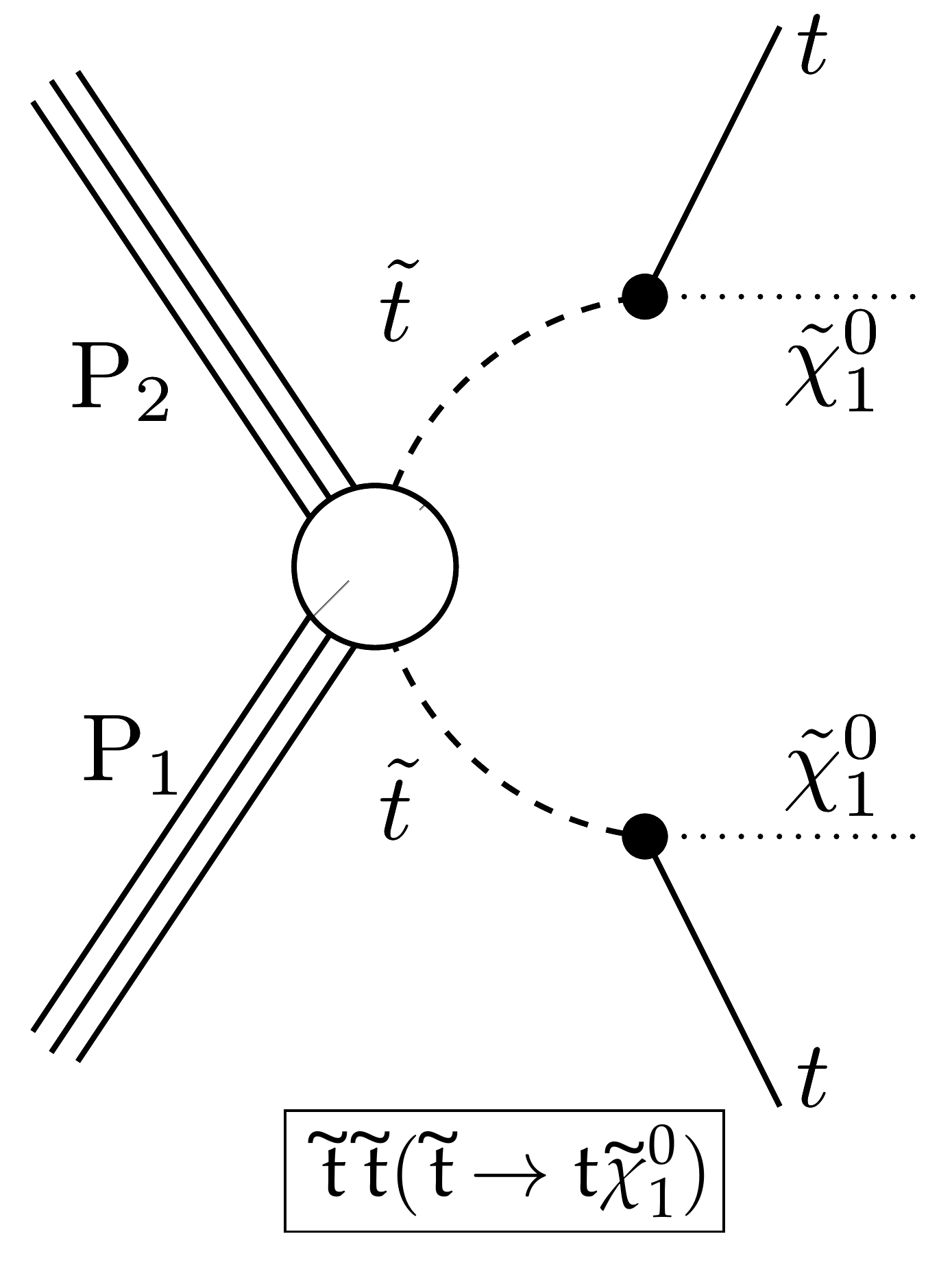}\hspace*{\fill}
\includegraphics[width=0.20\textwidth]{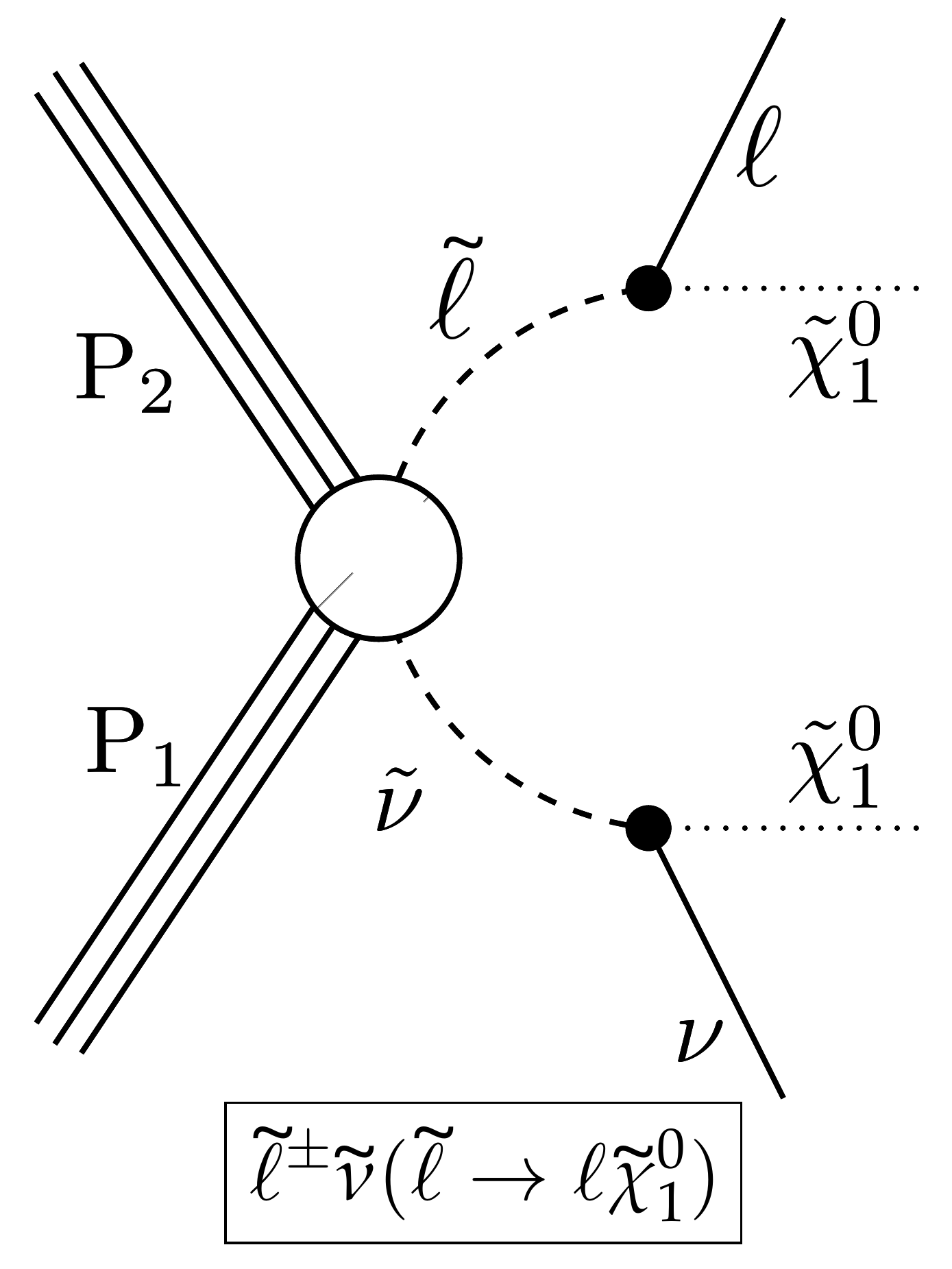}\hspace*{\fill}
\\
\caption{The twelve most common principal processes in the pMSSM, listed in order
of their frequency before the constraints of the CMS searches. Both on-shell and off-shell states are included.
 Indices of particle charge, flavor, and chirality are ignored in the construction, with the exception of the flavor of the third-generation squarks and quarks. Asterisks in the labels indicate where process names involving long decay chains have been abbreviated.}
\label{fig:diagrams1}
\end{figure*}

For the decomposition, signal events are analyzed at the generator
level for each model point, and the pair of SUSY particles most frequently produced directly
from the proton-proton interaction is taken as the production
mode for that model point. Then the principal (dominant) process for
that point is built as a tree diagram starting from the pair of SUSY mother
particles and following the decay modes with the highest branching
fractions until endpoints consisting of only SM particles
and LSPs are reached. Indices of particle charge, flavor,  and chirality are ignored in the
construction, with the exception of the flavor of the third-generation squarks and quarks.
Over 100 distinct principal processes are found
among the total 7200 studied points, of which the first twelve
are listed in Fig.~\ref{fig:diagrams1}. Many
of the principal processes are seen to correspond to common SMS scenarios, while others
depict more unusual scenarios with long decay chains.

\begin{figure*}[h]
  \centering
 \includegraphics[width=0.62\textwidth]{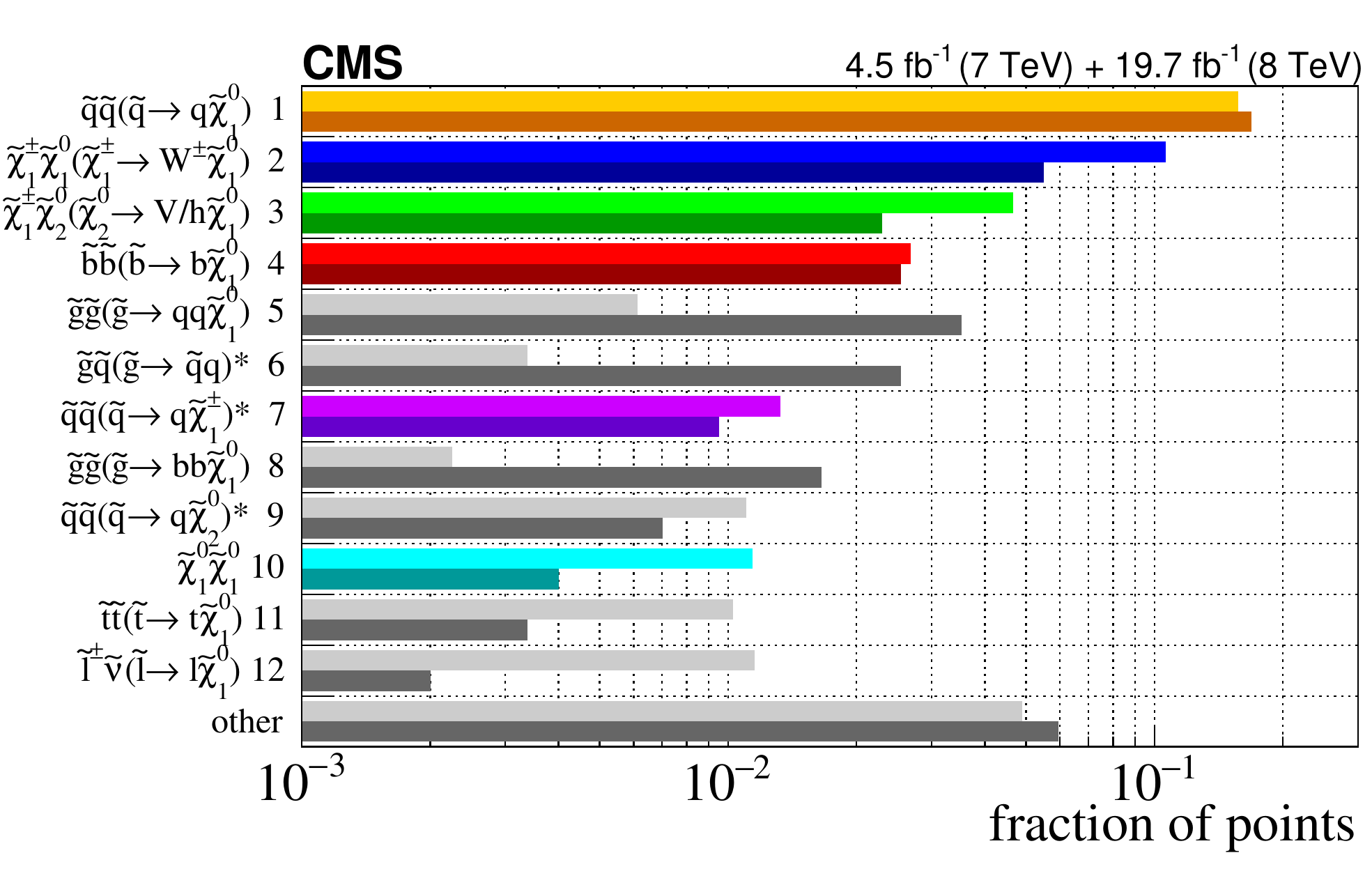}
 \includegraphics[width=0.35\textwidth]{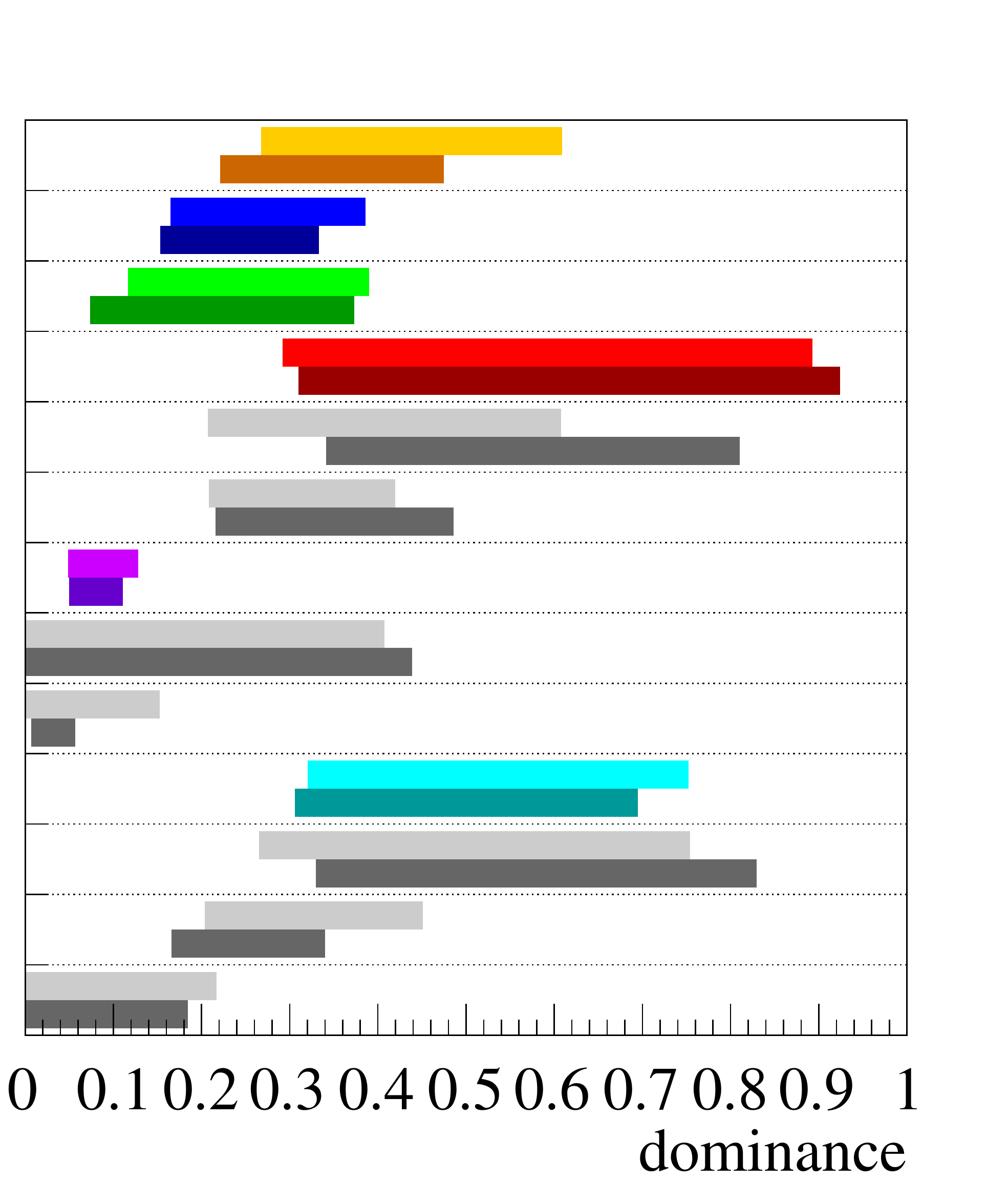}
    \caption{The left plot shows the fraction of excluded
  (dark) and nonexcluded (light) points out of all considered points,
  by principal process. Color is assigned to the processes that are most common after the constraints of the CMS searches, which are selected for further study. The dominance
  of principal processes, as defined in Eq.~\ref{eq:dominance}, is given in the right plot, where the bands show the
  RMS range of the dominance. }
    \label{fig:TopoHisto}
\end{figure*}

The distribution of principal processes for excluded and nonexcluded
points is given in Fig.~\ref{fig:TopoHisto} (left).  It is seen that processes
involving direct gluino production (5 and 8) are excluded with
a much higher
frequency than they survive, and those with EW
gaugino production (2, 3, and 10) survive with a higher frequency than they are
excluded. Processes with first-generation squark production (1 and 7) survive
and are excluded at similar rates, and processes with
slepton production (12) have exceptionally high survival rates.
These trends are likely attributable to the difference in the
production cross section between colored and noncolored particles for
a given SUSY mass scale.
The overflow bin (other), which contains many principal
processes, including modes of colored and noncolored particle
production, indicates a survival rate approximately equal to the
exclusion rate.
The dominance is defined for each model point
as the ratio of the cross section of the principal process to the total SUSY
production cross section at 8\TeV,
 \begin{equation}
  \text{dominance} \equiv \sigma^{8\TeV}_\text{principal}/\sigma^{8\TeV}_\text{tot},
  \label{eq:dominance}
\end{equation}
and is shown in Fig.~\ref{fig:TopoHisto} (right). Most values of the dominance are in the
range 0.05\textendash0.60. The excluded and nonexcluded values for the
dominance are seen to agree within the RMS of the distributions, indicating that the
presence of multiple event signatures within a single model
hypothesis does not
significantly impact our ability to exclude such a model point.

Dedicated searches exist that correspond to some of the most frequent principal processes, indicating areas where the SMS approach is likely well optimized. For example, points with principal processes 1, $\PSQ\PSQ(\PSQ \to\PQq\PSGczDo)$, and 4, $\PSQb\PSQb(\PSQb \to\PQb\PSGczDo)$, enjoy searches that target these processes explicitly.   A few principal processes have not been explicitly targeted by the host of CMS SUSY searches, including processes 2, $\PSGcpmDo\PSGcz(\PSGcpmDo \to\PW^{\pm}\PSGczDo)$, and 3, $\PSGcpmDo\PSGczDt(\PSGc \to{\rm V/h}\PSGczDo)$, the asymmetric EW gaugino production modes. New searches that target these or the other processes with insufficient coverage may serve to broaden the overall sensitivity to the pMSSM.

Next, we characterize the nonexcluded model space by the predicted
final states to shed light on what signatures may serve to target the nonexcluded points in Run 2.  We define a set of loose baseline physics objects and event
variables, at the generator level, as follows:
\begin{itemize}
\item Leptons: electrons, muons, or taus having a transverse momentum \pt greater than 5\GeV and an isolation less than 0.2. Here, isolation  =
$[(\Sigma_{i}{\pt}_i) - \pt]/\Sigma_{i}{\pt}_i$,
where the sums run over all detector-visible particles $i$ within a
$\Delta$R cone of 0.5 around the object, with $\Delta$R = $\sqrt{\smash[b]{(\Delta\eta)^2+(\Delta\phi)^2}}$, where $\eta$ is the pseudorapidity and $\phi$ is the azimuthal angle in radians;
\item Jets: particles clustered with the anti-\kt jet algorithm \cite{Cacciari:2008gp} with distance parameter 0.5. The jets are required to have a \pt greater than 20\GeV;
\item $\cPqb$-jets: jets matched to a b hadron within a $\Delta$R of 0.5;
\item \MET{}: the missing transverse energy, calculated as the magnitude
  of the vector sum of the transverse momenta of
 visible particles with $\pt > 5$\GeV;
\item \HT{}: the scalar sum of the \pt of the jets with
  a $\pt> 50$\GeV.
\end{itemize}

\begin{figure}[h]
  \centering
  \includegraphics[width=0.95\linewidth]{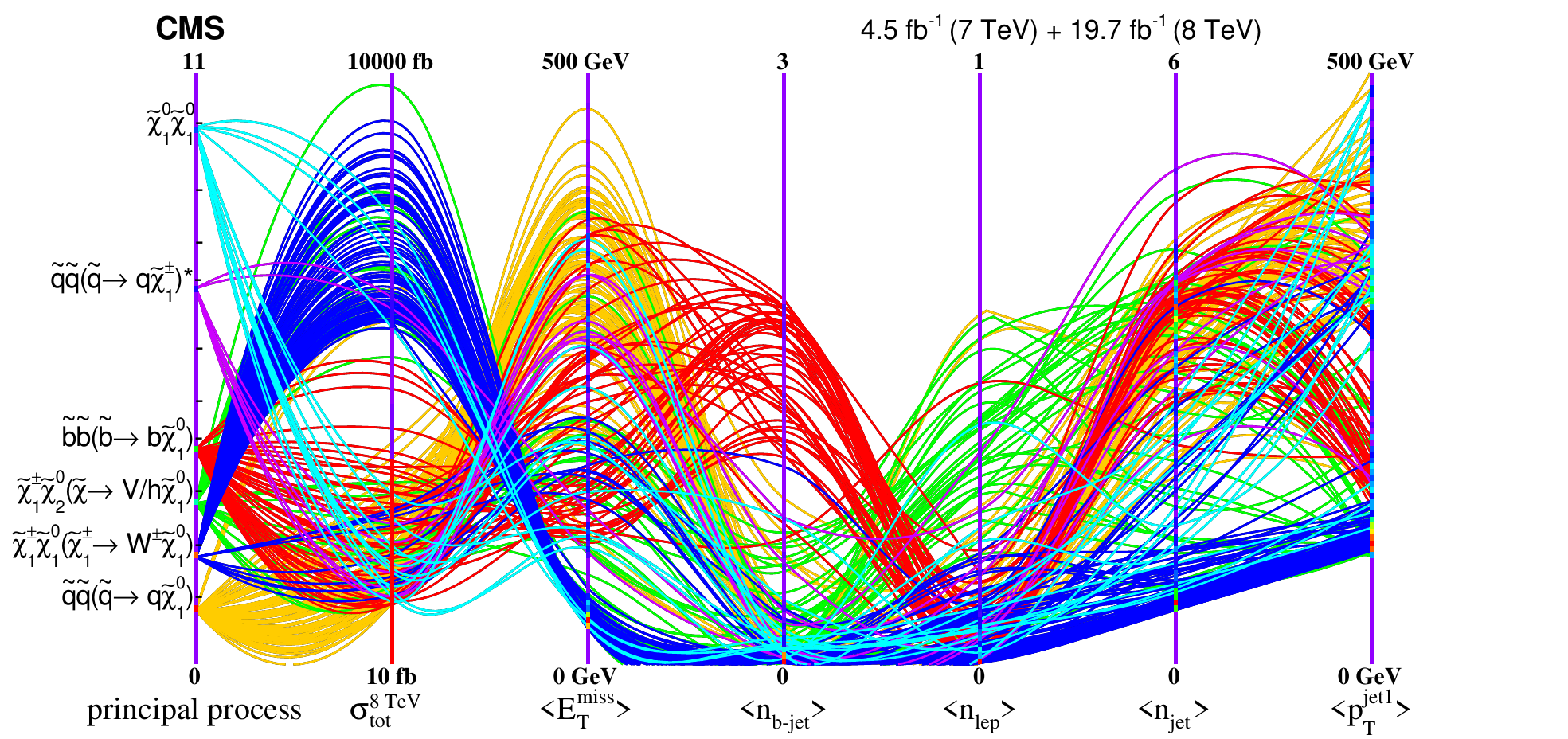}
    \caption{A parallel coordinates plot showing several hundred
      selected nonexcluded model points for the six most common
      principal processes, with seven key properties.
      From the left, the selected properties are: the principal
      process, the 8\TeV signal production cross section (in
      $\log_{10}$ scale), the average value of the \MET{}, the average number of
      $\cPqb$-jets, leptons, and jets, and finally, the average \pt
      momentum of the leading jet. Color is assigned based on the
      principal process. Orange codes for process 1, blue for
      process 2, green for 3, red for 4, violet for 7, and
      cyan for 10. The range of each axis is given at its lower and upper end. Lines
      arching toward higher vertical positions typically indicate more
      ``discoverable'' scenarios. }
    \label{fig:parcor}
\end{figure}

We use a parallel coordinates visualization technique that enables the
display of multiple dimensions. In Fig.~\ref{fig:parcor},
we show nonexcluded points corresponding to the six selected
principal processes (those denoted by color in Fig.~\ref{fig:parcor}).
Vertical axes are chosen to represent meaningful properties of the model
points, and each model point is represented as a curved line traversing the plot from left to right, intersecting each axis at the parameter value taken by the model point. The curvature of the lines is added to help distinguish between similar pMSSM points, but the trajectories of the lines between the axes do not carry physical information. A number of distinct scenarios are seen to have survived the CMS
analyses.  A minimum threshold of 20\unit{fb} has been applied to the 8\TeV signal
cross sections to limit the scope to those
points that could potentially still be probed with the Run 1 data set
using an expanded set of analyses and techniques.

The nonexcluded points associated with principal processes 1, $\PSQ\PSQ(\PSQ \to\PQq\PSGczDo)$, and 4, $\PSQb\PSQb(\PSQb \to\PQb\PSGczDo)$, are seen to give rise to large average \MET{}, jet multiplicities between 2 and 4, and moderate to low cross sections due the the large masses of the squarks. Given the
higher cross sections in Run 2, these high \MET{} scenarios will become increasingly more accessible.

Model points with principal processes 2, $\PSGcpmDo\PSGcz(\PSGcpmDo \to\PW^{\pm}\PSGczDo)$, and 3, $\PSGcpmDo\PSGczDt(\PSGc \to{\rm V/h}\PSGczDo)$, typically predict large cross sections, in the range between 100\unit{fb} and 1\unit{pb},
but a limited number of physical observables with discriminating
power, primarily due to compression in the mass spectrum between the
LSP and the other EW gauginos. These points peak low in the average
multiplicity of jets, leptons, and in average \MET{}. They could
potentially be probed with searches that involve events with initial
state radiation and soft boson decay products that are aligned with
the \MET{}. We note that process 2 is the principal process that
characterizes the pMSSM point with the largest $Z$-significance,
3.6. This model point corresponds to a $\PSGczDo$ mass of
around 200\GeV and a mass difference between the lightest chargino and
LSP of about 3\GeV, which are properties of many model points that survived the CMS analyses.

Points with principal processes 3, $\PSGcpmDo\PSGczDt(\PSGc \to{\rm V/h}\PSGczDo)$, and 10, $\PSGczDo\PSGczDo$, tend to follow the trend profiled
by process 2, $\PSGcpmDo\PSGcz(\PSGcpmDo \to\PW^{\pm}\PSGczDo)$, differing primarily in the lepton multiplicity and, in the case of at least one lepton, in the average \pt of the highest-\pt lepton (leading lepton). The close resemblance of
processes 10 and 2  is mostly due to the fact that the mass
difference between the $\PSGcpmDo$ and
$\PSGczDo$ is frequently very small (less than 3\GeV), causing the ensuing off-shell W
boson of process 2 to produce undetectably soft objects.

Points with principal processes 5, $\PSg\PSg(\PSg \to{\PQq\PQq}\PSGczDo)$, and 6, $\PSg\PSQ(\PSg \to\PSQ\PQq\PSGczDo)$, the most frequent modes
involving gluinos, are not highlighted in Fig.~\ref{fig:parcor}, since their frequency among nonexcluded points is relatively small. We note that several of the nonexcluded models with
very light gluino masses (less than 700\GeV) correspond to principal process 6, with mass differences between the $\PSg$ and LSP that range around 100\GeV. Sensitivity to these model points may be possible by considering final states with three or fewer jets and \MET{} thresholds that are lower than typically applied.

Points with principal process 7, $\PSQ\PSQ(\PSQ \to\PQq\PSGcpmDo)^{\text{*}}$, do not display distinct trends in the properties selected, which is partly due to these points having a
low dominance of around 0.1. Such model points have a diverse set of
secondary processes, which are not directly examined here.

A general observation about the model points in Fig.~\ref{fig:parcor}
is the significant anticorrelation of observables, which
manifests as the criss-crossing of lines between the axes. For example,
model points with very high average \MET{} tend to have very low cross sections,
and vice versa. This is a consequence of the fact that, no significant excess of events
having been observed in data, the surviving model points are those with
very few experimentally accessible observables; otherwise they would
have been excluded.

We note that the surviving pMSSM point with the lowest value of $m_{\PSg}$ (about 600\GeV) is not characterized by one of the twelve most frequent principle processes discussed above, but by processes involving gluino pair production, where each gluino decays into two light-flavor quarks and an EW gaugino, and where the EW gaugino subsequently decays into a vector boson and an LSP. The mass difference between the intermediate gaugino and the LSP is about 5\GeV, which, in most events, does not leave enough energy for the vector bosons to have decay products that are reconstructed in the detector.

\begin{figure}[htb]
  \centering
  \includegraphics[width=\cmsFigWidth]{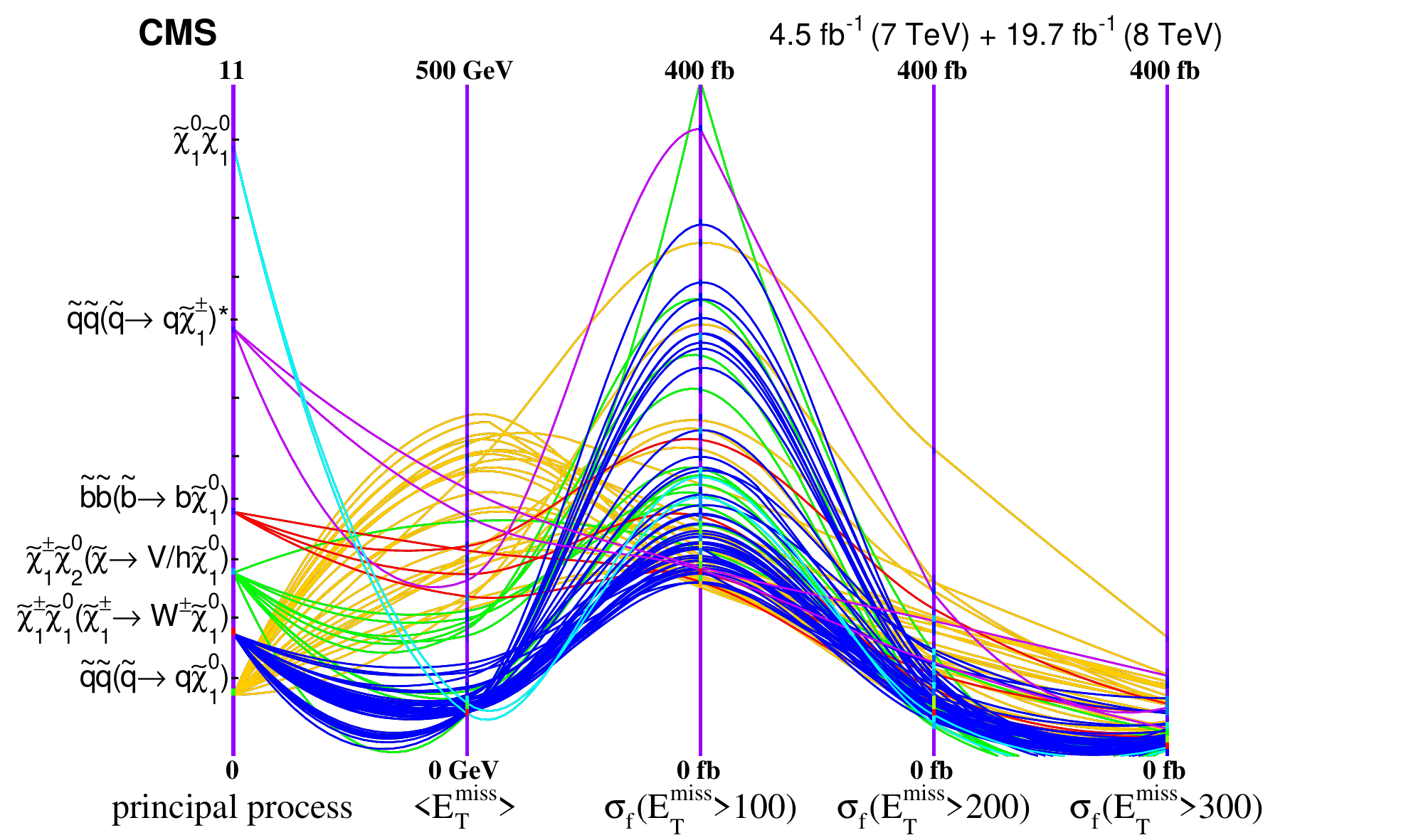}
    \caption{A parallel coordinates plot of the nonexcluded pMSSM points
      with the axes set as the principal process, the average \MET{} (in \GeV), and the
    fiducial cross section (in linear scale) for various thresholds on the
    \MET{}. All nonexcluded points corresponding to processes 1, 2, 3,
    4, 7, and 10 that have a fiducial cross section for $\MET>100$\GeV greater than 100\unit{fb}
    are shown. Color is
    assigned to values of the principal process in the same manner as
    in Fig.~\ref{fig:parcor}.}
    \label{fig:parcorMET}
\end{figure}

\begin{figure}[htb]
  \centering
         \includegraphics[width=\cmsFigWidth]{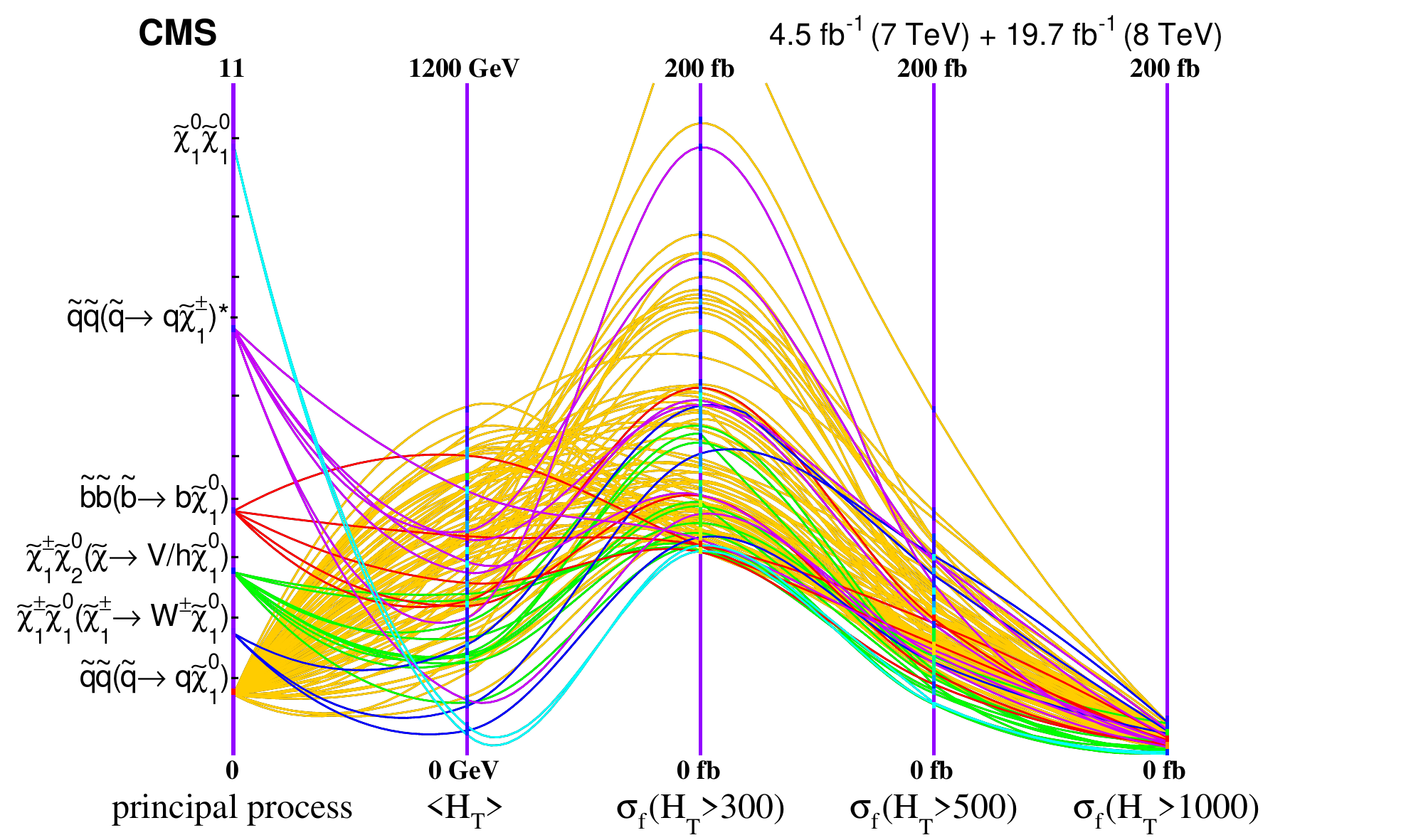}
    \caption{A parallel coordinates plot of the nonexcluded pMSSM
      points with the axes set as the principal process, the average
      \HT{} (in \GeV), and the
    fiducial cross section (in linear scale) for various thresholds on the
    \HT{}. All nonexcluded points corresponding to processes 1, 2, 3,
    4, 7, and 10 that have a fiducial cross section for $\HT >
300$\GeV greater than 300\unit{fb} are shown. Color is assigned to
    values of the principal process in the same manner as in
    Fig.~\ref{fig:parcor}. }
    \label{fig:parcorHT}
\end{figure}

\begin{figure}[htb]
  \centering
         \includegraphics[width=\cmsFigWidth]{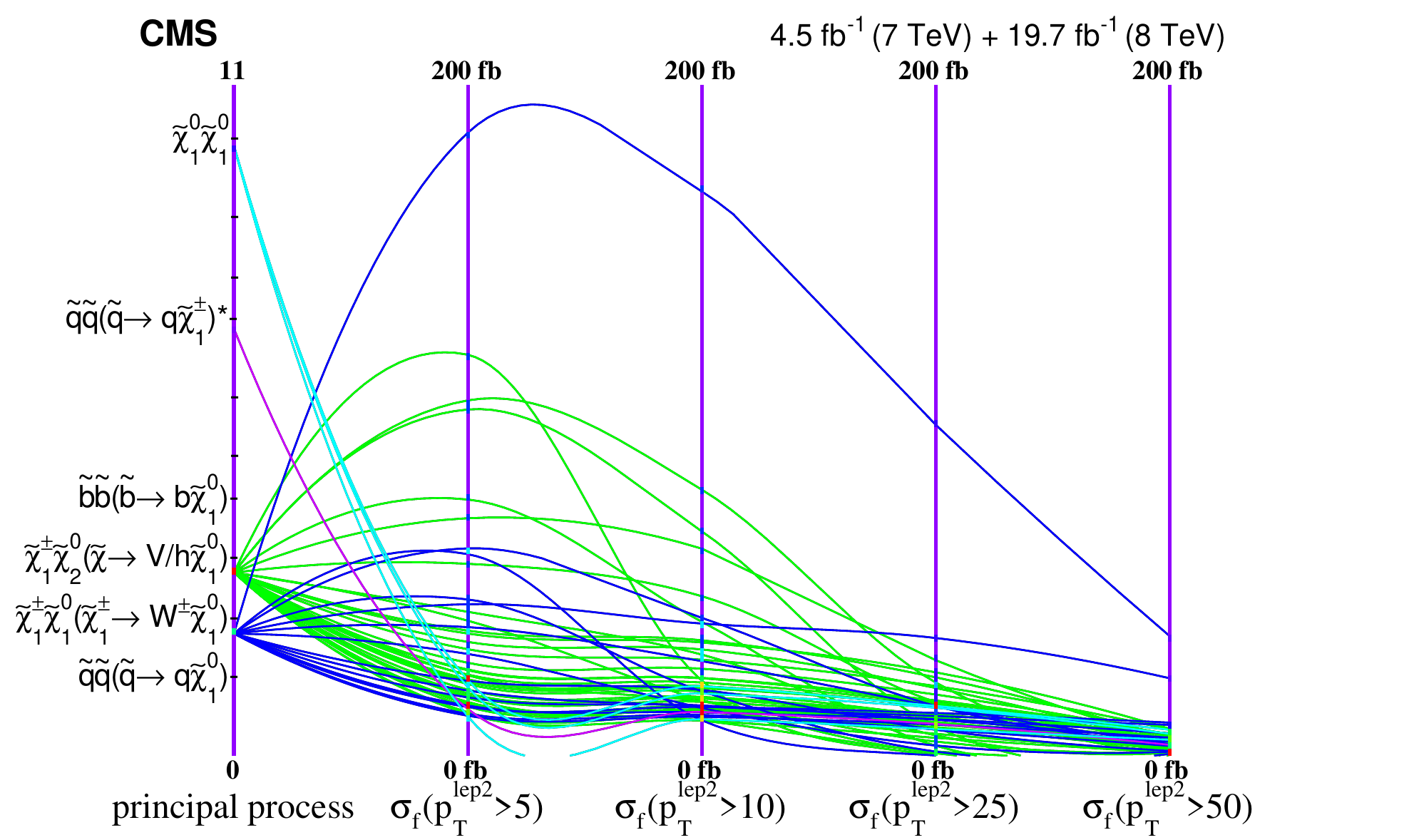}
    \caption{A parallel coordinates plot of the nonexcluded pMSSM
      points with the axes set as the principal process and the
    fiducial cross section (in linear scale) for various thresholds on the
    sub-leading lepton \pt (in\GeV). All nonexcluded points corresponding to processes 1, 2, 3,
    4, 7, and 10 that have a fiducial cross section for $\pt^{\text{lep2}}>
5$\GeV greater than 30\unit{fb} are shown. Color is
    assigned to values of the principal process in the same manner as
    in Fig.~\ref{fig:parcor}. }
    \label{fig:parcorLEP2}
\end{figure}

With over 50\%
of all nonexcluded points corresponding to cross sections of
greater than 10\unit{fb}, it is critical to further examine why these points were not accessed in Run 1.
We attempt to gain an
understanding by further characterizing the signal, evaluating fiducial cross sections corresponding to a
range of final-state observables. The fiducial cross section
$\sigma_{\text{f}}$ of a final-state is defined for each model point as
 \begin{equation}
   \sigma_{\text{f}} = \sigma_{\text{tot}}^{8\TeV}A,
\end{equation}
where {\it A} is the acceptance times signal efficiency
computed as the fraction of simulated signal events
passing a set of event-level criteria.
We examine a set of final-state observables that loosely correspond to trigger thresholds or
signal regions of the examined searches. Figures
\ref{fig:parcorMET}-\ref{fig:parcorLEP2} show the impact of adjusting
various thresholds on the fiducial cross sections of nonexcluded points.

Some principal processes can be associated with large fiducial
cross sections, depending on the final state considered. For example, points with mostly
first-generation squark production give rise to large
fiducial cross sections for events with high \HT{}, resulting in Fig.
\ref{fig:parcorHT} showing mostly orange-colored points; and points with production involving EW gauginos give rise to substantial
fiducial cross sections for events with a high multiplicity of soft leptons, which
explains the unaccompanied blue and green lines in Fig.~\ref{fig:parcorLEP2}.  Somewhat striking is the behavior of the \MET{} fiducial cross section
(Fig.~\ref{fig:parcorMET}), which can increase rapidly (by up to a
factor of ten) as the threshold
is relaxed from 200 to 100 \GeV. It is apparent that many of the
nonexcluded regions are not accessible with thresholds of 200\GeV,
a common criterion applied offline to achieve full efficiency
with the triggers. The fiducial cross section decreases noticeably as the
threshold is further increased from 200 to 300 \GeV. Similar behavior is seen for the \HT{} fiducial cross section
(Fig.~\ref{fig:parcorHT}). Fiducial cross
sections are quite large for these final states when a
threshold of 300$\GeV$ is applied, but fall off substantially for higher
thresholds.

Of course, a loosening of the object thresholds would increase the background yield as well as signal yield. Thorough analysis of specific backgrounds will be necessary
to select optimal values for kinematic thresholds and other analysis techniques to probe the most difficult points. However, the lesson that nonexcluded pMSSM models have large cross sections in background-rich kinematic regions is an open invitation for the development of new techniques that improve signal to background discrimination and background modeling.

\section{Summary}
\label{sec:concl}

The impact of a representative set of the 7 and 8\TeV CMS SUSY searches
on a potentially accessible subspace of the minimal supersymmetric
standard model (pMSSM) has been investigated. The subspace of the pMSSM is defined by restricting the ranges of the 19 pMSSM parameters to values that are either physically motivated or that correspond to models that are potentially accessible in the long-term LHC program. An additional restriction is imposed that the lightest chargino decay promptly or with a lifetime that leads to at most a short decay length in the detector.
The set of searches,
taken individually and in combination, include those with all-hadronic final
states, like-sign and opposite-sign
charged leptons, and multiple leptons in configurations sensitive to
electroweak production of superpartner particles.
They are found to exclude all analyzed pMSSM points with a gluino
mass less than 500\GeV (approximately 250 of the
7200 sample points),
and 98\% of scenarios in which the lightest colored supersymmetric particle is less than
300\GeV.
While the sensitivity of searches to top squarks extends up to $m_{\tilde{\text{t}}_1}\approx 700$\GeV, the overall impact on the top squark mass is small because
the region of highest sensitivity, $m_{\tilde{\text{t}}_1}\lesssim500\GeV$, is
already suppressed by the results of previous experiments, such as the measurement of
the $\PQb\to\PQs\PGg$
branching fraction.  Neutralino and chargino masses less than
$300\GeV$ are significantly disfavored, but not ruled out, by
the CMS data.
Measurements of the Higgs boson mass and signal strengths are included
in this study, but add little to the model constraints.

Approximately half of this potentially-accessible subspace of the pMSSM is excluded by the CMS data. Of the surviving points,
about half have cross sections greater than 10\unit{fb}, and some have
cross sections greater than 1\unit{pb}. Most high cross section points
  correspond to electroweak gaugino production with mass splittings between the
  second-lightest and the lightest SUSY particle less than 3 \GeV. Nonexcluded model points
  with low-mass gluinos correspond to processes involving
  intermediate electroweak gauginos that are nearly degenerate with the lightest SUSY particle. The
surviving points evade the experimental constraints largely because they
overlap with the kinematical parameter space of more copiously produced
standard model processes. Some of these may be probed by future searches
that target the nonexcluded processes detailed in Section
  \ref{sec:unexplored}, benefiting as well from the higher energy and luminosity of the
LHC.

\begin{acknowledgments}

\hyphenation{Bundes-ministerium Forschungs-gemeinschaft Forschungs-zentren} We congratulate our colleagues in the CERN accelerator departments for the excellent performance of the LHC and thank the technical and administrative staffs at CERN and at other CMS institutes for their contributions to the success of the CMS effort. In addition, we gratefully acknowledge the computing centers and personnel of the Worldwide LHC Computing Grid for delivering so effectively the computing infrastructure essential to our analyses. Finally, we acknowledge the enduring support for the construction and operation of the LHC and the CMS detector provided by the following funding agencies: the Austrian Federal Ministry of Science, Research and Economy and the Austrian Science Fund; the Belgian Fonds de la Recherche Scientifique, and Fonds voor Wetenschappelijk Onderzoek; the Brazilian Funding Agencies (CNPq, CAPES, FAPERJ, and FAPESP); the Bulgarian Ministry of Education and Science; CERN; the Chinese Academy of Sciences, Ministry of Science and Technology, and National Natural Science Foundation of China; the Colombian Funding Agency (COLCIENCIAS); the Croatian Ministry of Science, Education and Sport, and the Croatian Science Foundation; the Research Promotion Foundation, Cyprus; the Secretariat for Higher Education, Science, Technology and Innovation, Ecuador; the Ministry of Education and Research, Estonian Research Council via IUT23-4 and IUT23-6 and European Regional Development Fund, Estonia; the Academy of Finland, Finnish Ministry of Education and Culture, and Helsinki Institute of Physics; the Institut National de Physique Nucl\'eaire et de Physique des Particules~/~CNRS, and Commissariat \`a l'\'Energie Atomique et aux \'Energies Alternatives~/~CEA, France; the Bundesministerium f\"ur Bildung und Forschung, Deutsche Forschungsgemeinschaft, and Helmholtz-Gemeinschaft Deutscher Forschungszentren, Germany; the General Secretariat for Research and Technology, Greece; the National Scientific Research Foundation, and National Innovation Office, Hungary; the Department of Atomic Energy and the Department of Science and Technology, India; the Institute for Studies in Theoretical Physics and Mathematics, Iran; the Science Foundation, Ireland; the Istituto Nazionale di Fisica Nucleare, Italy; the Ministry of Science, ICT and Future Planning, and National Research Foundation (NRF), Republic of Korea; the Lithuanian Academy of Sciences; the Ministry of Education, and University of Malaya (Malaysia); the Mexican Funding Agencies (BUAP, CINVESTAV, CONACYT, LNS, SEP, and UASLP-FAI); the Ministry of Business, Innovation and Employment, New Zealand; the Pakistan Atomic Energy Commission; the Ministry of Science and Higher Education and the National Science Centre, Poland; the Funda\c{c}\~ao para a Ci\^encia e a Tecnologia, Portugal; JINR, Dubna; the Ministry of Education and Science of the Russian Federation, the Federal Agency of Atomic Energy of the Russian Federation, Russian Academy of Sciences, and the Russian Foundation for Basic Research; the Ministry of Education, Science and Technological Development of Serbia; the Secretar\'{\i}a de Estado de Investigaci\'on, Desarrollo e Innovaci\'on and Programa Consolider-Ingenio 2010, Spain; the Swiss Funding Agencies (ETH Board, ETH Zurich, PSI, SNF, UniZH, Canton Zurich, and SER); the Ministry of Science and Technology, Taipei; the Thailand Center of Excellence in Physics, the Institute for the Promotion of Teaching Science and Technology of Thailand, Special Task Force for Activating Research and the National Science and Technology Development Agency of Thailand; the Scientific and Technical Research Council of Turkey, and Turkish Atomic Energy Authority; the National Academy of Sciences of Ukraine, and State Fund for Fundamental Researches, Ukraine; the Science and Technology Facilities Council, UK; the US Department of Energy, and the US National Science Foundation.

Individuals have received support from the Marie-Curie program and the European Research Council and EPLANET (European Union); the Leventis Foundation; the A. P. Sloan Foundation; the Alexander von Humboldt Foundation; the Belgian Federal Science Policy Office; the Fonds pour la Formation \`a la Recherche dans l'Industrie et dans l'Agriculture (FRIA-Belgium); the Agentschap voor Innovatie door Wetenschap en Technologie (IWT-Belgium); the Ministry of Education, Youth and Sports (MEYS) of the Czech Republic; the Council of Science and Industrial Research, India; the HOMING PLUS program of the Foundation for Polish Science, cofinanced from European Union, Regional Development Fund; the Mobility Plus program of the Ministry of Science and Higher Education (Poland); the OPUS program, contract Sonata-bis DEC-2012/07/E/ST2/01406 of the National Science Center (Poland); the Thalis and Aristeia programs cofinanced by EU-ESF and the Greek NSRF; the National Priorities Research Program by Qatar National Research Fund; the Programa Clar\'in-COFUND del Principado de Asturias; the Rachadapisek Sompot Fund for Postdoctoral Fellowship, Chulalongkorn University (Thailand); the Chulalongkorn Academic into Its 2nd Century Project Advancement Project (Thailand); and the Welch Foundation, contract C-1845.

\end{acknowledgments}

\bibliography{auto_generated}
\cleardoublepage \appendix\section{The CMS Collaboration \label{app:collab}}\begin{sloppypar}\hyphenpenalty=5000\widowpenalty=500\clubpenalty=5000\textbf{Yerevan Physics Institute,  Yerevan,  Armenia}\\*[0pt]
V.~Khachatryan, A.M.~Sirunyan, A.~Tumasyan
\vskip\cmsinstskip
\textbf{Institut f\"{u}r Hochenergiephysik der OeAW,  Wien,  Austria}\\*[0pt]
W.~Adam, E.~Asilar, T.~Bergauer, J.~Brandstetter, E.~Brondolin, M.~Dragicevic, J.~Er\"{o}, M.~Flechl, M.~Friedl, R.~Fr\"{u}hwirth\cmsAuthorMark{1}, V.M.~Ghete, C.~Hartl, N.~H\"{o}rmann, J.~Hrubec, M.~Jeitler\cmsAuthorMark{1}, A.~K\"{o}nig, M.~Krammer\cmsAuthorMark{1}, I.~Kr\"{a}tschmer, D.~Liko, T.~Matsushita, I.~Mikulec, D.~Rabady, N.~Rad, B.~Rahbaran, H.~Rohringer, J.~Schieck\cmsAuthorMark{1}, R.~Sch\"{o}fbeck, J.~Strauss, W.~Treberer-Treberspurg, W.~Waltenberger, C.-E.~Wulz\cmsAuthorMark{1}
% ...
\vskip\cmsinstskip
% ...
1:~~Also at Vienna University of Technology, Vienna, Austria\\
% ...

\end{sloppypar}
\end{document}